\documentclass[11pt]{cernrep}
\usepackage{graphicx}
\usepackage{floatflt}
\usepackage{color}
\usepackage{colortbl}

\newcommand{\Pt}{{P_t}}
\newcommand{\dphi}{\Delta\phi}
\newcommand{\phigj}{\phi_{(\gamma,~jet)}}

\newcommand{\Ptgj}{$\Pt^{\gamma}$ and $\Pt^{jet}$~~}
\newcommand{\ptgj}{$~\Pt^{\gamma}-\Pt^{jet}~$}

\newcommand{\la}{\langle}
\newcommand{\ra}{\rangle}
\newcommand{\gpj}{~``$\gamma+jet$''~}
\newcommand{\gpp}{~``$\gamma+parton$''~}
\newcommand{\rrr}{\to} 

\newcommand{\pth}{\hat{p}_{\perp}^{\;min}}
\newcommand{\Db}{\Pt(O\!+\!\eta>4.2)}
\newcommand{\Ptg}{\Pt^{\gamma}}
\newcommand{\ptg}{$\Pt^{\gamma}$}
\newcommand{\Fptgj}{(\Pt^{\gamma}\!-\!\Pt^{Jet})/\Pt^{\gamma}}

\newcommand{\coltab}{0.69}

\newcommand{\aaa}{\hspace*{.39cm}}

\newcommand{\hmm}{\hspace*{-1.3mm}}

\newcommand{\tg}{\tilde{\gamma}}

\newcommand{\Gvc}{\footnotesize{$(GeV/c)$} }

\newcommand{\gJ}{(\Pt^{\gamma}\!-\!\Pt^{J})/\Pt^{\gamma}}
\newcommand{\gpart}{(\Pt^{\gamma}\! - \! \Pt^{part})/\Pt^{\gamma}}
\newcommand{\Jpart}{(\Pt^{J}\! - \! \Pt^{part})/\Pt^{J}}

\newcommand{\sgmgj}{\sigma(Db[\gamma,J])}
\newcommand{\sgmgp}{\sigma(Db[\gamma,part])}

\newcommand{\lt}{\!<\!}
\newcommand{\gt}{\!>\!}

\suppressfloats[!]

\def\baselinestretch{1.0}

\begin{document}


\vskip2cm

%
\noindent                                                                      
{\bf \Large \gpj process application for setting the absolute  scale of \\[4pt]
\hspace*{15mm} jet energy and determining the gluon distribution \\[4pt]
\hspace*{59mm} at the Tevatron in Run II.}

\thispagestyle{empty}

\vskip3cm

\normalsize

\centerline{\sf D.V.~Bandurin, N.B.~Skachkov}
~\\[-3mm]

\centerline{\it Laboratory of Nuclear Problems}
~\\[-5mm]
\centerline{\it Joint Institute for Nuclear Research, Dubna, Russia}
~\\[15mm]


\vskip3cm
\hspace*{6cm}{\bf Abstract}\\[15pt]

The consequences of application of new set of criteria, proposed in our
previous works, for the improvement of a jet energy calibration accuracy with
the process ``$p\bar{p}\to\gamma+jet+X$''  at Tevatron and for
a reduction of background events contribution are studied.
The efficiencies of the used selection criteria are estimated.
The distributions of these events over \ptg ~and $\eta^{jet}$ are presented.
The features of \gpj events in the central calorimeter region of
the D0 detector ($|\eta|\lt0.7$) are investigated.
 
It is also shown that the samples of \gpj events, selected with the cuts used for
the jet energy calibration,
may have the statistics sufficient for determining the gluon distribution function of
a proton in the region of $2\times 10^{-3}\lt x\lt 1.0$ and the values of $Q^2$ by one order higher than
that reached in the experiments at HERA.
 
Monte Carlo events produced by the PYTHIA 5.7 generator are used for this aim.

\thispagestyle{empty}

\newpage

\setcounter{page}{1}

\section{INTRODUCTION.} 

Setting an absolute energy scale for a jet, detected mostly by hadronic and
electromagnetic calorimeters (HCAL and ECAL), is an important task for any 
$p\bar{p}$ or $pp$  collider experiment (see e.g. [1--8]). 

The main goal of this work is to find out the selection criteria for ``$p\bar{p}\to\gamma+jet+X$''
events (we shall use in what follows the abbreviation \gpj for them) that would lead to 
the most precise determination of the transverse momentum of a jet (i.e. $\Pt^{jet}$) 
via assigning a
photon $\Ptg$ to a signal produced by a jet. Our study is based on the \gpj events generated by 
using PYTHIA 5.7 \cite{PYT}. Their analysis was done on the ``particle level'' 
(in the terminology of [1]), i.e. without inclusion of detector effects.
The information provided by this generator is analyzed to track
starting from the parton level (where parton-photon balance is supposed to take place in a case of 
initial state radiation absence) all possible sources that may lead to the \ptgj disbalance 
in a final state. We use here the methods applied in \cite{9}--\cite{BKS_GLU}  
(see also \cite{DMS}) and \cite{GMS}, \cite{GMS_NN} for analogous task at LHC energy.
The corresponding cuts on physical variables,  introduced in \cite{9}--\cite{BKS_P5},
are applied here. Their efficiency is estimated at the particle level of simulation
at Tevatron energy with account of D0 detector geometry. 

We consider here the case of the Tevatron Run~II luminosity 
$L=10^{32}~ cm^{-2}s^{-1}$. It will be shown below that its value is quite sufficient 
for selecting  the event samples of large enough volume for application strict 
cuts as well as of new physical variables introduced in \cite{9}--\cite{BKS_P5}.

Section 2~ is a short introduction into the physics connected with the discussed problem.
General features of \gpj processes are presented here.
We review the possible sources of the $\Pt^{\gamma}$ and $\Pt^{jet}$ disbalance and the ways of
selecting those events where this disbalance has a minimal value
on the particle level.

In Section 3.1 we give the definitions  are given for the transverse momenta of
different physical objects 
that we suppose to be important for studying the physics connected with a jet calibration
procedure. Values of these transverse momenta 
enter into the $\Pt$-balance equation that reflects the
total $\Pt$ conservation law for the $p\bar{p}$-collision event as a whole.

Section 3.2 describes the criteria we have chosen to select \gpj events
for the jet energy  calibration procedure. The ``cluster'' (or mini--jet) suppression criterion 
($\Pt^{clust}_{CUT}$) which was formulated in an evident form in our previous publications 
\cite{9}--\cite{BKS_GLU} is used here
\footnote{We use here, as in \cite{BKS_P1}--\cite{BKS_GLU}, for most application the PYTHIA's
default jetfinder LUCELL  as well as UA1 taken from the CMS program of fast simulation CMSJET 
\cite{CMJ} for defining jets in an event.}. 
(Its important role for selection of events with a good balance of \Ptgj
will be illustrated in Sections 5--8.)
\footnote{The analogous  third jet cut thresholds 
$E_T^3$ (varying from 20 to 8 $GeV$) for improving a single jet energy resolution
in di-jet events were used in \cite{Bert}.}
These clusters have a physical meaning of a part 
of another new experimentally measurable quantity,
introduced in \cite{9}--\cite{BKS_GLU} for the first time, namely,
the sum of $\bf{\Pt}$ of those particles that are {\it out} of the \gpj system
(denoted as $\Pt^{out}$) and are {\it detectable}
in the whole pseudorapidity $\eta$ region covered by the detector
\footnote{$|\eta|\lt4.2$ for D0}.
The vector and scalar forms of the total $\Pt$ balance equation, used for the $p\bar{p}-$event
as a whole, are given in Sections 3.1 and 3.3 respectively.

Another new thing is a use of a new physical object, proposed also in \cite{9}--\cite{BKS_GLU}
and named an ``isolated jet''. This jet is contained in the cone of radius $R=0.7$ 
in the $\eta-\phi$ 
space and does not have any noticeable $\Pt$ activity in some ring around.
The width of this ring is taken to be of $\Delta R=0.3$ (or approximately of the width of 3 calorimeter towers).
In other words, we will select a class of events having  a total $\Pt$ activity inside 
the ring around this ``isolated jet'' within $3-5\%$ of jet $\Pt$.
It will be shown in Sections 6, 7 and Appendix 2  that the number of events
with such a clean topological structure would not be small at Tevatron energy and 
$L=10^{32}~ cm^{-2}s^{-1}$.

Section 4 is devoted to the estimation of a size of
a non-detectable neutrino contribution to a jet.
The correlation of the upper cut value, imposed onto $\Pt^{miss}$
\footnote{see (7) for definition}, 
with the mean value 
of $\Pt$ of neutrinos belonging to the jet $\Pt$
is considered. The detailed results of this section are presented
in the tables of Appendix 1. They also include 
the ratios of the gluonic events $qg\to q+\gamma$ containing the information about
the gluon distribution inside a proton. In the same tables
the expected number of events
(at $L_{int}= 300 ~pb^{-1}$) having charm ($c$) and beauty ($b$) quarks 
in the initial state of the gluonic subprocess are also given.

Since the jet energy calibration is rather a practical than an academic task,
in all the following sections we present the rates obtained with the cuts 
varying from strict to weak because their choice would be
a matter of step-by-step statistics collection during the data taking.

Section 5 includes the results of studying the dependence of the initial state radiation (ISR) $\Pt$-spectrum 
 on the cut imposed on the clusters $\Pt$ ($\Pt^{clust}_{CUT}$) and on the angle 
between the transverse momenta vectors of a jet and a photon.
We also present the rates for four different types of \gpj events,
in which jet fits completely in one definite region 
of the calorimeter: in Central Calorimeter (CC) with $|\eta|\lt0.7$ or in
Intercryostat Calorimeter (IC) with $0.7<|\eta|\lt1.8$ or in End Calorimeter (EC) with 
$1.8\lt|\eta|\lt2.5$ or, finally, in Forward Calorimeter (FC) with $2.5\lt|\eta|\lt4.2$.

In Section 6 our analysis is concentrated on  the ``$\gamma+1~jet$''
events having a jet entirely contained within the central calorimeter region.
The dependence of spectra of different physical variables
\footnote{mostly those that have a strong influence on the \ptgj balance in an event.}
(and among them those appearing in the $\Pt$ balance equation of event as a whole) 
on $\Pt^{clust}_{CUT}$ is shown there.

The dependence of the number of events (for $L_{int}=300~pb^{-1}$) on $\Pt^{clust}_{CUT}$ 
as well as the dependence on it of the fractional $\Fptgj$ disbalance
is studied in Section 7. The details of this study are presented in the tables of
Appendix 2 that together with the corresponding Figs.~10--12  can serve to justify
the variables and cuts introduced in Section 3. 


In Section 8 we present an estimation of the efficiency of background suppression
(that was one of the main guidelines to establish the selection rules proposed in Section 3) for
different numerical values of cuts. 

The importance of the simultaneous use of the above-mentioned new parameters 
$\Pt^{clust}_{CUT}$ and $\Pt^{out}_{CUT}$ and also of the ``isolated jet'' criterion for background 
suppression (as well as for improving the value of the \Ptgj balance)
is demonstrated in Tables 8--11 of Section 8 as well as in the tables of Appendix 3
for various $\Ptg$ intervals. 

The tables of Appendix 3 include  
a fractional disbalance values $\Fptgj$ that are found with an additional 
(as compared with tables of Appendix 2) 
account of the $\Pt^{out}$ cut. They contain the final and {\it first main} 
result (as they include the background contribution) 
of our study of setting an absolute scale of a jet energy 
at the particle level defined by generation with PYTHIA.


Section 9 contains the {\it second main} result of our  study of \gpj events 
at Tevatron energy. Here we investigate a possibility of using the same sample
of the topologically clean \gpj events, obtained with the described cuts,
for determining the gluon distribution in a proton (as it was done earlier for LHC energy in
\cite{BKS_GLU}, \cite{DMS}). The kinematic plot presented
here shows what a region of $x$ and $Q^2$ variables 
can be covered at Tevatron energies with a sufficient number of events for this aim. The comparison
with the kinematic regions covered by other experiments
where parton distributions were studied is also shown in the same plot (see Fig.~17).

About the Summary. We tried to write it in a way allowing a dedicated reader,
who is interested in result rather than in method, to pass directly to it after this sentence.

Since the results presented here were obtained with the PYTHIA simulation,
we are planning to carry out analogous estimations with another event generator like HERWIG, for example,
in subsequent papers.

\section{GENERALITIES OF THE \gpj PROCESS.}
\it\small
\hspace*{9mm}
Useful variables are introduced here for studying their effects on the initial and final state
radiation basing on the simulation in the framework of PYTHIA. Other effects of non-perturbative nature
like primordial parton $k_{~t}$ effect, parton-to-jet hadronization that may
lead to \ptgj disbalance within the physical models used in PYTHIA are also discussed.
\rm\normalsize
\vskip4mm

\subsection{Leading order picture.}        

\setcounter{equation}{1}

The idea of absolute jet energy scale setting 
calibration) by means of the physical process ``$p\bar{p}\rrr \gamma+jet+X$''
was realized many times in different experiments
(see [1--8] and references therein).
It is based on the parton picture where two partons ($q\bar{q}$ or
$qg$), supposed to be moving in different colliding nucleons with
zero transverse momenta
(with respect to the beam line), produce a photon called the ``direct photon''.
This process is described by the leading order (LO) Feynman
diagrams shown in Fig.~1 
\footnote{for the explanation of the numeration of lines see
Section 2.2}
for the ``Compton-like'' subprocess \\[-20pt]
\begin{eqnarray}
\hspace*{6.04cm} qg\to q+\gamma \hspace*{7.5cm} (1a)
\nonumber
\end{eqnarray}
\vspace{-4mm}
and for the ``annihilation'' subprocess 
~\\[-8pt]
\begin{eqnarray}
\hspace*{6.02cm} q\overline{q}\to g+\gamma.  \hspace*{7.4cm} (1b)
\nonumber
\end{eqnarray}

As the initial partons were supposed to have zero transverse momenta,
$\Pt$ of the ``$\gamma$+parton'' system produced in the final state
should be also equal to zero, i.e. one can write the following $\Pt$ balance equation for 
photon and final parton \\[-15pt]
\begin{eqnarray}
{{\bf{\Pt}}}^{\gamma+part}={{\bf{\Pt}}}^{\gamma}+{{\bf{\Pt}}}^{part} = 0.
\end{eqnarray}
One could expect that the transverse momentum
of the jet produced by the final state parton ($q$ or $g$) with 
${{\bf{\Pt}}}^{part}=-{{\bf{\Pt}}}^{\gamma}$ will
be close in magnitude with a reasonable precision 
to the transverse momentum of the final state photon,
 i.e. ${\bf{\Pt}}^{jet}\approx-{\bf{\Pt}}^{\gamma}$.
Thus, in principle, having a well-calibrated photon energy scale one can determine 
a jet energy scale. That is the a main idea of the procedure.
But a more detailed analysis leads to some features needed to be taken into account
and to a photon--jet $\Pt$ balance equation in a more complex form.
~\\[-6.1cm]
\begin{center}
\begin{figure}[h]
  \hspace{10mm} \includegraphics[width=13cm,height=8.3cm]{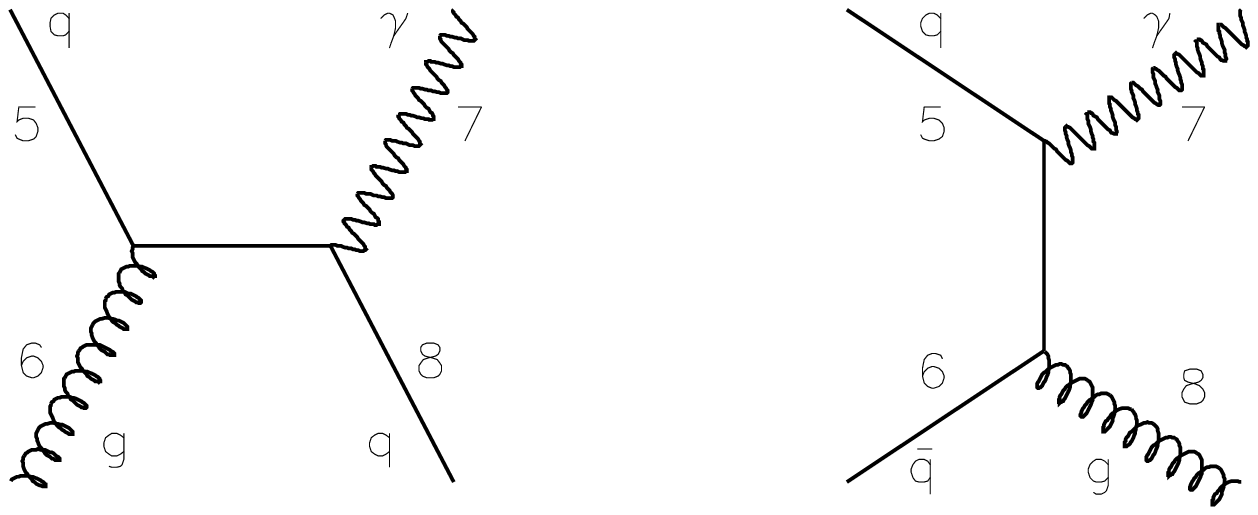} 
\vskip-11mm
\hspace*{42mm} {\small(a)} \hspace*{47mm} {\small(b)}
 \vskip-2mm
\caption{\hspace*{0.0cm} Some of the leading order Feynman diagrams for direct
photon production.} \label{fig1:LO}
\end{figure}
\vskip-10.5mm
\end{center}

\subsection{Initial state radiation.}                           

Since we believe in the perturbation theory, the leading
order (LO) picture described above is expected to be dominant and
determine the main contribution to the cross section. The Next-to-Leading Order
(NLO) approximation (see some of the NLO diagrams in Figs.~2 and 4) introduces some
deviations from a rather straightforward LO-motivated idea of a jet energy calibration.
A gluon ~radiated in the initial state (ISR), as it is seen from Fig.~2, 
can have its own non-zero transverse momentum
$\Pt^{gluon}\equiv \Pt^{ISR}\neq 0$. Apart of a problem of appearance of extra jets
(or mini-jets and clusters), that will be discussed in what follows, it leads
to the non-zero transverse momenta of partons that appear in the initial state
of fundamental $2\rrr2$ QCD subprocesses (1a) and (1b). As a result of
the transverse momentum conservation there arises a disbalance between 
the transverse momenta of a photon $\Pt^{\gamma}$ and of a parton $\Pt^{part}$ produced 
in the fundamental $2\to 2$ process $5+6\to 7+8$ shown in Fig.~2 (and in Fig.~3)
and thus, finally, the disbalance between $\Ptg$ and $\Pt$ of a jet produced by this parton.
~\\[-6cm]
\begin{center} \begin{figure}[h]
  \hspace{15mm} \includegraphics[width=13cm,height=8.3cm,angle=0]{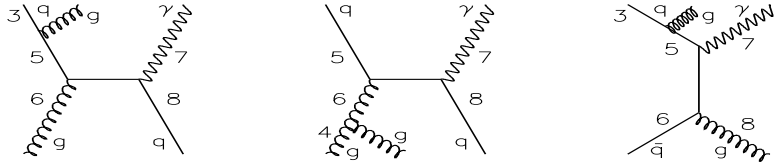}
  \vspace{-12mm}
  \caption{\hspace*{0.0cm} Some of Feynman diagrams of direct photon
production including gluon radiation in the initial state.}
    \label{fig2:NLO}
  \end{figure}
\end{center}
\vspace{-1.0cm}

Following \cite{BKS_P1}--\cite{BKS_P5} and \cite{GPJ} we choose  the modulus of the vector sum of
the transverse momentum vectors ${\bf{\Pt}^5}$ and ${\bf{\Pt}^6}$
of the incoming  into $2\rrr 2$ fundamental QCD subprocesses $5+6\to 7+8$
partons (lines 5 and 6 in Fig.~2) and the sum of their modulus
as two quantitative measures \\[-5pt]
\vspace{-7mm}
\begin{eqnarray}
\Pt^{5+6}=|{\bf{\Pt}^5}+{\bf{\Pt}^6}|, \rm{} \qquad \Pt{56}=|{\bf{\Pt}^5}|+|{\bf{\Pt}^6}|
\end{eqnarray}
to estimate the $\Pt$ disbalance caused by ISR
\footnote{The variable $\Pt^{5+6}$ was used in analysis in \cite{9}--\cite{BKS_P1}.}. 
The modulus of the vector sum \\[-5mm]
\begin{eqnarray}
\Pt^{\gamma+jet}=|{\bf{\Pt}^{\gamma}}+{\bf{\Pt}^{jet}}|
\end{eqnarray}
was also used as an estimator of the final state  $\Pt$
disbalance in the \gpj system in \cite{BKS_P1}--\cite{BKS_P5}.

The numerical notations in the Feynman diagrams  (shown in Figs. 1 and 2)
and in formula (3)  are chosen to be in correspondence with those
used in the PYTHIA event listing for description of the parton--parton subprocess
displayed schematically in Fig.~3. The ``ISR'' block describes the initial
state radiation process that can take place before the fundamental
hard $2\to 2$ process.
\begin{center}
  \begin{figure}[h]
  \vspace{-0.8cm}
   \hspace{3cm} \includegraphics[width=10cm,height=5cm,angle=0]{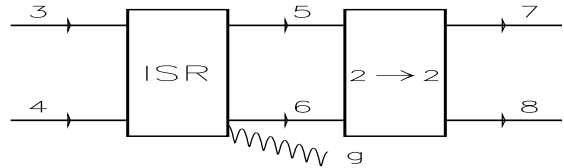}
  \vspace{-2.5cm}
 \caption{\hspace*{0.0cm} PYTHIA ``diagram'' of~ $2\to2$ process (5+6$\to$7+8)
following the block (3+4$\to$5+6) of initial state radiation (ISR), drawn here to illustrate the PYTHIA 
event listing information.}
    \label{fig4:PYT}
  \end{figure}
\end{center}

~\\[-2.3cm]

\subsection{Final state radiation.}                            
\begin{center}
\begin{figure}[htbp]
  \hspace{15mm} \includegraphics[width=13cm,height=8.4cm,angle=0]{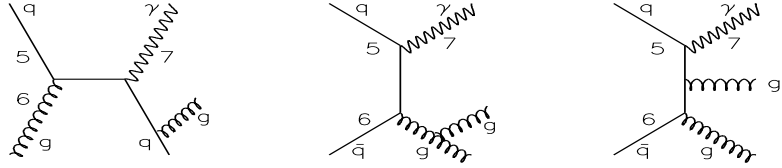}
  \vspace{-12mm}
  \caption{\hspace*{0.0cm} Some of Feynman diagrams of direct photon
production including gluon radiation in the final state.}
    \label{fig4:NLO}
  \end{figure}
\end{center}

Let us consider fundamental subprocesses in which there is
no initial state radiation but instead  final state radiation
(FSR) takes place. These subprocesses are described in the quantum
field theory by the NLO diagrams like those shown in Fig.~\ref{fig4:NLO}. It
is clear that appearance of an extra gluon leg in the final
state may lead to appearance of additional jets (or clusters) in an event as it happens 
in the case of ISR described above.
So, to suppress FSR (manifesting itself as some extra
jets or clusters) the same tools as for reducing ISR should be
used. But due to the string model of fragmentation used in
PYTHIA it is much more difficult to deduce basing on the PYTHIA event listing information
the variables (analogous to (3) and (4)) to describe the disbalance between 
$\Pt$ of a jet parent parton and $\Ptg$. That is why, keeping in mind a close
analogy of the physical pictures of ISR and FSR (see
Figs.~\ref{fig2:NLO} and \ref{fig4:NLO}), we shall concentrate 
in the following sections on the initial state radiation
supposing it to serve in some sense as a quantum field theory perturbative
model of the final state radiation mechanism.
\\[-6cm]
\vspace{-1.0cm}

\subsection{Primordial parton $k_t$ effect.}                   

Now after considering the disbalance sources connected with the
perturbative corrections to the leading order diagrams let us
mention the physical effects of a non-perturbative nature. 
A possible non-zero value of the intrinsic transverse parton
velocity inside a colliding proton may be another source of the
$\Pt^{\gamma}$ and $\Pt^{part}$ disbalance in the final state.
Nowadays this effect can be described mainly in the
phenomenological way. Its reasonable value is supposed to lead to
the value $k_t \leq \,1.0 ~GeV/c$. Sometimes in the literature a total
effect of ISR and of the intrinsic parton transverse momentum is denoted by a common
symbol ``$k_t$''. Here we follow the approach and the
phenomenological  model used in PYTHIA where these two sources of the \Ptgj
disbalance, having different nature, perturbative and non-perturbative, 
can be switched on separately by different keys 
\footnote{Variables MSTP(61) for ISR and PARP(91), PARP(93), MSTP(91) for
intrinsic parton transverse momentum $k_t$ (see \cite{PYT})}. 
In what follows we shall keep the value 
of $k_t$ mainly to be fixed  by the PYTHIA default value $\langle k_t \rangle=0.44~ GeV/c$. 
The dependence of the disbalance between $\Pt^{\gamma}$ and $\Pt^{jet}$
on possible variation of $k_t$ was discussed in detail in \cite{BKS_P5,D0_Note}.
 The general conclusion from there is that any variation of $k_t$ within
 reasonable boundaries (as well as slightly beyond them) does not produce a
 large effect in the case when the initial state radiation is
 switched on. The latter makes a dominant contribution.

 \subsection{Parton-to-jet hadronization.}                      

Another non-perturbative effect that leads to the \ptgj disbalance
is connected with hadronization (or fragmentation into hadrons) of the parton
produced in the fundamental $2\to 2$ subprocess into a jet. 
The  hadronization of a parton into a jet 
is described in PYTHIA within the Lund string fragmentation model. 
The mean values of the fractional $(\Pt^{jet}-\Pt^{parton})/\Pt^{parton}$ disbalance
is presented in the tables of Appendix 2 for UA1
jetfinding algorithm. 
Is is seen that a hadronization effect has a sizable contribution into \ptgj disbalance.

\section{CHOICE OF MEASURABLE PHYSICAL VARIABLES FOR THE \gpj
PROCESS AND THE CUTS FOR BACKGROUND REDUCTION.}               

\it\small
\hspace*{9mm}
The classification of different physical objects that participate in \gpj events and that may give
a noticeable contribution into the total $\Pt$-balance in an event as a whole is done.

Two new physical observables, namely, $\Pt$ of a cluster and $\Pt$ of all detectable particles beyond 
\gpj system, as well as the definition of isolated jet, proposed for studying \ptgj disbalance in 
\cite{9}--\cite{BKS_P5}, are discussed.

The selection cuts for physical observables of \gpj events are given.

The  $\Pt$-balance equation for the event as a whole is written in scalar form that allow to express
the \ptgj disbalance in terms of the considered physical variables.
\rm\normalsize
\vskip4mm

Apart from (1a) and (1b), other QCD subprocesses with large cross sections, 
by orders of magnitude larger than the
cross sections of (1a) and (1b), can also lead to high $\Pt$ photons
and jets in final state. So, we face the problem of selecting
signal \gpj events from a large QCD background.
Here we shall discuss a choice of physical variables that would be
useful, under some cuts on their values, for separation of the
desirable processes with direct photon (``$\gamma^{dir}$'') 
from the background events. A possible ``$\gamma^{dir}-$candidate''
may originate from the $\pi^0,~\eta,~\omega$ and $K^0_s$ meson decays 
 or may be caused by a bremsstrahlung photon 
or by an electron (see Section 8).

We take the D0 ECAL size to be limited by$|\eta|\!\!\leq \!\!2.5$ and 
the calorimeter to be limited by $|\eta| \leq 4.2$ and
to consist of CC, IC, EC, FC parts, where $\eta = -ln~(tan~(\theta/2))$ is a pseudorapidity
defined in terms of a polar angle $\theta$ counted from the beam line. In a plane
transverse to the beam line the azimuthal angle $\phi$ defines 
directions of ${\bf{\Pt}^{Jet}}$ and ${\bf{\Pt}^{\gamma}}$. \\

\subsection{Measurable physical variables and the $\Pt$ vector balance equation.}

In $p\bar{p}\to \gamma + jet + X$ events we are going to study
the main physical object will be a high $\Pt$ jet to be detected
in the $|\eta|\lt4.2$ region and a direct photon registered by the
ECAL up to $|\eta|\lt2.5$. In these events there will be a set of
particles mainly caused by beam remnants, i.e. by spectator parton
fragments, that are flying mostly in the direction of a non-instrumented
volume ($|\eta|>4.2$) in the detector. Let us denote the total transverse
momentum of these non-observable particles ($i$) as \\[-4mm]
\begin{equation}
\sum\limits_{i
\in |\eta| \gt4.2} {\bf{\Pt}^i} \equiv {\bf{\Pt}^{|\eta| \gt4.2}}.  \label{eq:sel1}
\end{equation} 

Among the particles  with $|\eta|\lt4.2$ there may also be neutrinos.
We shall denote their total momentum as
\begin{equation}
\sum\limits_{i \in |\eta|\lt4.2} {\bf{\Pt}_{(\nu)}^i}
 \equiv {\bf{\Pt}_{(\nu)}}.
\label{eq:sel2}
\end{equation}
\vspace{-2mm}

\noindent
A sum of transverse momenta of these two kinds of non-detectable particles
will be denoted as $\Pt^{miss}$
\footnote{This value is a part of true missing $\Pt$ in an experiment that includes the detector
effects (see [1, 2]).}:
\vspace{-3mm}
\begin{eqnarray}
 {\bf{\Pt}^{miss}} = {\bf{\Pt}_{(\nu)}} + {\bf{\Pt}^{|\eta|>4.2}}.
\end{eqnarray}

\vspace{-2mm}

A high-energy jet may also contain neutrinos that may carry
a part of the total jet energy. The average
values of this energy can be estimated from a simulation.

From the total jet transverse momentum ${\bf{\Pt}^{Jet}}$  we shall
separate the part that, in principle, can be detected in
the ECAL+HCAL calorimeter system and in the muon system. Let us denote
this detectable part as ${\bf{\Pt}^{jet}}$ (small ``j''!).
 So, we shall
present the total jet transverse momentum ${\bf{\Pt}^{Jet}}$ as a sum of three
parts:  

1. ${\bf \Pt}^{Jet}_{(\nu)}$, containing the
contribution of neutrinos that belong to the jet, i.e.
a non-detectable part of jet $\Pt$ ($i$ - neutrino):
\vspace{-4mm}
\begin{eqnarray}
 {\bf \Pt}^{Jet}_{(\nu)} = \sum\limits_{i \in Jet} {\bf \Pt}_{(\nu)}^i.
\end{eqnarray}
~\\[-20pt]

2. ${\bf \Pt}^{Jet}_{(\mu)}$, containing the
contribution of jet muons to ${{\bf \Pt}^{Jet}}$ ($i$ - muon):
\vspace{-2mm}
\begin{eqnarray}
 {{\bf \Pt}^{Jet}_{(\mu)}} = \sum\limits_{i \in Jet} {{\bf \Pt}_{(\mu)}^i}.
\end{eqnarray}
~\\[-20pt]

These muons make a weak signal in the calorimeter
but their energy can be measured, in principle, in the muon system (in the
region of $|\eta|\lt2.5$ in the case of D0 geometry).
Due to the absence of the muon system and the tracker
beyond the $|\eta|\lt2.5$ region,
there exists a part of $\Pt^{Jet}$ caused by muons with $|\eta|>2.5$.
We denote this part as $\Pt^{Jet}_{(\mu,|\eta|>2.5)}$. It is non-detectable part and
can be considered as an analogue of $\Pt^{Jet}_{(\nu)}$. 

As for both points 1 and 2, let us say in advance that
the estimation of the average values of neutrino and muon
contributions to $\Pt^{Jet}$ (see Section 4 and Tables 1--3 of Appendix 1) have shown
that they are quite small: about $0.30\%$ of $\la\Pt^{Jet}\ra_{all}$ is due to neutrinos 
and about $0.33\%$ of $\la\Pt^{Jet}\ra_{all}$ is due to muons
where ``$all$'' means averaging over all events including those without neutrinos 
and/or muons in jets. So, they together may cause  approximately about $0.63\%$ of 
the \Ptgj disbalance if muon signal is lost.

3. Finally, as we have mentioned before,
we use ${{\bf \Pt}^{jet}}$ to denote the part of ${{\bf \Pt}^{Jet}}$ which
includes all detectable particles of the jet
\footnote{We shall consider the issue of charged particles
contribution with small $\Pt$ 
into the total jet $\Pt$ while discussing the results of the full GEANT
 simulation (with account of the magnetic field effect) in our forthcoming
papers.} , i.e. the sum of $\Pt$ of jet particles that may produce a signal
in the calorimeter ($calo$) and muon system ($\mu$):
\vspace{-2mm}
\begin{eqnarray}
 {{\bf \Pt}^{jet}} =  {{\bf \Pt}^{Jet}_{(calo)}} +{{\bf \Pt}^{Jet}_{(\mu)}} ,
\quad |\eta^{\mu}|\lt2.5.
\end{eqnarray}

Thus, in the general case we can write for any $\eta$ values:
\vspace{-1.5mm}
\begin{eqnarray}
 {{\bf \Pt}^{Jet}}
={{\bf \Pt}^{jet}}
+{{\bf \Pt}^{Jet}_{(\nu)}}
+{{\bf \Pt}^{Jet}_{(\mu,|\eta^{\mu}|>2.5)}}.
\end{eqnarray}

In the case of $p\bar{p}\to \gamma + jet + X$ events
the particles detected in the $|\eta|\lt4.2$ region may originate
from the fundamental subprocesses (1a) and (1b) corresponding to LO diagrams shown in Fig.~1, 
as well as from the processes corresponding to NLO diagrams (like those in Figs.~2, 4 that include ISR and FSR),
and also from the ``underlying'' event [1], of course.


So, for any event we separate the particles in the $|\eta|\lt4.2$ region into two subsystems. 
The first one consists of the particles belonging to the 
``$\gamma +Jet$'' system (here ``$Jet$'' denotes the jet with the highest $\Pt$, greater
$30 ~GeV/c$, having the total transverse momentum ${{\bf \Pt}^{\gamma +Jet}}$ 
(large ``Jet'', see (4)). The second subsystem involves all other ($O$) particles
beyond the ``$\gamma +Jet$'' system in the  region, covered by the detector, i.e. $|\eta|\lt4.2$.
The total transverse momentum of this
$O$-system are denoted as $\Pt^{O}$ and it is a sum of
$\Pt$  of additional mini-jets (or clusters) and $\Pt$ of
single hadrons, photons and leptons in the $|\eta| \lt 4.2$ region. Since a part of
neutrinos are also present among these leptons, 
the difference of ${{\bf \Pt}_{(\nu)}}$ and ${{\bf \Pt}^{Jet}_{(\nu)}}$
gives us the transverse momentum \\[-20pt]
\begin{eqnarray}
 {{\bf \Pt}^{O}_{(\nu)}} = {{\bf \Pt}_{(\nu)}} - {{\bf \Pt}^{Jet}_{(\nu)}} \quad
|\eta^{\nu}|\lt4.2,
\end{eqnarray}

\noindent
carried out by the neutrinos that do not belong to the jet but are
contained in the $|\eta|\lt4.2$ region.

We denote by ${{\bf \Pt}^{out}}$ a part of ${{\bf \Pt}^O}$ that can be measured, in principle, in the detector.
Thus, ${{\bf \Pt}^{out}}$ is a sum of $\Pt$ of other mini-jets or, generally,
clusters (with $\Pt^{clust}$ smaller than $\Pt^{Jet}$) and $\Pt$ of single
hadrons ($h$), photons ($\gamma$) and electrons ($e$) with $|\eta| \lt 4.2$
and muons ($\mu$) with $|\eta^\mu| \lt 2.5$ that are out of the \gpj system.
For simplicity these mini-jets and clusters will be called ``clusters''
\footnote{As was already mentioned in  Introduction, these clusters are found
by the LUCELL jetfinder with the same value of the cone radius as for jets: $R^{clust}=R^{jet}=0.7$.}.
So, for our \gpj events ${{\bf \Pt}^{out}}$ is the following sum 
(all $\{h,~\gamma,~e,~\mu \} \not\in$ Jet):\\[-17pt]
\begin{eqnarray}
{{\bf \Pt}^{out}} =
{{\bf \Pt}^{clust}}
+{{\bf \Pt}^{sing}_{(h)}}
+{{\bf \Pt}^{nondir}_{(\gamma)}}
+{{\bf \Pt}^{}_{(e)}}
+{{\bf \Pt}^{O}_{(\mu, |\eta^\mu|\lt2.5)}}; \quad  |\eta|\lt4.2.
\end{eqnarray}

\noindent
And thus, finally, we have:\\[-15pt]
\begin{eqnarray}
{{\bf \Pt}^{O}} =
{{\bf \Pt}^{out}}+{{\bf \Pt}^{O}_{(\nu)}}+{{\bf \Pt}^{O}_{(\mu, |\eta^\mu|>2.5)}}.
\label{eq:out}
\end{eqnarray}

\noindent
With these notations we come to the following vector form \cite{BKS_P1} of
the $\Pt$- conservation law for the ``$\gamma + Jet$'' event
(where $\gamma$ is a direct photon) as a whole (supposing that the jet and the photon are contained
in the corresponding detectable regions):\\[-1pt]
\begin{equation}
{{\bf \Pt}^{\gamma}} +
{{\bf \Pt}^{Jet}} +
{{\bf \Pt}^{O}} +
{{\bf \Pt}^{|\eta|>4.2}} = 0
\label{eq:vc_bal}
\end{equation}

\noindent
with last three terms defined correspondingly by (11), (15) and (5) respectively.

\subsection{Definition of selection cuts for physical variables.}

\noindent
1. We shall select the events with one jet and one ``$\gamma^{dir}$-candidate''
(in what follows we shall designate it as $\gamma$ and call the
``photon'' for brevity) 
\footnote{only in Section 8, devoted to the backgrounds, we shall denote
$\gamma^{dir}$-candidate by $\tilde{\gamma}$}
with\\[-11pt]
\begin{equation}
\Pt^{\gamma} \geq 40~ GeV/c~\quad {\rm and} \quad \Pt^{Jet}\geq 30 \;GeV/c.
\label{eq:sc1}
\end{equation}
In the simulation the ECAL signal is considered as a candidate for a direct photon
if it fits inside one D0 calorimeter tower having size $0.1\times0.1$ in 
the $\eta-\phi$ space.

For most of our applications in Sections 4, 5 and 6 mainly the PYTHIA
jetfinding algorithm LUCELL will be used
\footnote{Comparison with the UA1 and UA2 jetfinding algorithms was presented in
\cite{D0_Note, BKS_P3,BKS_P4}}.
The jet cone radius R in the $\eta-\phi$ space counted from the jet initiator cell (ic) is
taken to be $R_{ic}=((\Delta\eta)^2+(\Delta\phi)^2)^{1/2}=0.7$.

\noindent
2. To suppress the contribution of background processes, i.e. to select mostly the events with ``isolated''
direct photons and to discard the events with fake ``photons'' (that
may originate as $\gamma^{dir}$-candidates from meson decays, for instance), we restrict

a) the value of the scalar sum of $\Pt$ of hadrons and other particles surrounding
a ``photon'' within a cone of $R^{\gamma}_{isol}=( (\Delta\eta)^2+(\Delta\phi)^2)^{1/2}=0.7$
(``absolute isolation cut'')
\footnote{We have found that $S/B$ ratio with $R^{\gamma}_{isol}\!=\!0.7$ is in about 1.5 times better
than with $R^{\gamma}_{isol}\!=\!0.4$ what is accompanied by only $10\%$ of additional loss of the number of
signal events.}
\\[-7pt]
\begin{equation}
\sum\limits_{i \in R} \Pt^i \equiv \Pt^{isol} \leq \Pt_{CUT}^{isol};
\label{eq:sc2}
\end{equation}
\vspace{-2.6mm}

b) the value of a fraction (``fractional isolation cut'')\\[-7pt]
\begin{equation}
\sum\limits_{i \in R} \Pt^i/\Pt^{\gamma} \equiv \epsilon^{\gamma} \leq
\epsilon^{\gamma}_{CUT}.
\label{eq:sc3}
\end{equation}

\noindent
3. We accept only the events having no charged tracks (particles)
with $\Pt>5~GeV/c$ within the $R=0.4$ cone around the $\gamma^{dir}$-candidate.

\noindent
4. 
To suppress the background events with photons resulting from
$\pi^0$, $\eta$, $\omega$
and $K_S^0$ meson decays, we require the absence of a high $\Pt$ hadron
in the tower containing the $\gamma^{dir}$-candidate:\\[-10pt]
\begin{equation}
\Pt^{hadr} \leq 7~ GeV/c.
\label{eq:sc5}
\end{equation}

\noindent
At the PYTHIA level of simulation this cut may effectively take into account 
the imposing of an upper cut on the HCAL energy in the cells behind
the ECAL signal cells fired by the direct photon. 
In real experimental conditions one can require that a fraction of the photon energy, 
deposited in ECAL to be greater than some value ($\approx 0.95-0.96$ as it is now at D0).


\noindent   
5. We select the events with the vector ${{\bf \Pt}^{Jet}}$ being ``back-to-back'' to
the vector ${{\bf \Pt}^{\gamma}}$ (in the plane transverse to the beam line)
within $\dphi$ defined by the equation:\\[-12pt]
\begin{equation}
\phigj=180^\circ \pm \Delta\phi,
\label{eq:sc7}
\end{equation}
where $\phigj$ is the angle
between the \Ptgj vectors: 
${{\bf \Pt}^{\gamma}}{{\bf \Pt}^{Jet}}=\Pt^{\gamma}\Pt^{Jet}\cdot cos(\phigj)$,
$\Pt^{\gamma}=|{{\bf \Pt}^{\gamma}}|,~~\Pt^{Jet}=|{{\bf \Pt}^{Jet}}|$.
The cases $\Delta\phi \leq 17^\circ, 11^\circ, 6^\circ$ are considered in this paper
($6^\circ$ is approximately one D0 calorimeter tower  size in $\phi$).

\noindent 
6. As we have already mentioned in Section 3.1, one can
expect reasonable results of the jet energy calibration procedure
modeling and subsequent practical realization
 only if one uses a set of selected events with small $\Pt^{miss}$ 
(see (7) and (\ref{eq:sc_bal})). 
So, we also use the following cut:\\[-17pt]
\begin{eqnarray}
\Pt^{miss}~\leq \Pt^{miss}_{CUT}.
\label{eq:sc11}
\end{eqnarray}
The aim of the event selection with small $\Pt^{miss}$ is quite obvious: 
we need a set of events with a reduced $\Pt^{Jet}$ uncertainty due to a possible 
presence of a non-detectable particle contribution to a jet and due to
the term $\Pt^{|\eta|>4.2}$ (see (7) and (\ref{eq:sc_bal})).

The influence of $\Pt^{miss}_{CUT}$ on the selection of events with a reduced value
of the total sum of neutrino contribution into $\Pt^{Jet}$ is studied in Section 4.

\noindent
7. The initial and final state radiations (ISR and FSR) manifest themselves most clearly
as some final state mini-jets or clusters activity.
To suppress it, we impose a new cut condition that was not formulated in
an evident form in previous experiments: we choose the \gpj events
that do not have any other
jet-like or cluster high $\Pt$ activity  by selecting the events with the values of
$\Pt^{clust}$ (the cluster cone $R_{clust}(\eta,\phi)=0.7$), being lower than some threshold
$\Pt^{clust}_{CUT}$ value, i.e. we select the events with\\[-10pt]
\begin{equation}
\Pt^{clust} \leq \Pt^{clust}_{CUT}
\label{eq:sc8}
\end{equation}
($\Pt^{clust}_{CUT}=15, 10, 5 ~GeV/c$ are most effective as will be shown in Sections 6--8).
Here, in contrast to \cite{BKS_P1}--\cite{BKS_P5}, the clusters are found
by one and the same jetfinder LUCELL while three different jetfinders UA1, UA2 and LUCELL
are used to find the jet ($\Pt^{Jet}\geq 30~ GeV/c$) in the event.

\noindent
8. Now we pass to another new quantity (proposed also for the first time in \cite{BKS_P1}--\cite{BKS_P5}) 
that can be measured at the experiment.
We limit the value of the modulus of the vector sum of ${{\bf \Pt}}$ of all
particles, except those of the \gpj system, that fit into the region $|\eta|\lt4.2$ covered by
the ECAL and HCAL, i.e., we limit the signal in the cells ``beyond the jet and photon'' region
by the following cut:
\begin{equation}
\left|\sum_{i\not\in Jet,\gamma-dir}\!\!\!{{\bf \Pt}^i}\right| \equiv \Pt^{out} \leq \Pt^{out}_{CUT},
~~|\eta^i|\lt4.2.
\label{eq:sc9}
\end{equation}

\noindent
The importance of $\Pt^{out}_{CUT}$ and $\Pt^{clust}_{CUT}$
for selection of events with a good balance of \Ptgj and for
the background reduction will be demonstrated in Sections 7 and 8.

Below the set of selection cuts 1 -- 8 will be referred to as
``Selection 1''. The last two of them, 7 and 8, are new criteria \cite{BKS_P1}
not used in previous experiments. 

9. In addition to them one more new object, introduced in 
\cite{BKS_P1} -- \cite{BKS_P5} and named an ``isolated jet'',  will be used in our analysis.
i.e. we shall require the presence of a ``clean enough'' (in the sense of limited $\Pt$
activity) region inside the ring of $\Delta R=0.3$ width (or approximately of a size of 
three calorimeter towers) around the jet.  Following this picture, we restrict the ratio of the scalar sum
of transverse momenta of particles belonging to this ring, i.e.\\[-5pt]
\begin{equation}
\Pt^{ring}/\Pt^{jet} \equiv \epsilon^{jet}\leq\epsilon^{jet}_0, \quad {\rm where ~~~~ }
\Pt^{ring}=\!\!\!\!\sum\limits_{\footnotesize i \in 0.7\lt R \lt1.0} \!\!\!\!|{{\bf \Pt}^i}|.
\label{eq:sc10}
\end{equation}
~\\[-4pt]
($\epsilon^{jet}_0$ is chosen to be $3-5\%$, see Sections 7 and 8).

The set of cuts 1 -- 9 will be called in what follows ``Selection 2''.

The exact values of the cut parameters $\Pt^{isol}_{CUT}$,
$\epsilon^{\gamma}_{CUT}$, $\epsilon^{jet}$, $\Pt^{clust}_{CUT}$, $\Pt^{out}_{CUT}$
 will be specified below, since they may be
different, for instance, for various $\Pt^{\gamma}$ intervals
(being looser for higher  $\Pt^{\gamma}$).

\subsection{The scalar form of the $\Pt$ balance equation and 
the jet energy calibration procedure.}
%

Let us rewrite the basic $\Pt$-balance equation (\ref{eq:vc_bal}) of 
Section 3.1 with the notations introduced here in the scalar form
more suitable for the following applications: \\[-15pt]
\begin{eqnarray}
\frac{\Pt^{\gamma}-\Pt^{Jet}}{\Pt^{\gamma}}=(1-cos\dphi) 
+ \Db/\Pt^{\gamma}, \label{eq:sc12}
\label{eq:sc_bal}
\end{eqnarray}
where
$\Db\equiv ({{\bf \Pt}^{O}}+{{\bf \Pt}^{|\eta|>4.2)}})\cdot \bf{n^{Jet}}$ ~~~~ with ~~
$\bf{n^{Jet}}={{\bf \Pt}^{Jet}}/\Pt^{Jet}$.

As will be shown in Section 7, the first term on the
right-hand side of equation (\ref{eq:sc_bal}), i.e. $(1-cos\dphi)$ is negligibly
small as compared with the second term 
\footnote{in a case of Selection 1}
and tends to decrease fast with  growing $\Pt^{Jet}$. 
So, in this case the main contribution to the $\Pt$ disbalance in the
\gpj system is caused by the term $\Db/\Pt^{\gamma}$.

$\Pt^{Jet}$ can be easily expressed from equation (\ref{eq:sc_bal}) through: \\[-15pt]
\begin{eqnarray}
\Pt^{Jet} =\alpha \cdot \Pt^{\gamma}
\label{eq:sc_bal2}
\end{eqnarray}
with $\alpha$ defined as~ $\alpha = cos\dphi - \Db/\Pt^{\gamma}$.

Having defined in every selected event $\Pt^{Jet}$ from equation (\ref{eq:sc_bal2})
one can determine calibration coefficients $\{C_i\}$ via minimizing of
a standard deviation of the function:\\[-10pt]
\begin{eqnarray}
F=\sum_{j=1}^{N_{event}}  \left( 
\frac{ \Pt^{Jet}-\sum_{i=1}^{N_l} C_i \Pt^{i,c} }{\Delta \Pt^{Jet}} 
\right)^2
\label{eq:minvar}
\end{eqnarray}
In this expression $N_l$ is a number of calorimeter layers
\footnote{$N_l=8$ at D0 (4 for ECAL and 4 for HCAL).},
$\Pt^{i,c}$ is energy deposition in the $i-$th calorimeter layer and $\Delta \Pt^{Jet}$ is
the error on $\Pt^{Jet}$ caused by uncertainty in $\alpha$ ($\Delta \alpha$) 
and uncertainty due to limited accuracy of $\Pt^{\gamma}$ determination 
($\Delta \Pt^{\gamma}$)
\footnote{For instance, in the central region of D0 calorimeter ($|\eta|<0.9$)
electron/photon energy resolution can be written
through $\sigma/E = 23\%/\sqrt(E) \oplus 20\%/E \oplus 0.4\%$.}.
So, one can write (see (\ref{eq:sc_bal2})):\\[-15pt]
\begin{eqnarray}
\Delta \Pt^{Jet} = \Delta \alpha \oplus \Delta \Pt^{\gamma}  
\end{eqnarray}

Obtained in this way the calibration coefficients $\{C_i\}$ in the selected \gpj events
for every bin of $\eta^{jet}$ and calorimeter $\Pt^{jet}$
then should be applied to energy depositions in each layer
$\Pt^{i,c}$ of a found jet in any event
to reconstruct a jet transverse momentum at the particle level
The accuracy of such a reconstruction will directly depend on the accuracy of the coefficients $\{C_i\}$.
The latter, in their turn,  are caused by the error of $\Delta \Pt^{Jet}$ (see (\ref{eq:minvar}))
\footnote{Other possibility, based on the usage of artificial
neural networks (ANN), was also considered (see  \cite{QCD_talk3} and \cite{MassANN}). 
In this approach one can obtain a better energy resolution of the reconstructed jet
but it requires a bigger statistics for ANN training. The calibration coefficients $\{C_i\}$
in this case will be replaced by set of ANN weights $\{w_{ij}\}$ and function (\ref{eq:minvar}) --- 
by a more
complicated expression.}.

Having determined relatively perfectly  a photon energy scale the $\Delta \Pt^{\gamma}$,
the $\Delta \Pt^{Jet}$ uncertainty will be mainly defined by $\Delta \alpha$ 
(namely by term $\Db/\Pt^{\gamma}$ of equation (\ref{eq:sc_bal})).

\section{ESTIMATION OF A NON-DETECTABLE PART OF $\Pt^{Jet}$ AND $\Pt^{miss}$ SPECTRA.}

\it\small
\hspace*{9mm}
It is shown that by imposing an upper cut on  the missing transverse momentum
$\Pt^{miss}\lt10~GeV/c$ one can reduce  the correction to the measurable part of $\Pt^{jet}$ due 
to neutrino contribution down to the value of
$\Delta_\nu=\la\Pt^{Jet}_{(\nu)}\ra_{all\; events}=0.1~GeV/c$
in all intervals of $\Ptg$. 
\rm\normalsize
\vskip4mm

In Section 3.1 we have divided the transverse momentum of a jet,
i.e. $\Pt^{Jet}$, into two parts, a detectable $\Pt^{jet}$ and non-detectable
($\Pt^{Jet}-\Pt^{jet}$), consisting of $\Pt^{Jet}_{(\nu)}$
and $\Pt^{Jet}_{(\mu,|\eta|>2.5)}$ (see (11)).
In the same way, according to equation (15), we divided
the transverse momentum $\Pt^O$ of ``other particles'', that are out 
of $\gamma^{dir}+jet$ system,
into a detectable part $\Pt^{out}$
and a non-detectable part consisting of the sum of $\Pt^O_{(\nu)}$
and $\Pt^O_{(\mu,|\eta|>2.5)}$ (see (15))
%
%

We shall estimate here what part of $\Pt^{Jet}$ may be carried out by non-detectable particles 
(mainly neutrinos originating from weak decays)
\footnote{In \cite{BKS_P5} and \cite{QCD_talk} it was  shown that main source of high $\Pt$ 
neutrinos in background processes 
are $W^\pm$ decays, which also contain $e^\pm$ that in its turn may fake direct photons.}.

We shall consider the case of switched-on decays of $\pi^{\pm}$ and $\;K^{\pm}$ mesons 
\footnote{According to the PYTHIA default agreement, $\pi^{\pm}$ and $\;K^{\pm}$ mesons are stable.}.
Here $\pi^{\pm}$ and $K^{\pm}$ meson decays are allowed inside the solenoid volume with the barrel 
radius $R_B=80~ cm$ and the distance from the interaction vertex to Endcap along the $z$-axis 
$L=130~cm$ (D0 geometry).

\begin{figure}[htbp]
\vspace{-1.9cm}
\hspace{-10mm} \includegraphics[width=19cm,height=19.7cm]{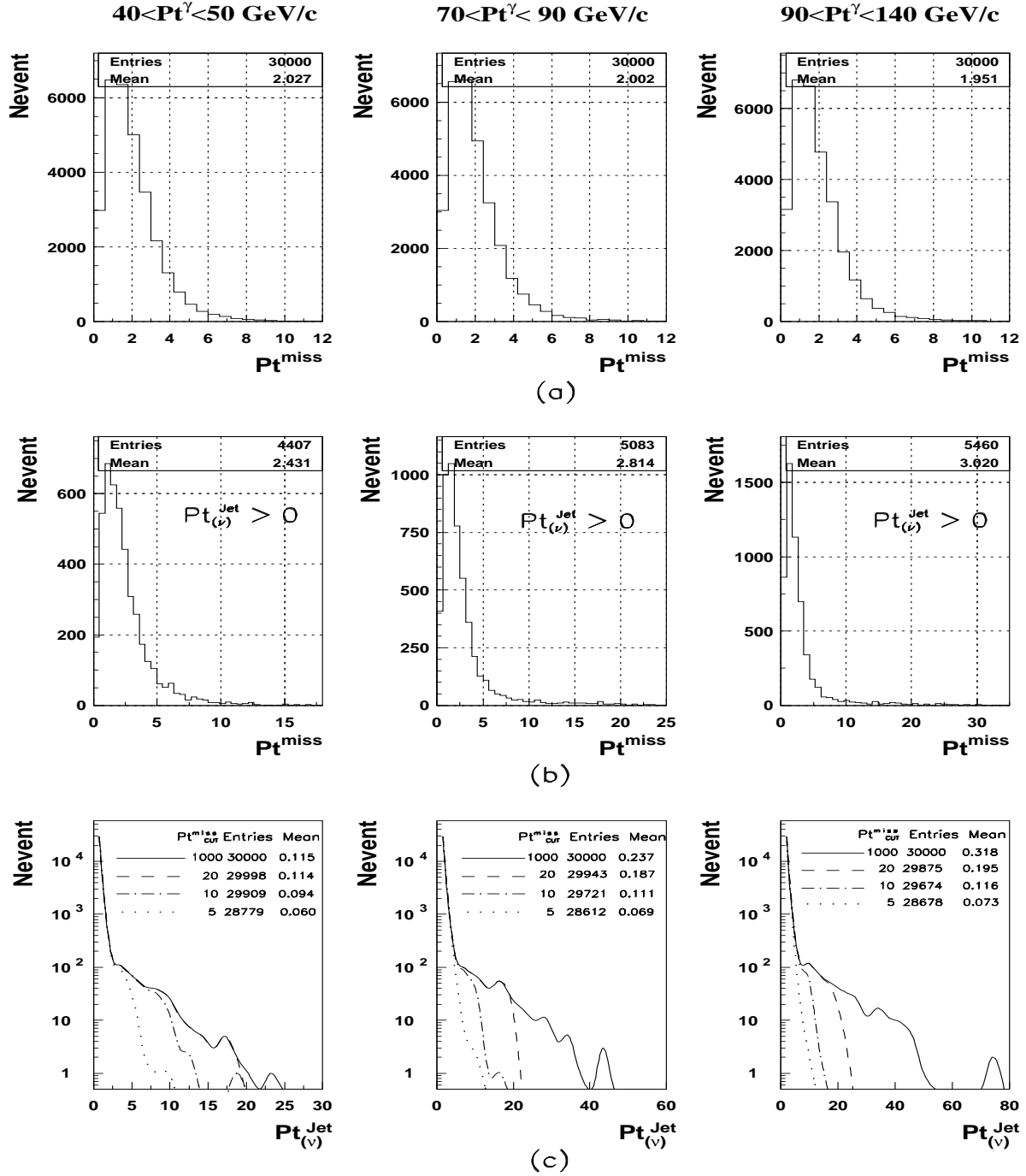}
\vspace{-0.5cm}
\caption{\hspace*{0.0cm}~ a) $\Pt^{miss}$ spectra in all events; ~~
b) $\Pt^{miss}$ spectra in events having jets with non-zero $\Pt$
neutrinos,~ i.e. 
$\Pt^{Jet}_{(\nu)}\gt0$;~ c) $\Pt^{Jet}_{(\nu)}$ spectra
and their mean values dependence on the values 
of $\Pt^{miss}_{CUT}$ in various $\Ptg(\approx\Pt^{Jet})$ intervals.
$\pi^{\pm}$ and $K^{\pm}$ meson decays are allowed inside the solenoid of $R=80~cm$ and $L=130~cm$
($\Pt^{clust}_{CUT}=30 ~GeV/c$).}
\label{fig20-22}
\end{figure}

For this aim we shall use the bank of the signal \gpj events, i.e. caused by subprocesses
(1a) and (1b), generated for three $\Ptg$ intervals:
$40\lt\Ptg\lt50$, $70\lt\Ptg\lt90$ and $90\lt\Ptg\lt140~GeV/c$
and selected with conditions (16)--(23) (Selection 1) and the following cut values:
\begin{equation}
 \Pt_{CUT}^{isol}=4 ~GeV/c,~~\epsilon^{\gamma}_{CUT}=7\%,
~~\dphi\lt17^\circ,~~\Pt^{clust}_{CUT}=30~ GeV/c.
\end{equation}
Here the cut $\Pt^{clust}_{CUT}=30~ GeV/c$ has the meaning of a very weak restriction
on mini-jets or clusters activity.
No restriction was imposed on the $\Pt^{out}$ value.
The results of analysis of these events, based on the application of LUCELL jetfinder,
are presented in Fig.~\ref{fig20-22}.

The first row of Fig.~\ref{fig20-22} contains  $\Pt^{miss}$ spectra in the \gpj events 
for different
$\Ptg$ intervals and demonstrates (to a good accuracy) their practical independence on $\Ptg$.

In the second row of Fig.~\ref{fig20-22} we present the spectra of
$\Pt^{miss}$ for those events (denoted as $\Pt^{Jet}_{(\nu)}>0$) which
contain jets having neutrinos, i.e. having  a non-zero $\Pt^{Jet}_{(\nu)}$
component of  $\Pt^{Jet}$. 
These figures show a very weak dependence of the $\Pt^{miss}$ spectrum
on the direct photon $\Pt^{\gamma}$ 
Comparison of the
number of entries in the second row plots of Fig.~\ref{fig20-22} with
those in the first row allows to conclude that the part of events with the jet
having the non-zero neutrinos contribution is about $15-18\%$.

%
%

The effect of imposing general $\Pt^{miss}_{CUT}$  in each event of our
sample is shown in the third row of Fig.~\ref{fig20-22}. The upper cut
 $\Pt^{miss}_{CUT}=1000 ~GeV/c$
means the absence of any upper limit for $\Pt^{Jet}_{(\nu)}$.
The most important illustrative fact that in the absence of any restriction
on $\Pt^{miss}$ the total neutrino $\Pt$ inside the jet
averaged over all events can be as large as $\Pt^{Jet}_{(\nu)}\approx 0.32~GeV/c$
at $90\lt\Ptg\lt140 ~GeV/c~  GeV/c$ (see the right-hand plot of the third row in 
Fig.~\ref{fig20-22}).
In the $40\lt\Pt^{Jet}\lt50~GeV/c$ interval, 
we have already
a very small mean value of $\Pt^{Jet}_{(\nu)}$ equal to $0.12 ~GeV/c$
even without imposing any $\Pt^{miss}_{CUT}$. From the same plots of the third row of
Fig.~\ref{fig20-22} we see that with general $\Pt^{miss}_{CUT}=10~GeV/c$
the average correction due to neutrino contribution is $0.1~GeV/c$
in all three intervals of $\Ptg$. 

At the same time, as it was demonstrated in \cite{BKS_P5} and \cite{QCD_talk},
this cut essentially reduces  the admixture of the $e^\pm$-events, in which $e^\pm$, mainly 
originating from 
the $W^\pm\to e^\pm\nu$ weak decays, may fake the direct photon signal. 
These events are characterized by big values of $\Pt^{miss}$ (it is higher, on the average, 
by about one order of magnitude than in the signal ``$\gamma^{dir}+jet$'' events) that may worsen 
the jet calibration accuracy.

The situation, analogous to neutrino, holds for the $\Pt^{Jet}_{(\mu)}$ contribution.

The detailed information about the values of non-detectable
$\Pt^{Jet}_{(\nu)}$ averaged over all events
(no cut on $\Pt^{miss}$ was used) as well as about mean $\Pt$ values
of muons belonging to jets $\la \Pt^{Jet}_{(\mu)}\ra$ is presented
in Tables 1--3 of Appendix 1 for the sample of events with jets which are
entirely contained in the central region of the calorimeter
($|\eta^{jet}|\lt0.7$) and found  by UA1 jetfinder.           
In these tables the ratio of the number of events with non-zero $\Pt^{Jet}_{(\nu)}$
to the total number of events is denoted by $R^{\nu \in Jet}_{event}$ and
the ratio of the number of events with non-zero $\Pt^{Jet}_{(\mu)}$
to the total number of events is denoted by $R^{\mu \in Jet}_{event}$.

The quantity $\Pt^{miss}$ in events with $\Pt^{Jet}_{(\nu)}\gt0$ is denoted in these tables as
$\Pt^{miss}_{\nu\in Jet}$ and is given there for three $\Ptg$ intervals
($40\lt\Ptg\lt50$, $70\lt\Ptg\lt90$ and $90\lt\Ptg\lt140$)
and $\Pt^{clust}_{CUT}=30, 20,15,10,5 ~GeV/c$)
%
\footnote{Note that the values of $\Pt^{miss}$ and $\Pt^{miss}_{\nu\in Jet}$ in the plots of 
Fig.~\ref{fig20-22}
are slightly different from those of Appendix 1 as the numbers in from Fig.~\ref{fig20-22}
were found for events in the whole $|\eta|\lt4.2$ region.}



\section{EVENT RATES FOR DIFFERENT $\Pt^{\gamma}$ AND $\eta^{jet}$ INTERVALS.}

\it\small
\hspace*{9mm}
The number of \gpj events distribution over $\Pt^{\gamma}$ and $\eta^{\gamma}$ is shown here.
It is found that in each interval of the $\Delta\Pt^{\gamma}=10~ GeV/c$ width 
the rates decrease by a factor more than 2.
The number of events with jets which transverse momentum are completely (or with $5\%$ accuracy)
contained in CC, IC, EC and FC regions are presented in Tables \ref{tab:sh1}, \ref{tab:sh2}
for integrated luminosity $L_{int}=300~pb^{-1}$.
\rm\normalsize
\vskip4mm

\subsection{Dependence of distribution of the number of events
on the ``back-to-back'' angle $\phigj$ and on $\Pt^{ISR}$. }            

The definitions of the physical variables introduced in
Sections 2 and 3  allow to study
a possible way to select the events with a good $\Pt^{\gamma}$ and $\Pt^{Jet}$ balance.
Here we shall be interested to get (by help of PYTHIA generator and the theoretical models 
therein) the form of the spectrum of the variable $\Pt{56}$ (see (3))
(which is approximately proportional to $\Pt^{ISR}$ up to the value of intrinsic parton
transverse momentum $k_t$ inside a proton) at different values of $\Ptg$.
For this aim four samples of \gpj events were generated by using PYTHIA
with 2 QCD subprocesses (1a) and (1b) being included simultaneously. 
In what follows we shall call these events as 
``signal events''. The generations were done with the values of the PYTHIA parameter 
CKIN(3)($\equiv\pth$) equal to  20, 25, 35 and 45 $GeV/c$ in order to cover
four $\Pt^{\gamma}$ intervals: 40--50, 50--70, 70--90 and 90--140 $GeV/c$, respectively
\footnote{$\la k_t\ra$ was taken to be fixed at the PYTHIA default value, i.e. $\la k_t\ra=0.44\,GeV/c$}.
Each sample in these $\Ptg$ intervals had a size of $5\cdot 10^6$ events.
The cross sections for the two subprocesses were found to be as given in Table~\ref{tab:cross4}.\\[-9mm]
\begin{table}[h]
\begin{center}
\caption{The cross sections (in $microbarns$) of the $qg\to q+\gamma$ and $q\overline{q}\to g+\gamma$ subprocesses
for four $\Pt^{\gamma}$ intervals.}
\normalsize
\vskip.1cm
\begin{tabular}{||c||c|c|c|c|}                  \hline \hline
\label{tab:cross4}
Subprocess& \multicolumn{4}{c|}{$\pth$ ($GeV/c$)} \\\cline{2-5}
   type   & 20 & 25 & 35  & 45 \\\hline \hline
$qg\to q+\gamma$           & 0.97$\cdot10^{-2}$& 4.78$\cdot10^{-3}$& 1.36$\cdot10^{-3}$& 4.95$\cdot10^{-4}$ \\\hline
$q\overline{q}\to g+\gamma$& 0.20$\cdot10^{-2}$& 0.96$\cdot10^{-3}$& 0.35$\cdot10^{-3}$& 1.56$\cdot10^{-4}$ \\\hline
Total                      & 1.17$\cdot10^{-2}$& 5.75$\cdot10^{-3}$& 1.71$\cdot10^{-3}$& 6.51$\cdot10^{-4}$ \\\hline
\end{tabular}
\end{center}
\vskip-7mm
\end{table}

 For our analysis  we used Selection 1   (see (16)--(23))
 and the values of cut parameters (32).

\begin{figure}{htbp}
\vskip-25mm
\hspace{1mm} \includegraphics[height=13.0cm,width=15.5cm]{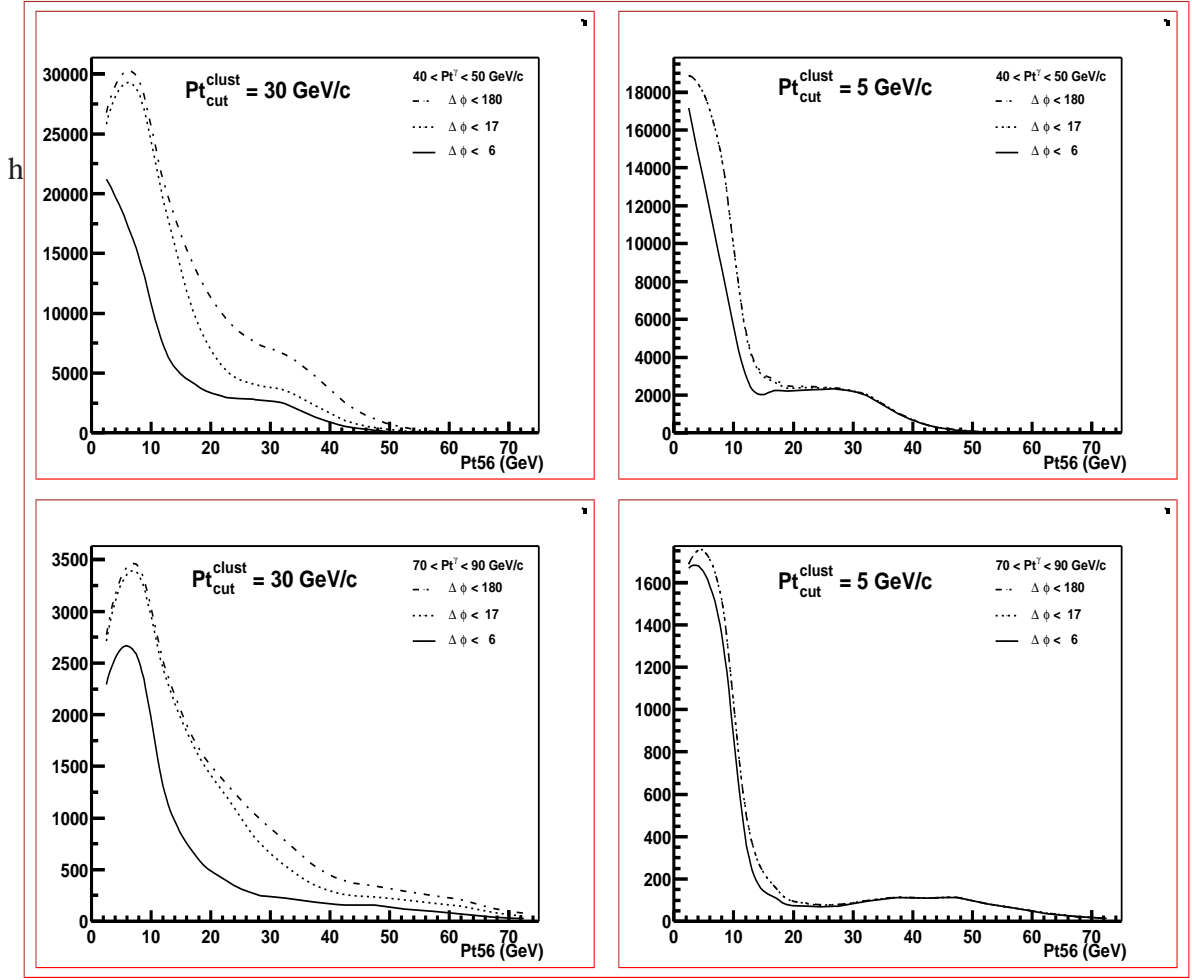}
\caption{A dependence of the number of events (at $L_{int}=300~ pb^{-1}$)
on $\dphi_{max}$ and $\Pt^{clust}_{CUT}$ for two $\Pt^{\gamma}$ intervals: 
$40\leq \Pt^{\gamma}\leq 50 ~GeV/c$~ for two upper plots and 
$70\leq \Pt^{\gamma}\leq 90 ~GeV/c$~ for two bottom plots.}
\label{pt56}
\end{figure}

In Fig.~\ref{pt56} we present the $\Pt56$ spectra for
two most illustrative cases of
$\Pt^{\gamma}$ intervals $40\lt\Pt^{\gamma}\lt50 ~GeV/c$ (two upper plots) and
$70\lt\Pt^{\gamma}\lt90 ~GeV/c$ (two bottom plots). The distributions of the
 number of events for the integrated luminosity $L_{int}=300\,pb^{-1}$
in different $\Pt56$ intervals  and for different ``back-to-back'' angle intervals
$\phigj=180^\circ \pm \dphi~$ ($\dphi\leq17^\circ$ and $6^\circ$ as well as without 
any restriction on $\dphi$, i.e. for the whole $\phi$ interval $\dphi\leq180^\circ$)
\footnote{The value $\Delta\phi=6^\circ$ approximately
coincides with one D0  HCAL tower size in the $\phi$-plane.
}
are given there. The LUCELL jetfinder was used for determination of jets and clusters
\footnote{More details connected with UA1 jetfinder application can be found in Section 7
and Appendix 2 for a jet contained in CC region.}.
Left column of Fig.~\ref{pt56} correspond to $\Pt^{clust}\lt30\,GeV/c$ 
and serve as an illustration since it is rather a weak cut condition, while
right column of Fig.~\ref{pt56}  correspond to a more
restrictive selection cut $\Pt^{clust}_{CUT}=5\,GeV/c$.

Tables \ref{tab:pt56-1} and \ref{tab:pt56-2} show the number of events
(at $L_{int}=300~ pb^{-1}$) left after application different cuts on the angle $\dphi$
for two values of $\Pt^{clust}_{CUT}$.
In the case of weak restriction  $\Pt^{clust}\lt30 ~GeV/c$
we can see that for the $40\leq \Pt^{\gamma}\leq 50 ~GeV/c$
interval about 75$\%$ of events are concentrated
in the $\Delta\phi\lt17^\circ$ range, while 41$\%$ of events are
in the $\Delta\phi\lt6^\circ$ range.
As for $70\leq \Pt^{\gamma}\leq90\, GeV/c$: 
about 86$\%$ of events have
$\Delta\phi\lt17^\circ$ and 50$\%$ of them have $\Delta\phi\lt6^\circ$.

A tendency of distributions of the number of signal \gpj events to be
very rapidly concentrated in a rather narrow
back-to-back angle interval $\Delta\phi\lt17^\circ$ as $\Pt^{\gamma}$ grows
becomes more distinct with a more restrictive cut on the cluster $\Pt$.
 From Table \ref{tab:pt56-2} we see that in the first interval
 $40\leq \Pt^{\gamma}\leq 50\, GeV/c~$ more than $99\%$ of the events, selected 
with  $\Pt^{clust}_{CUT}= 5\,GeV/c$, have $\Delta\phi\lt17^\circ$, 
while $~76\%$ of them are in the $\Delta\phi\lt6^\circ$ range. 
It should be mentioned that after application of cut $\Pt^{clust}_{CUT}=5 ~GeV/c$ 
only about $40\%$ of events remain as compared with a case of $\Pt^{clust}_{CUT}=30 ~GeV/c$.
For $70\leq \Pt^{\gamma}\leq 90\, GeV/c$ more than  $90\%$ of the events, 
subject to the cut $\Pt^{clust}_{CUT}= 5 ~GeV/c$, have $\Delta\phi\lt6^\circ$. 
It means that while suppressing cluster or mini-jet activity 
\begin{table}[htbp]
\begin{center}
\vskip-16mm
\small
\caption{A dependence of the number of events on $\dphi_{max}$ and on
$\Pt^{\gamma}$ for $L_{int}=300~ pb^{-1}$,}
\vskip-3pt
{\footnotesize $\Pt^{clust}_{CUT}=30 ~GeV/c$.}
\vskip0.2cm
\begin{tabular}{||c||r|r|r|r||} \hline \hline
\label{tab:pt56-1}
 $\Pt{56}$  &\multicolumn{4}{c||}{ $\dphi_{max}$} \\\cline{2-5}
 $(GeV/c)$  &\aaa $180^\circ$\aaa&\aaa$17^\circ$\aaa&\aaa$11^\circ$\aaa&\aaa$6^\circ$\aaa \\\hline\hline
 40 -- 50  &    110691 &     82913 &     68921 &     44830 \\\hline   
 50 -- 70  &     71075 &     55132 &     45716 &     29692 \\\hline 
 70 -- 90  &     14853 &     12727 &     10919 &      7418 \\\hline   
 90 -- 140 &      5887 &      5534 &      4974 &      3655 \\\hline 
\end{tabular}
\vskip4mm
\small
\caption{A dependence of the number of events  on $\dphi_{max}$ and on
$\Pt^{\gamma}$ for $L_{int}=300~ pb^{-1}$,}
\vskip-3pt
{\footnotesize $\Pt^{clust}_{CUT}=5 ~GeV/c$.}
\vskip0.2cm
\begin{tabular}{||c||r|r|r|r||} \hline \hline
\label{tab:pt56-2}
 $\Pt{56}$  &\multicolumn{4}{c||}{ $\dphi_{max}$} \\\cline{2-5}
 $(GeV/c)$  &\aaa $180^\circ$\aaa&\aaa$17^\circ$\aaa&\aaa$11^\circ$\aaa&\aaa$6^\circ$\aaa \\\hline\hline
 40 -- 50  &  37576 &     37235 &     35473 &     27025 \\\hline  
 50 -- 70  &  19056 &     19017 &     18651 &     15149 \\\hline      
 70 -- 90  &   3773 &      3773 &      3755 &      3387 \\\hline 
 90 -- 140 &   1525 &      1525 &      1524 &      1468 \\\hline     
\end{tabular}
\end{center}
\end{table}  

\noindent    
by imposing $\Pt^{clust}_{CUT}=5 ~GeV/c$
we can select the sample of events with a clean ``back-to-back'' (within 17$^\circ$ in $\phi$) 
topology of $\gamma$ and jet orientation
\footnote{Unfortunately, as it will be discussed below and is seen in Fig.~\ref{pt56},
it does not mean that $\Pt^{clust}_{CUT}$ allows to suppress completely the ISR.
(see also the event spectra over $\Pt^{clust}$ in Fig.~7 of the following Section 6.)}.

Thus one can conclude that PYTHIA simulation predicts that
at Tevatron  energies most of the \gpj events (more than $75\%$)
may have the vectors ${{\bf \Pt}^{\gamma}}$ and ${{\bf \Pt}^{jet}}$ being back-to-back
within $\Delta\phi\lt17^\circ$ after imposing $\Pt^{clust}_{CUT}=30~GeV/c$.  
The cut $\Pt^{clust}_{CUT}=5~GeV/c$ significantly improves this tendency
\footnote{A growth of \ptg produces the same effect, as is seen
from Tables \ref{tab:pt56-1} and \ref{tab:pt56-2} and
will be demonstrated in more detail in Section 6 and Appendix 2.}.

It is worth mentioning that this picture
reflects the predictions of one of the generators
based on the approximate  LO values for
the cross section. It may change if the
next-to-leading order or soft physics
\footnote{We thank E.~Pilon and J.~Ph.~Jouliet for the information
about new Tevatron data on this subject and for clarifying the importance
of NLO corrections and soft physics effects.}
effects are included.


From Fig.~\ref{pt56} one can see that in the case when there are no restrictions
on $\Pt^{clust}$ the $\Pt56$ spectrum becomes a bit wider for larger values of
$\Pt^{\gamma}$.

At the same time, one can conclude from the comparison of left and right upper plots
that the width of the most populated part of the $\Pt56$ (or $\Pt^{ISR}$) 
spectrum reduces by about 40$\%$ with restricting $\Pt^{clust}_{CUT}$. So, for $\dphi_{max}=17^\circ$ we
see that it drops from $0\!\lt\!\Pt{56}\!\lt\!20\;GeV/c$  for  $\Pt^{clust}_{CUT}\!=\!30\;GeV/c$ 
to a $\!$ narrower interval of 
$0\!\lt\!\Pt{56}\!\lt\!10-12\,GeV/c$  for the $\Pt^{clust}_{CUT}=5\,GeV/c$.
At higher $\Ptg$ intervals (two bottom plots) for the same value $\dphi_{max}=17^\circ$ the reduction factor 
of the $\Pt56$ spectrum width is more than two
(from $0\lt\Pt56\lt30~ GeV/c$ for $\Pt^{clust}_{CUT}=30\,GeV/c$
to $0\lt\Pt56\lt12-15 ~GeV/c$ for $\Pt^{clust}_{CUT}=5 ~GeV/c$). 

Thus, we can summarize that the PYTHIA  generator predicts
an increase in the $\Pt^{ISR}$ spectrum with growing $\Pt^{\gamma}$ and this increase
can be substantially reduced by imposing a restrictive cut on 
$\Pt^{clust}$
\footnote{for more details see Sections 6 and 7}.
Unfortunately, $\Pt^{ISR}$ cannot be
completely suppressed  by $\Delta\phi$ and $\Pt^{clust}$ cuts alone
\footnote{In Sections 7, 8 the effect of the additional $\Pt^{out}_{CUT}$ will be discussed.}.
%
%
%
That is why we prefer to use the $\Pt$ balance equation for
the event as a whole (see equations (\ref{eq:vc_bal}) and (\ref{eq:sc_bal}) 
of Sections 3.1 and 3.3), i.e. an equation
that takes into account the ISR and FSR effects,
rather than balance equation (2) for fundamental processes (1a) and (1b) as discussed in Section 2.1
\footnote{
In Section 6 we shall study a behavior of each term that enter equation (\ref{eq:sc_bal})
in order to find the criteria that 
would allow to select events with a good balance of \Ptgj.}.
%


\subsection{$\Pt^{\gamma}$ and $\eta^{\gamma}$ dependence of event rates.}

~\\[-12mm]

Here we shall present
the number of events for different $\Pt^{\gamma}$ and $\eta^{\gamma}$ intervals as
predicted by PYTHIA simulation with weak cuts defined mostly by (32)
with only change of $\Pt^{clust}_{CUT}$ value from 30 to 10$~GeV/c$.
The lines of Table \ref{tab:pt-eta} correspond to  $\Pt^{\gamma}$ intervals
and the columns to $\eta^{\gamma}$ intervals. The last column of this table contains the total number
of events  (at $L_{int}=300\,pb^{-1}$) in the whole ECAL $\eta^{\gamma}$-region
$|\eta^{\gamma}|\lt2.5$ for a given $\Pt^{\gamma}$ interval.
We see that the number of events  decreases fast
with growing $\Pt^{\gamma}$ (by more than 50$\%$ for each subsequent interval). 

\def\baselinestretch{0.99}
\begin{table}[htbp]
~\\[-12mm]
\small
\begin{center}
\caption{Rates for $L_{int}=300\,pb^{-1}$ for different $\Pt^{\gamma}$ intervals and $\eta^{\gamma}$
($\Pt^{clust}_{CUT}= 10 \, GeV/c$ and $\Delta\phi \leq 17^\circ$).}
\vskip0.2cm
\begin{tabular}{||c||r|r|r|r|r|r|r||r||} \hline \hline
\label{tab:pt-eta}
$\Pt^{\gamma}$ &\multicolumn{7}{c||}{$|\eta^{\gamma}|$~~ intervals}
&all ~~$\eta^{\gamma}$ \\\cline{2-9}
$(GeV/c)$ & 0.0-0.4 & 0.4-0.7 & 0.7-1.1 & 1.1-1.4 & 1.4-1.8 & 1.8-2.1 & 2.1-2.5 &0.0-2.5
 \\\hline \hline
 40 -- 50  &   10978  &  11232  &  10604  &  10337  &   9662  &   8051 &    5806 &   66679 \\\hline  
 50 -- 60  &    4483  &   4210  &   4489  &   3938  &   3624  &   2814 &    1562 &   25121 \\\hline  
 60 -- 70  &    2028  &   1732  &   1890  &   1587  &   1442  &    984 &     607 &   10270 \\\hline  
 70 -- 80  &     949  &    931  &    937  &    753  &    637  &    392 &     170 &    4770 \\\hline  
 80 -- 90  &     508  &    513  &    469  &    363  &    309  &    180 &      62 &    2405 \\\hline  
 90 --100  &     302  &    287  &    252  &    201  &    149  &     80 &      25 &    1295 \\\hline  
100 --120  &     285  &    280  &    257  &    189  &    125  &     61 &      11 &    1207 \\\hline  
120 --140  &     134  &    121  &     98  &     63  &     38  &      9 &       1 &     465 \\\hline\hline    
 40 --140  &   19662  &  19302  &  18992  &  17427  &  15986  &  12571 &    8245 &  112216 \\\hline\hline    
\hline
\end{tabular}
~\\[10mm]
\small
\caption{Selection 1. $\Delta \Pt^{jet} / \Pt^{jet} = 0.00$~~~
($\Pt^{clust}_{CUT}= 10 \, GeV/c, ~~\Delta\phi \leq 17^\circ$ and $L_{int}=300\,pb^{-1}$ ).}
\vskip0.2cm
\begin{tabular}{||c||c|c||c|c||c|c||c|c||} \hline \hline
\label{tab:sh1}
$\Pt^{\gamma}$&$CC$&$CC\!\to \! IC$&$IC$&$IC\!\to \! CC,EC$&$EC$ &$EC\!\to\! IC,FC$&$FC$&$FC\!\to\! EC$  
\\\hline \hline
 40 -- 50  &    9965  &  13719  &   8152 &   22225  &    617 &    8854  &    554  &   1912\\\hline
 50 -- 60  &    4009  &   5597  &   3104 &    8791  &    207 &    2766  &    109  &    413\\\hline
 60 -- 70  &    1754  &   2515  &   1339 &    3615  &     71 &     979  &     14  &     93\\\hline
 70 -- 80  &     930  &   1195  &    651 &    1593  &     21 &     348  &      1  &     23\\\hline
 80 -- 90  &     503  &    596  &    328 &     811  &      9 &     136  &      0  &      6\\\hline
 90 -- 100 &     283  &    352  &    165 &     421  &      3 &      59  &      0  &      1\\\hline
100 -- 120 &     263  &    351  &    137 &     389  &      2 &      37  &      0  &      0\\\hline
120 -- 140 &     118  &    143  &     50 &     142  &      1 &       7  &      0  &      0\\\hline\hline
 40 -- 140 &   17822  &  24462  &  13927 &   37988  &    930 &   13184  &    678  &   2448 \\\hline\hline   
\end{tabular}
\vskip.4cm
\caption{Selection 1. $\Delta \Pt^{jet} / \Pt^{jet} \leq 0.05$~~~
($\Pt^{clust}_{CUT}= 10 \, GeV/c, ~~\Delta\phi \leq 17^\circ$ and $L_{int}=300\,pb^{-1}$ ).}
\vskip0.2cm
\begin{tabular}{||c||c|c||c|c||c|c||c|c||} \hline \hline
\label{tab:sh2}
$\Pt^{\gamma}$&$CC$&$CC\!\to \! IC$&$IC$&$IC\!\to \! CC,EC$&$EC$ &$EC\!\to\! IC,FC$&$FC$&$FC\!\to\! EC$  
\\\hline \hline
 40 -- 50  &   17951 &    5733 &   20631 &    9746 &    4174 &    5296 &    1280 &    1186\\\hline
 50 -- 60  &    7466 &    2141 &    8313 &    3583 &    1403 &    1570 &     253 &     269\\\hline
 60 -- 70  &    3405 &     863 &    3553 &    1401 &     492 &     558 &      39 &      68\\\hline
 70 -- 80  &    1699 &     426 &    1667 &     577 &     179 &     190 &       6 &      17\\\hline
 80 -- 90  &     902 &     197 &     838 &     301 &      75 &      71 &       3 &       4\\\hline
 90 --100  &     528 &     107 &     440 &     146 &      31 &      31 &       0 &       0\\\hline
100 --120  &     537 &      98 &     384 &     142 &      19 &      20 &       0 &       0\\\hline
120 --140  &     223 &      37 &     143 &      48 &       5 &       3 &       0 &       0\\\hline\hline
 40 --140  &   32701 &    9603 &   35971 &   15943 &    6377 &    7738 &    1582 &    1545\\\hline\hline 
\end{tabular}
\end{center}
\end{table}   

\subsection{Estimation of \gpj event rates for different calorimeter regions.} 
Since a jet is a wide-spread object, the $\eta^{jet}$ dependence of rates
for different $\Pt^{\gamma}$ intervals will be presented in a different way than in Section 5.2.
Namely, Tables \ref{tab:sh1}--\ref{tab:sh2} include the rates of 
events ($L_{int}=300\,pb^{-1}$) 
for different $\eta^{jet}$ intervals, covered by the 
central, intercryostat, end and forward (CC, IC, EC and FC) parts of the calorimeter and 
for different $\Ptg(\approx\Pt^{Jet})$ intervals.

No restrictions on other parameters are used. The first column of Table 5 $CC$ gives the
number of events with the jets (found by the LUCELL jetfinding algorithm of PYTHIA),
all particles of which are comprised entirely (100$\%$)
\footnote{at the particle level of simulation!} 
in the CC part and there is a
$0\%$ sharing of  $\Pt^{jet}$ ($\Delta \Pt^{jet}=0$) between the CC and
the neighboring IC part of the calorimeter. The second columns of the 
tables $CC\!\to\!IC$ contain the number of events in which $\Pt$ of a jet is shared
between the CC and IC regions. The same sequence of restriction
conditions takes place in the next columns. Thus, the 
$IC, EC$ and $FC$ columns include the number of events with jets
entirely contained in these regions, while the $EC\!\to\!IC,FC$ column gives the number of  events where
the jet covers both the EC and IC or EC and FC regions. From these tables we can see what number of events
can be, in principle, most suitable for the precise jet energy absolute scale setting,
 carried out separately for the CC, EC and FC parts of the calorimeter in different $\Ptg$ 
intervals. 

The selection cuts are as in (32) but $\Pt^{clust}_{CUT}=10~GeV/c$

Less restrictive conditions, when up to $5\%$ of the jet $\Pt$
are allowed to be shared between the CC, EC and FC parts of the calorimeter, is given in 
Table \ref{tab:sh2}.
Tables \ref{tab:sh1}  and \ref{tab:sh2} correspond to the case of Selection 1
\footnote{The cost of passing to Selection 2 (defined in Section 3.2 with $\epsilon^{jet}\lt3\%$)
is a reduction of the number of events by factor equal to 2.}.

 From last summarizing line of Table \ref{tab:sh1} we see 
that for the entire interval $40\lt\Pt^{\gamma}\lt140\; GeV/c$~ PYTHIA predicts 
around 18000 events for CC and around 1000 events for EC at integrated luminosity $L_{int}=300~pb^{-1}$.

\section{INFLUENCE OF THE $\Pt^{clust}_{CUT}~$ PARAMETER
ON THE PHOTON AND JET $\Pt$ BALANCE AND ON THE INITIAL STATE RADIATION
SUPPRESSION.}

\it\small
\hspace*{9mm}
The influence of $\Pt^{clust}_{CUT}$ parameter (defining the upper limit on $\Pt$ of clusters or mini-jets 
in the event) on the variables characterizing the \ptgj balance  is studied.
\rm\normalsize
\vskip3mm

In this section we shall study the specific sample of events considered in 
the previous section that may be most suitable for the jet energy calibration in the CC region,
with jets entirely (100$\%$) contained in this region, i.e.
having 0$\%$ ~sharing of $\Pt^{jet}$ 
\footnote{at the PYTHIA particle level of simulation}
with IC.
 Below we shall call them {\it ''CC-events''}. The $\Ptg$ spectrum for this particular
set of events for $\Pt^{clust}_{CUT}=10 ~GeV/c$ was presented in the first column (CC) 
of Table \ref{tab:sh1}. Here we shall use three different jetfinders, LUCELL from PYTHIA
and UA1, UA2 from CMSJET \cite{CMJ}. The  $\Pt^{clust}$ distributions for generated events found by 
the all three jetfinders in two $\Pt^{\gamma}$ intervals, $40\lt\Pt^{\gamma}\lt50~GeV/c$
and $\!$ $70\lt\Pt^{\gamma}\lt90~GeV/c$, are shown in Fig.~7 for $\Pt^{clust}_{CUT}=30\;GeV/c$ and
$\dphi\leq 17^\circ$.

It is interesting to note an evident similarity of the $\Pt^{clust}$ spectra with $\Pt56$ spectra
shown in Fig.~6 (see also Figs.~8, 9), 
what support our intuitive picture of ISR and cluster connection described in  Section 2.2. \\[-0.6cm]

Here we shall study in more detail correlation of $\Pt^{clust}$ with $\Pt^{ISR}$ mentioned above.
The averaged value of intrinsic parton transverse momentum 
will be fixed at $\langle k_t \rangle= 0.44~ GeV/c$.
%

The banks of 1-jet \gpj events gained from the results of PYTHIA
generation of $5\cdot10^6$  signal \gpj events in each of four $\Pt^{\gamma}$
intervals (40--50, 50--70, 70--90, 90--140 $GeV/c$)
\footnote{they were discussed in Section 5}
will be used here. The observables defined in Sections 3.1 and 3.2  will be
restricted here by cuts of Selection 1  (16)--(23)
and the cut parameters defined by (32).

We have chosen two of these intervals  to illustrate the influence of the
$\Pt^{clust}_{CUT}$ parameter on the distributions of physical variables,
 that enter the balance equation (\ref{eq:sc_bal}). These distributions are shown in Fig.~\ref{fig:40luc}
($40<\Pt^{\gamma}<50~ GeV/c$) and Fig.~\ref{fig:70luc} ($70<\Pt^{\gamma}<90~ GeV/c$). 
 In these figures, in addition to three variables $\Pt56$,
$\Pt^{|\eta|>4.2}$, $\Pt^{out}$, already explained in Sections 2.2, 3.1 and 3.2,
we present distributions of two other variables, $\Db$~ and $(1-cos\dphi)$, which
define the right-hand side of the \ptgj balance equation (\ref{eq:sc_bal}).
The distribution of the $\gamma$-jet back-to-back  angle $\dphi$ (see (22))
is also presented in Figs.~\ref{fig:40luc}, \ref{fig:70luc}. \\[10pt]

\begin{flushleft}
\begin{figure}[htbp]
 \vskip-19mm
 \hspace{-.5cm} \includegraphics[height=74mm,width=9.7cm]{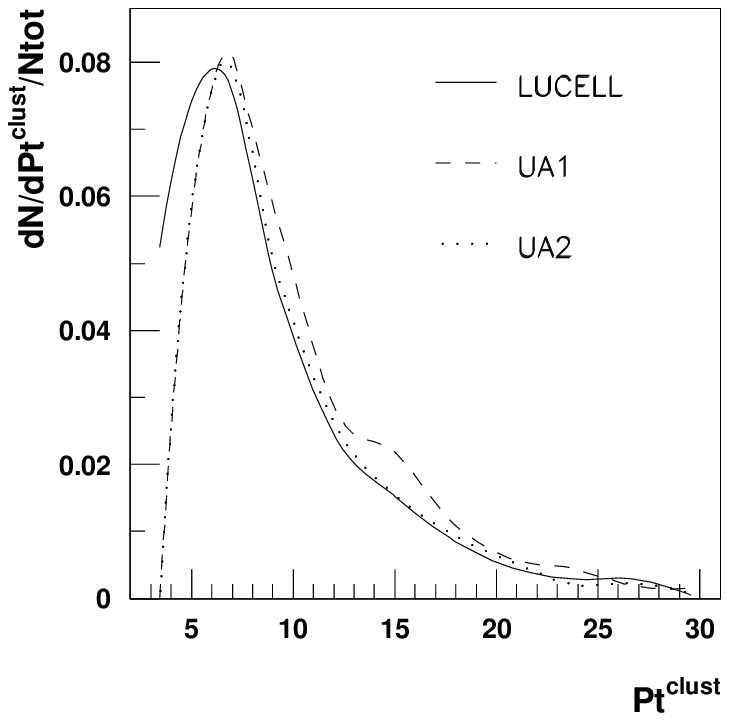}
    \label{fig9}
   \nonumber
  \end{figure}
\end{flushleft}
\begin{flushright}
\begin{figure}[htbp]
 \vskip-93.7mm
  \hspace{7.5cm} \includegraphics[height=74mm,width=9.7cm]{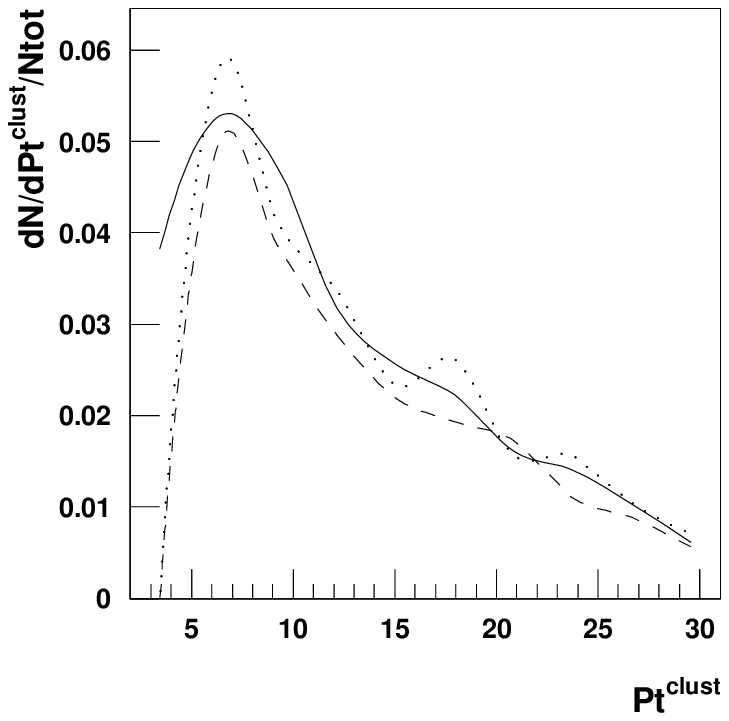}
   \nonumber
\label{fig7}
\vskip-12mm

\hspace*{4.2cm} (a) \hspace*{7.5cm} (b)\\[7pt]
\hspace*{0.3cm}{\footnotesize Fig.~7: $\Pt^{clust}$ distribution in \gpj
events from two $\Pt^{\gamma}$ intervals:
(a) $40<\Pt^{\gamma}<50\, GeV/c$ ~and \\
\hspace*{1.3cm} (b) $70<\Pt^{\gamma}<90\,GeV/c$~ with the same cut
$\Pt^{clust}_{CUT}=30\;GeV/c$ ($\dphi\leq 17^\circ$).}\\[-14mm]
\end{figure}
\end{flushright}

~\\[-7mm]

The ISR describing variable $\Pt56$ (defined by (3))
and both components of
(see (\ref{eq:sc_bal})), $(1-cos\dphi)$ and $\Db/\Ptg$, as well as $\Pt^{out}$ and  $\dphi$,
show a tendency to become smaller (the mean values and the widths)
with the restriction of the upper limit on the 
$\Pt^{clust}$ value (see Figs.~\ref{fig:40luc}, \ref{fig:70luc}).
It means that a jet energy calibration accuracy may increase with decreasing
$\Pt^{clust}_{CUT}$, which justifies the intuitive choice of this new variable in Section 3.
The origin of this improvement becomes clear from the $\Pt{56}$ density plot, which demonstrates 
a decrease of $\Pt{56}$ (or $\Pt^{ISR}$) values with decrease of $\Pt^{clust}_{CUT}$.
In Section 2.3 we gave arguments why it may also influence FSR.

Comparison of Fig.~\ref{fig:40luc} (for $~40\!<\Pt^{\gamma}\!<50 ~GeV/c$) and Fig.~\ref{fig:70luc}
(for $~70\!<\Pt^{\gamma}\!<90 ~GeV/c$) also shows that the values of $\Delta\phi$ as a degree of
back-to-backness of the photon and jet $\Pt$ vectors in the $\phi$-plane
decreases with increasing $\Pt^{\gamma}$. At the same time $\Pt^{out}$ and $\Pt^{ISR}$ distributions 
become slightly wider. It is also seen that
the $\Pt^{|\eta|>4.2}$ distribution practically does not depend on
$\Pt^{\gamma}$ and $\Pt^{clust}$
\footnote{see also Appendix 2 and Fig.~2}.

It should be mentioned that the results presented in Figs.~\ref{fig:40luc} and \ref{fig:70luc} were
 obtained with the LUCELL jetfinder of PYTHIA
\footnote{The results obtained with all jetfinders and
\ptgj ~balance will be discussed in Section 7 in more detail.}.

\begin{center}
\setcounter{figure}{7}
\begin{figure}[htbp]
\vspace{-3.0cm}
  \hspace{-0mm} \includegraphics[width=16cm]{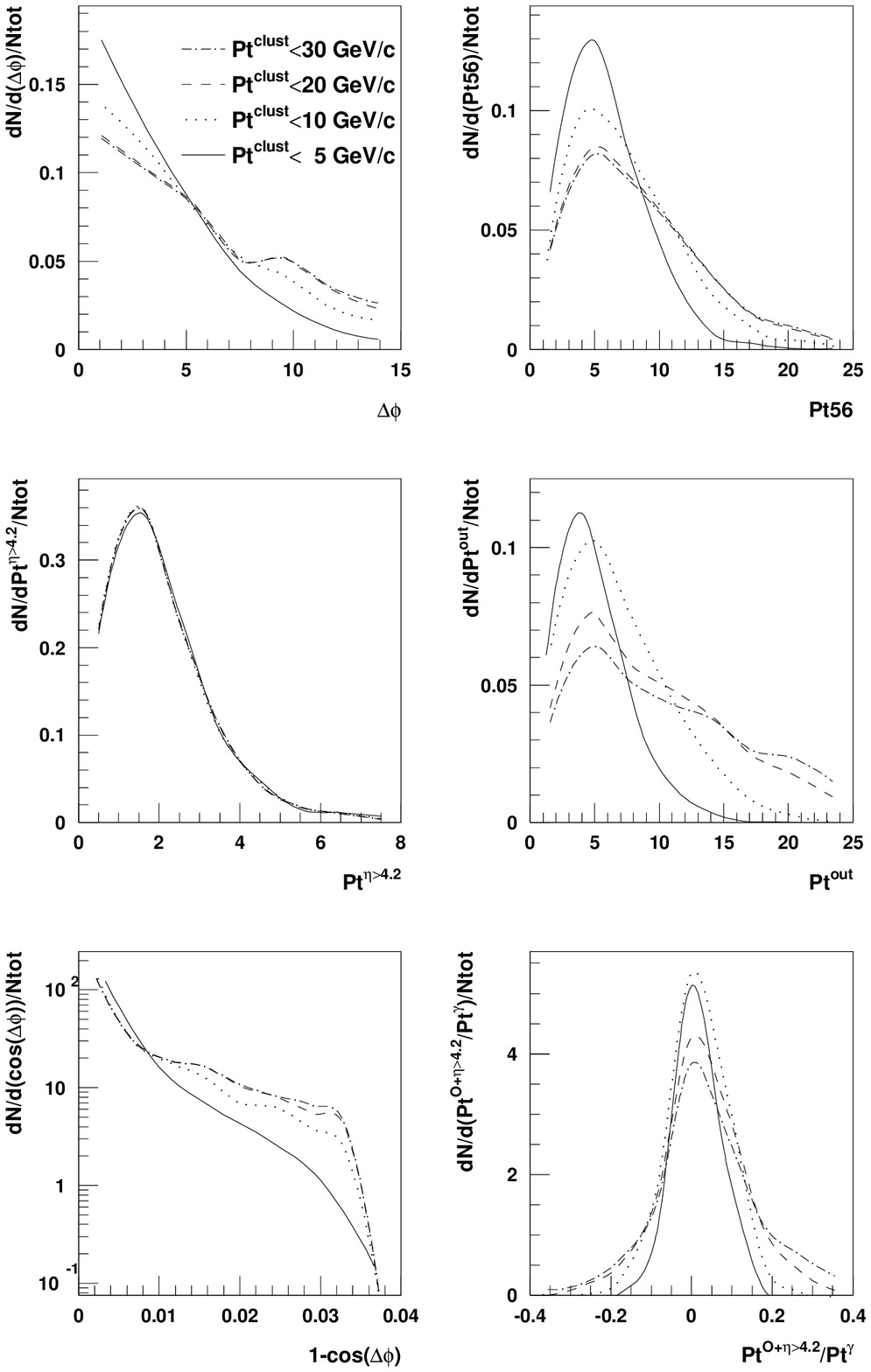} 
  \vspace{-0.5cm}
\caption{\hspace*{0.0cm} LUCELL algorithm, $\dphi<17^\circ$;~~
$40<\Pt^{\gamma}<50\, GeV/c$. Selection 1.}
\label{fig:40luc}
\end{figure} 
\begin{figure}[htbp]
 \vspace{-3.0cm}
  \hspace{-0mm} \includegraphics[width=16cm]{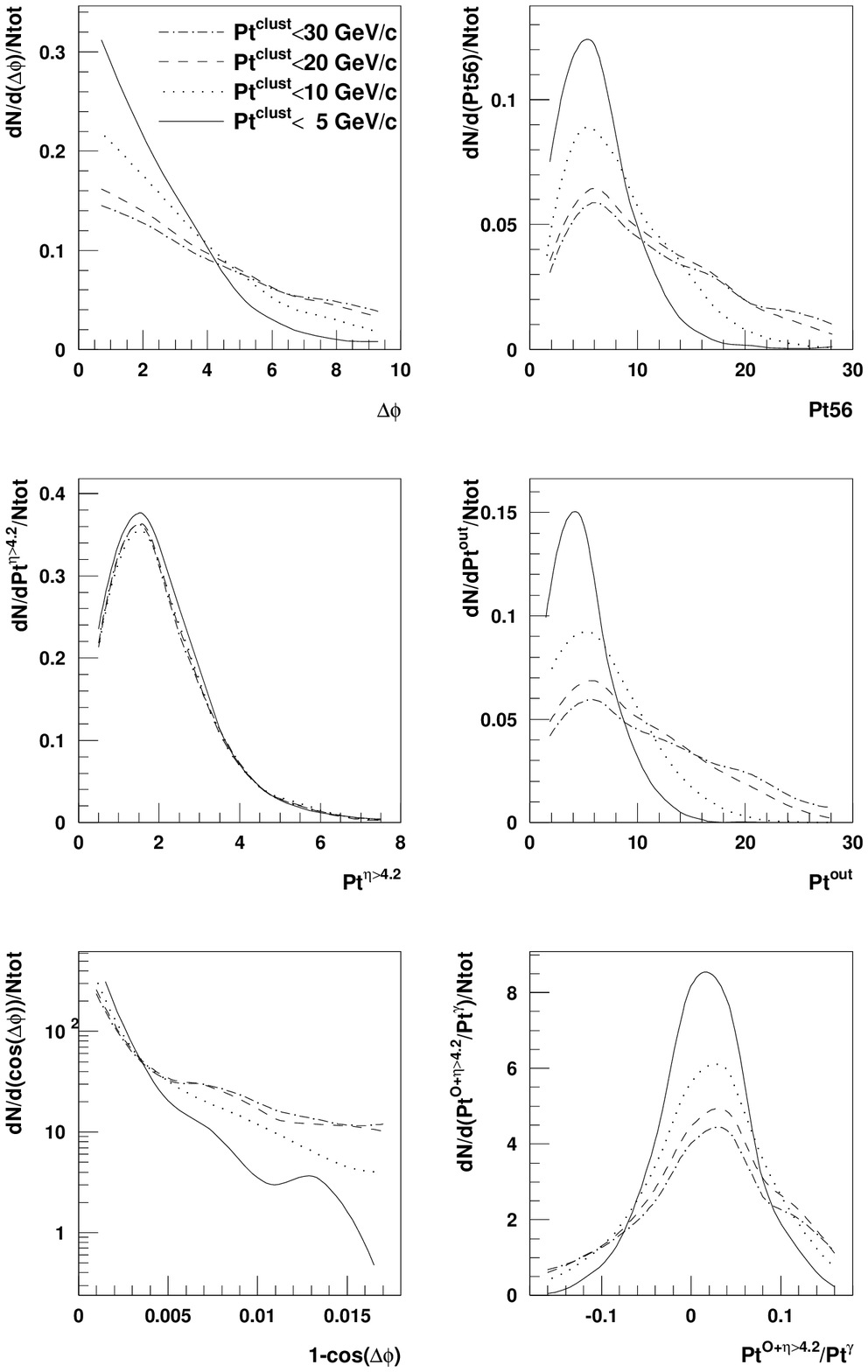}
  \vspace{-0.5cm}
\caption{\hspace*{0.0cm} LUCELL algorithm, $\dphi<17^\circ$;~~
$70<\Pt^{\gamma}<90\, GeV/c$. Selection 1.}
\label{fig:70luc}
\end{figure} 
\end{center}

\section{DEPENDENCE OF THE $\Pt$-DISBALANCE IN THE \gpj SYSTEM 
ON $\Pt^{clust}_{CUT}$ and $\Pt^{out}_{CUT}$ PARAMETERS.}

\it\small
\hspace*{9mm}
It is shown that with Selection 2 (that leads to about twice reduction of the number of events $Nevent$ for
$\Ptg\lt70~GeV/c$ and to about $30-40\%$ loss of them at $\Ptg\gt70~GeV/c$) one can select 
at the particle level the events with a value of the fractional $\Fptgj$ disbalance better than $1\%$.
The number of events (at $L_{int}=300~pb^{-1}$)
and other characteristics of \gpj events are presented in tables of Appendix 2 for interval
$40<\Pt^{\gamma}<140~ GeV/c$.
\rm\normalsize
\vskip4mm

Earlier we have introduced physical variables
for studying \gpj events (Section 3) and discussed what cuts
for them may lead to a decrease in the disbalance of $\Pt^{\gamma}$ and $\Pt^{Jet}$. 
One can make these cuts to be tighter if more events would be collected
during data taking.

Here we shall study in detail the dependence of the $\Pt$ disbalance
 in the \gpj system on $\Pt^{clust}_{CUT}$ and $\Pt^{out}_{CUT}$  values. 
For this aim we shall use the same samples of events as in Section 5 that were generated
by using PYTHIA with 2 QCD subprocesses (1a) and (1b) and collected 
to cover three $\Ptg$ intervals: 40--50, 70--90, 90--140 $GeV/c$.

\begin{figure}
\vskip-10mm
\hspace{-2mm} \includegraphics[width=16cm]{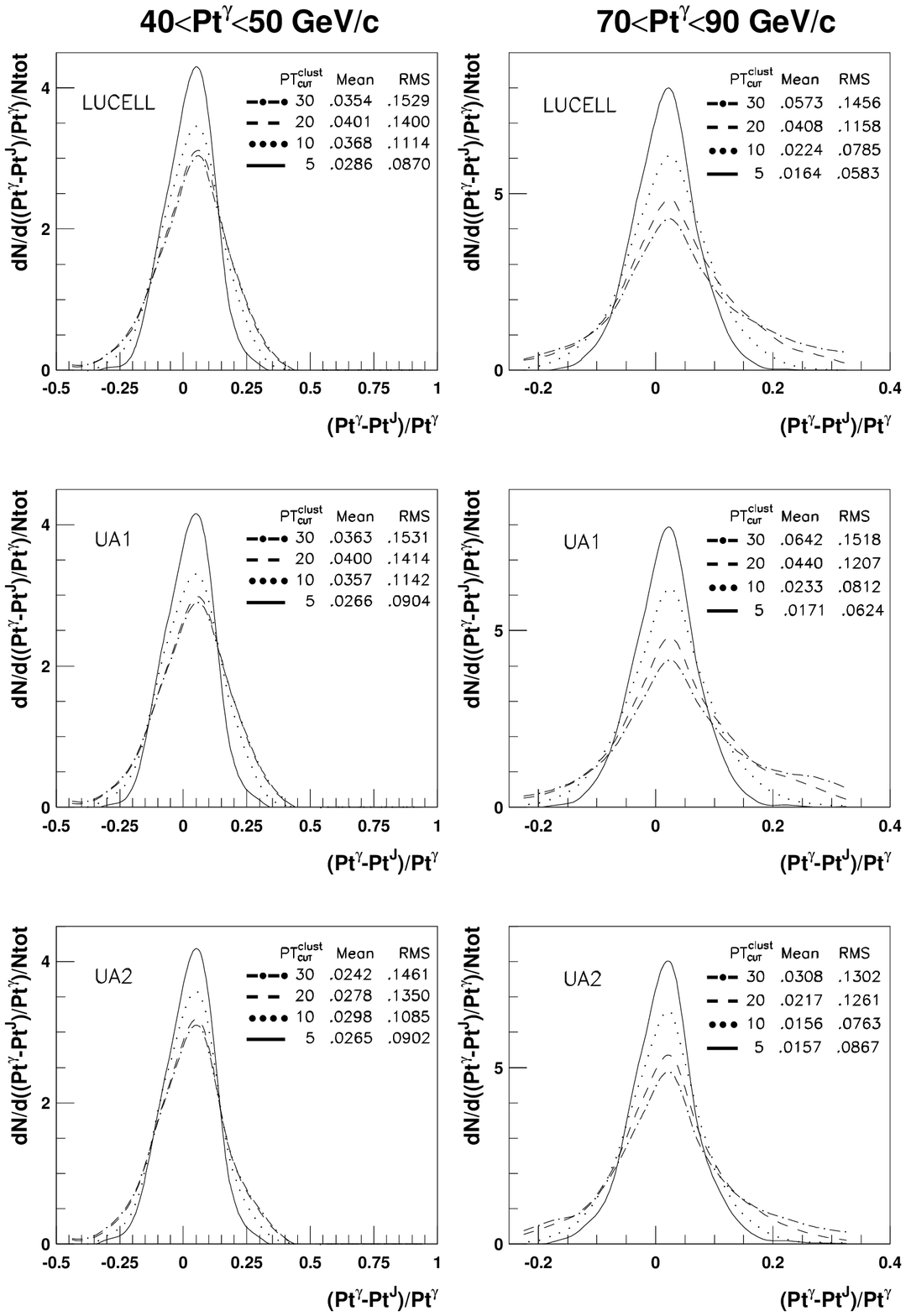} 
\caption{A dependence of $(\Pt^{\gamma}-\Pt^{J})/\Pt^{\gamma}$ on
$\Pt^{clust}_{CUT}$ for LUCELL, UA1 and UA2 jetfinding algorithms and two
intervals of \ptg. ~~The mean and RMS of the distributions are displayed on
the plots. $\dphi\lt17^\circ$. $\Pt^{out}$ is not limited. Selection 1.}
\label{fig:j-clu}
\vskip25mm 
\end{figure}

The normalized event distributions over $\Fptgj$ for two most illustrative $\Ptg$ intervals
$40\lt\Ptg\lt50$ and $70\lt\Ptg\lt90 ~GeV/c$ are shown for a case of $\dphi\leq17^\circ$ 
in Fig.~\ref{fig:j-clu} in different plots for three jetfinders LUCELL, UA1 and UA2. These plots demonstrate
the dependence of the mean and mean square deviations on $\Pt^{clust}_{CUT}$ value.

More details on $\Pt^{clust}_{CUT}$ dependence of different important
features of \gpj events 
are presented in tables of Appendix 2. They include the information about 
a topology of events and mean values of most important variables that characterize
$\Ptg-\Pt^{Jet}$ disbalance.
%
%

This information can be useful as a model guideline
while performing jet energy calibration procedure and also may serve for fine tuning
of PYTHIA parameters while comparing its predictions with the collected real data.

Appendix 2 contains the tables for events with $\Pt^{\gamma}$ varying within three intervals:
$40\lt\Ptg\lt50, ~70\lt\Ptg\lt90$ and $90\lt\Ptg\lt140 ~GeV/c$.
$\Delta \phi$ is limited there by $\Delta \phi < 17^\circ$. 
Tables 1--3 correspond to the events passed Selection 1 with a jet found by UA1 algorithm.
Tables 4--6 correspond to the events passed Selection 2.
The latter allows to select events with the ''isolated jet'', i.e. events
with the total $\Pt$ activity in the $\Delta R = 0.3$ ring around the jet not 
exceeding $3\%$ of jet $\Pt$ (see Section 3.2)
\footnote{In contrast to the case of LHC energies, where we required in Selection 2 
$\epsilon^{jet}\leq 6-8\%$ for $40\lt\Ptg\lt 50$ (see \cite{GPJ}),  
at Tevatron energies, due to less $\Pt$ activity in the space beyond the jet,
one can impose the tighter cut $\epsilon^{jet}\leq 3\%$.}.

The columns in all tables correspond to five different
values of cut parameter $\Pt^{clust}_{CUT}=30,\ 20,\ 15,$ $10$ and $5 ~GeV/c$.
The upper lines contain the expected numbers $N_{event}$ of ``CC events''
(i.e. the number of signal \gpj events in which the jet is entirely fitted into the CC region 
of the calorimeter; see Section 5) for the integrated luminosity $L_{int}=300\;pb^{-1}$. 

In the next four lines of the tables we put the values of $\Pt56$,
$\Delta \phi$, $\Pt^{out}$ and $\Pt^{|\eta|>4.2}$
defined by formulae (3), (22), (24) and (5) respectively and
averaged over the events selected with a chosen $\Pt^{clust}_{CUT}$ value.

From the tables we see that the values of $\Pt56$, $\Delta \phi$, $\Pt^{out}$ decrease fast
with decreasing $\Pt^{clust}_{CUT}\,$, while the averaged values of
$\Pt^{|\eta|>4.2}$ show very weak dependence on it (practically constant).
%
%

The following three lines present the average values of the variables
$\gpart$,
$\Jpart$,
$\gJ$ (here J$\equiv$Jet)
that serve as measures of the $\Pt$ disbalance in the \gpp and \gpj
systems as well as a measure of the parton-to-hadrons (Jet) fragmentation effect. 

The lines 9, 10 include the averaged
values of $\Db/\Pt^{\gamma}$ and $\,(1-cos(\dphi))$ quantities that appear on 
the right-hand side of \ptgj balance equation (\ref{eq:sc_bal}).

After application of cut $\dphi\lt17^\circ$
the value of $\left<1-cos(\dphi)\right>$ becomes smaller than the value of
$\left<\Db/\Pt^{\gamma}\right>$ in the case of Selection 1 and tends to decrease faster with
growing energy. So, we can again conclude that the main contribution 
into the $\Pt$ disbalance in the \gpj system, as defined by equation (\ref{eq:sc_bal}), comes from
the term $\Db/\Pt^{\gamma}$. With Selection 2 the contribution of
$\left<\Db/\Pt^{\gamma}\right>$ reduces with growing $\Pt^{clust}$
to the level of $\left<1-cos(\dphi)\right>$ and even to smaller values.

We have estimated separately the contributions of two terms  
${{\bf \Pt}^{O}}\cdot \bf{n^{Jet}}$ and ${{\bf \Pt}^{|\eta|>4.2}}\cdot \bf{n^{Jet}}$ 
that enter $\Db$ (see (\ref{eq:sc_bal})). 

Firstly from tables it is easily seen that $\Pt^{|\eta|>4.2}$ has practically the same value
in all $\Ptg$ intervals and it does not depend neither on $\dphi$ nor on $\Pt^{clust}$ values
being equal to $2 ~GeV/c$ up to a good precision
\footnote{Let us emphasize that it is a prediction of PYTHIA.}.

A mean value of $|{{\bf \Pt}^{|\eta|>4.2}}\cdot \bf{n^{Jet}}|$ contribution does not 
exceed $\approx 0.15~ GeV/c$ and a width (RMS) of the corresponding distribution 
contributes only $11-12\%$ to the total
width of the $\Db$ distribution. So, a mean and a width of $\Db$ are caused mainly by measurable term
${{\bf \Pt}^{O}}\cdot \bf{n^{Jet}}$
\footnote{A contribution of ${{\bf \Pt}^{O}_{(\nu)}}$ and ${{\bf \Pt}^{O}_{(\mu, |\eta^\mu|>2.5)}}$ 
(see (\ref{eq:out})) in the selected event samples is a negligibly small.}. 
Below in this section the cuts
on the value of $\Pt^{out}$ is applied to select events with better \Ptgj balance.

The following two lines contain the averaged values of the standard deviations
{\small $\sgmgj$} and {\small $\sgmgp$} of $\gJ (\equiv Db[\gamma,J])$ and
$\gpart(\equiv Db[\gamma,part])$ respectively.
These two variables drop approximately by about $50\%$ 
(and even more for $\Pt^{\gamma}>70 ~GeV/c$)
with restricting from $\Pt^{clust}_{CUT}=30 ~GeV/c$ to $5 ~GeV/c$
for all $\Pt^{\gamma}$ intervals.

The last lines of the tables present the number of generated events 
left after cuts.

Two features are clearly seen from these tables
\footnote{As was shown in \cite{D0_Note,BKS_P3} a transition from $\dphi\leq180^\circ$ to $\dphi\leq17^\circ$
supposed to be most effective in low $\Ptg$ intervals, does not affect the $\Fptgj$ disbalance strongly 
as compared with ``jet isolation'' criterion or cut on $\Pt^{clust}$ and $\Pt^{out}$}:\\
(1) in events with $\dphi\lt17^\circ$ the fractional disbalance on the {\it parton-photon} level $\gpart$\\
\hspace*{5mm} reduces to about $1\%$ (or even less) after imposing $\Pt^{clust}\lt 10~GeV/c$. It means that 
$\Pt^{clust}_{CUT}=$ \\
\hspace*{5mm} $10~GeV/c$  is really effective for ISR suppression as it was supposed in Section 3.1.\\
(2) {\it parton-to-jet} hadronization/fragmentation effect, that includes also FSR, can be estimated by
\hspace*{5mm}  the value of the following ratio $\Jpart$. It always has a negative value. It means \\
\hspace*{5mm} that a jet loses some part of the parent parton transverse momentum $\Pt^{part}$. It is seen that in  \\ 
\hspace*{5mm} the case of Selection 1 this effect gives a big contribution into \Ptgj disbalance even  \\ 
\hspace*{5mm} after application of $\Pt^{clust}_{CUT}=10~GeV/c$. The value of the fractional $\Jpart$ disbal- \\
\hspace*{5mm} ance does not vary strongly with $\Pt^{clust}_{CUT}$ in the cases of Selections 2 and 3.

We also see from the tables  that more restrictive cuts on
the observable $\Pt^{clust}$ lead to a decrease in the values of $\Pt56$ variable
(non-observable one) that serves, according to (3), as
a measure of the initial state radiation transverse momentum $\Pt^{ISR}$,
i.e. of the main source of the $\Pt$ disbalance in
the fundamental $2\to 2$ subprocesses (1a) and (1b).
Thus, variation of $\Pt^{clust}_{CUT}$ from $30 ~GeV/c$ to
$5 ~GeV/c$ (for $\dphi<17^\circ$) leads to suppression of the $\Pt56$ value
(or $\Pt^{ISR}$) approximately by $40\%$ for $40<\Pt^{\gamma}<50 ~GeV/c$
and by $\approx 60\%$ for $\Pt^{\gamma} \geq 90 ~GeV/c$.

In the first two intervals with $\Ptg\lt90~GeV/c$~ the decrease in $\Pt^{clust}_{CUT}$ leads to
some decrease in the $(\Pt^{\gamma}-\Pt^J)/\Pt^{\gamma}$
ratio (see Tables 1,2 of Appendix 2 and Fig.~10). 
In the case of $90<\Pt^{\gamma}<140 ~GeV/c$ the mean value of
$(\Pt^{\gamma}\!-\!\Pt^J)/\Pt^{\gamma}$ drops from $4.2\%$ to
$1.1\%$ (see Table 3 of Appendix 2).
After we pass to Selection 2 (Table 4--6 of Appendix 2) this disbalance becomes of the 
$1\%$ level and smaller but at the cost of statistics loss (by about $40-60\%$).
Tables 4--6 clearly show the prediction of PYTHIA about
the best level of jet calibration precision that can be achieved
after application of Selection 2.

{\normalsize \it Thus, to summarize the results presented in tables of  Appendix 2,
we want to underline that only
after imposing the jet isolation requirement (see Tables 4--6 of Appendix 2)
the mean values of $\Pt^{\gamma}$ and $\Pt^{Jet}$ disbalance, i.e.
$(\Pt^{\gamma}\!-\!\Pt^J)/\Pt^{\gamma}$, for all $\Ptg$ intervals 
are contained inside the $1\%$ window for any $\Pt^{clust}\leq20~GeV/c$.
The reduction of $\Pt^{clust}$ leads to lower values of mean square deviations 
of the photon-parton $Db[\gamma,part]$ and of photon-jet $Db[\gamma,J]$ balances. 
}

The Selection 2  (with $\Pt^{clust}_{CUT}=10 ~GeV/c$, for instance) 
leaves after its application the following number of events 
with jets {\it entirely contained} (see Section 5) {\it in the CC region} 
at $L_{int}=300 ~pb^{-1}$:\\
(1) about 4000 for  $40<\Pt^{\gamma}<50 ~GeV/c$,~~~~
(2) about 3000 for  $50<\Pt^{\gamma}<70 ~GeV/c$, \\
(3) about 850 for  $70<\Pt^{\gamma}<90 ~GeV/c$ and 
(4) about 500  for the $90<\Pt^{\gamma}<140 ~GeV/c$. 


So, we can say
that Selections 2, besides improving the \ptgj balance value,
is also important for selecting events with a clean jet topology and for rising the confidence level 
of a jet determination.

Up to now we have been studying the influence of the $\Pt^{clust}_{CUT}$
parameter on the balance. Let us see, in analogy with Fig.~\ref{fig:j-clu},
what effect is produced by $\Pt^{out}_{CUT}$ variation
\footnote{This variable enters into the expression $\Db/\Pt^{\gamma}$,
which makes a dominant contribution to the right-hand side of $\Pt$ balance
equation (\ref{sc_bal}), as we mentioned above.}.

\begin{figure}
\vskip-10mm
  \hspace*{-2mm} \includegraphics[width=16cm]{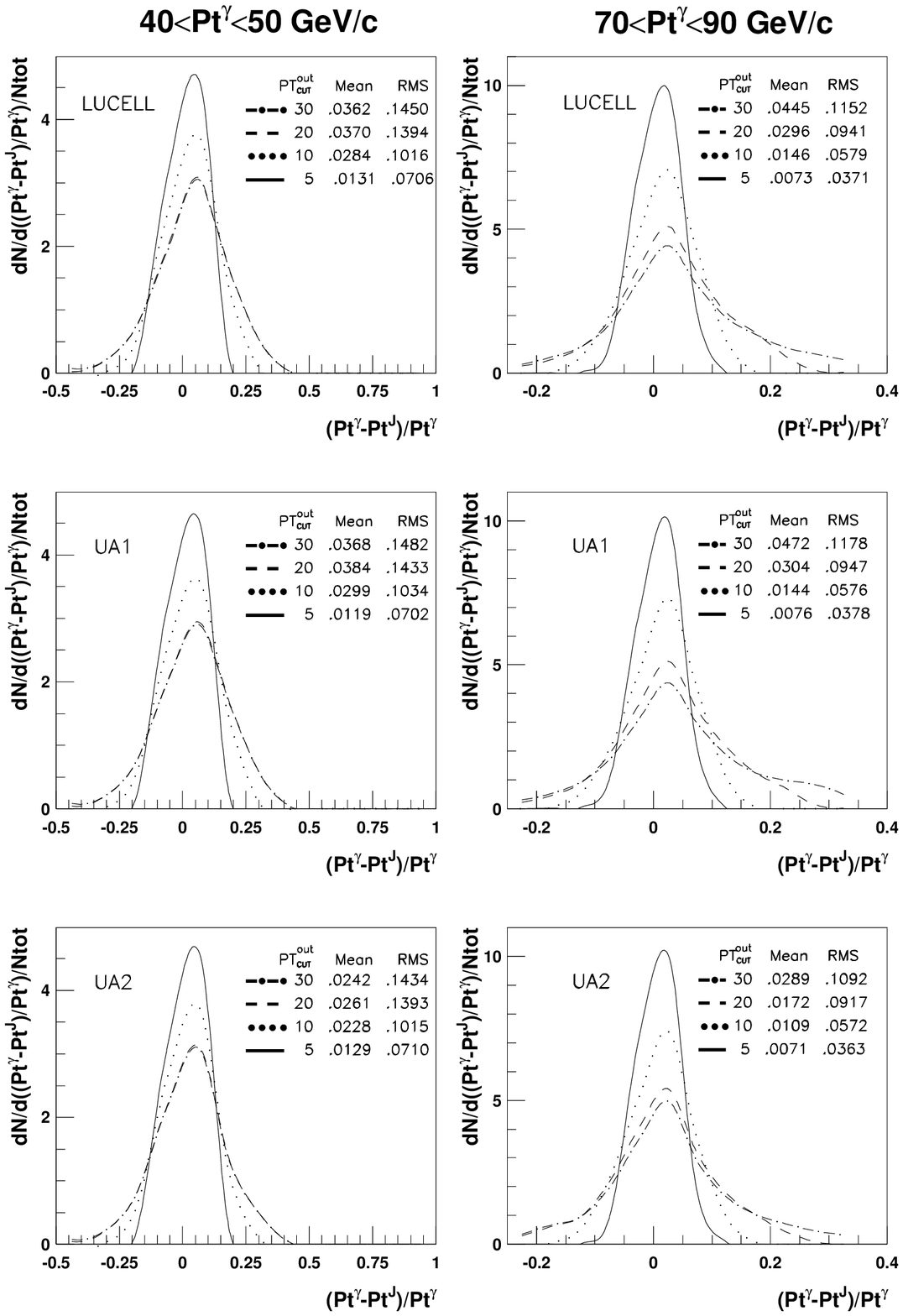} 
\caption{A dependence of $(\Pt^{\gamma}-\Pt^{J})/\Pt^{\gamma}$ on
$\Pt^{out}_{CUT}$ for LUCELL, UA1 and UA2 jetfinding algorithms and two
intervals of \ptg.  ~~The mean and RMS of the distributions are displayed on
the plots. $\dphi\leq17^\circ$, $\Pt^{clust}_{CUT}=30 ~GeV/c$. Selection 1.}
\label{fig:j-out}
\vskip25mm 

\end{figure}

If we  $\Pt^{out}_{CUT}=5~ GeV/c$, keeping $\Pt^{clust}$ practically unbound
by $\Pt^{clust}_{CUT}=30~ GeV/c$, then, as can be seen from Fig.~\ref{fig:j-out}, the mean
and RMS values of the $(\Pt^{\gamma}\!-\!\Pt^J)/\Pt^{\gamma}$
in the case of the LUCELL algorithm for $40<\Pt^{\gamma}<50~GeV/c$ 
decrease from $3.6\%$ to $1.3\%$ and from $14.5\%$
to $7.1\%$, respectively. For $70<\Pt^{\gamma}<90~GeV/c$ 
the mean and RMS values drop from $4.5\%$  to $0.7\%$ and from $11.5\%$ to $3.7\%$ respectively.
From these plots we also may conclude that variation of $\Pt^{out}_{CUT}$ improves the 
$\Pt$-disbalance, in fact, almost in the same way as the variation of
$\Pt^{clust}_{CUT}$. It is not surprising as the  cluster $\Pt$
activity is a part of the $\Pt^{out}$ activity. 

\begin{figure}
\vskip-10mm
  \hspace*{-2mm} \includegraphics[width=16cm]{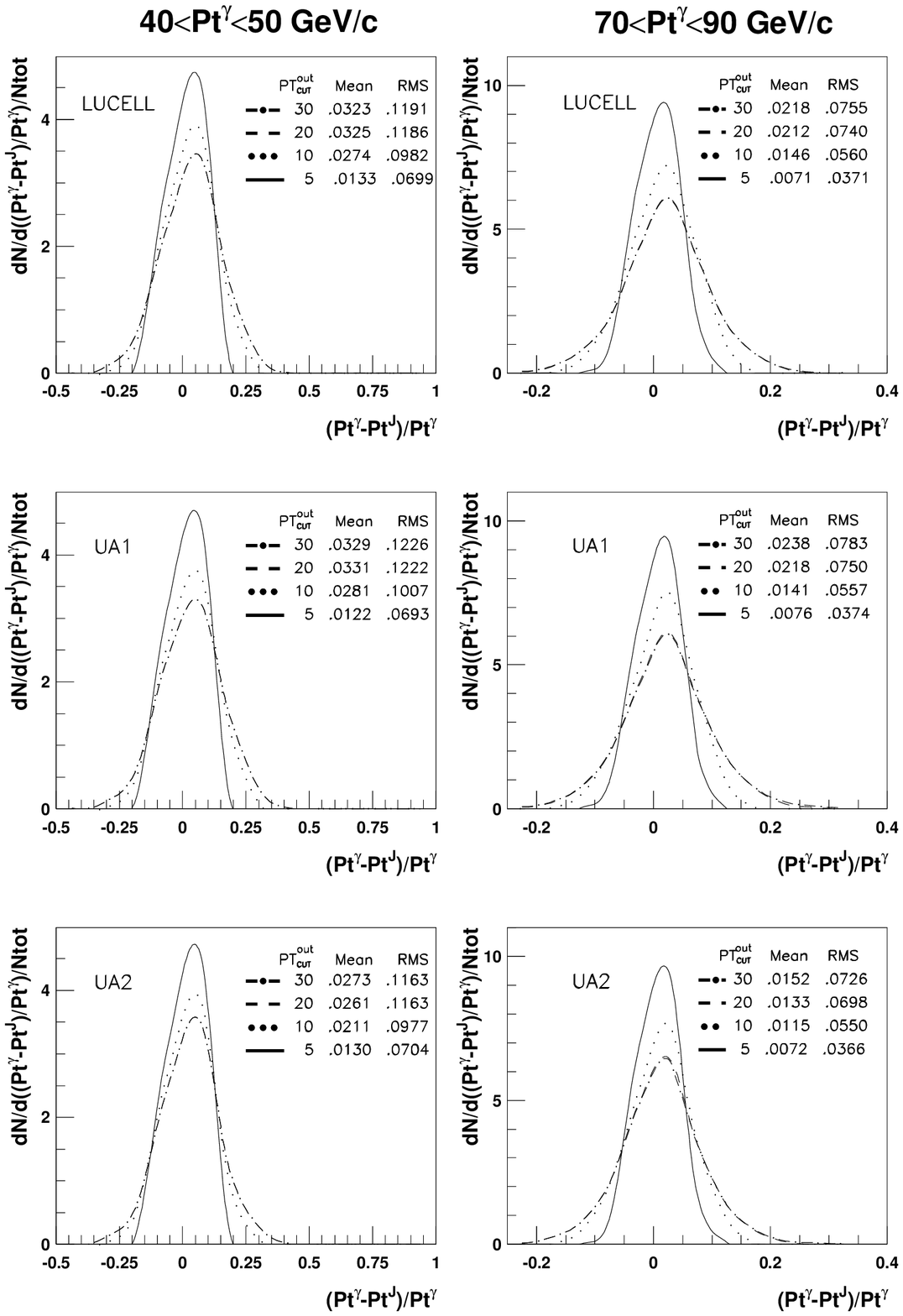} 
\caption{A dependence of $(\Pt^{\gamma}-\Pt^{J})/\Pt^{\gamma}$ on
$\Pt^{out}_{CUT}$ for LUCELL, UA1 and UA2  jetfinding algorithms and two
intervals of \ptg.  ~~The mean and RMS of the distributions are displayed on
the plots. $\dphi\leq17^\circ$, $\Pt^{clust}_{CUT}=10 ~GeV/c$. Selection 1.}
\label{fig:j-out-c}
\vskip25mm 
\end{figure}

The influence of the $\Pt^{out}_{CUT}$ variation on the
distribution of $(\Pt^{\gamma}\!-\!\Pt^J)/\Pt^{\gamma}$ is shown in
Fig.~\ref{fig:j-out-c} for Selection 1 with the fixed value
$\Pt^{clust}_{CUT}=10 ~GeV/c$. In this case the mean value of
$(\Pt^{\gamma}\!-\!\Pt^J)/\Pt^{\gamma}$ drops
from $3.2\%$ to $1.3\%$ for LUCELL and from $2.7\%$ to $1.3\%$ for UA2
algorithms for the $40<\Pt^{\gamma}<50~ GeV/c$ interval. At the same time
RMS value changes from $12\%$ to $7\%$ for all algorithms.
For interval $70<\Pt^{\gamma}<90~ GeV/c$ the mean value of fractional disbalance
$(\Pt^{\gamma}\!-\!\Pt^J)/\Pt^{\gamma}$ decrease to to $1.2-1.4\%$ at $\Pt^{out}_{CUT}=10 ~GeV/c$
and to less then $1\%$ at $\Pt^{out}_{CUT}=5 ~GeV/c$.
Simultaneously, RMS decreases to about $3.7\%$ for all three jetfinders.

More detailed study of $\Pt^{out}_{CUT}$ influence on the $\Fptgj$ disbalance
will be continued in the following Section 8 (see also Appendix 3).

{\it 
So, we conclude basing on the analysis of PYTHIA simulation (as a model)
 that the new cuts $\Pt^{clust}_{CUT}$ and $\Pt^{out}_{CUT}$ introduced in Section 3
as well as introduction of a new object, the ``isolated jet'', are found as those that may be
very efficient tools to improve the jet calibration accuracy. }
Their combined usage for this aim and for the background suppression will be a subject of a further
more detailed study in Section 8. 



\newpage


\section{ESTIMATION OF BACKGROUND SUPPRESSION CUTS EFFICIENCY.}       

\it\small
\hspace*{9mm}
The relative efficiency of ``hadronic'' cuts that are added to ``photonic'' ones, used to suppress the background
in the case of inclusive photon measurement, is estimated at the particle level. It is shown that an imposing of new cuts
on $\Pt$ of ``clusters'' ($\Pt^{clust}_{CUT}$) and on $\Pt$ activity in the region out of \gpj system 
($\Pt^{out}_{CUT}$) and also an application of ``jet isolation'' criterion would allow  to achieve
further (after ``photonic'' cuts) fourteen-fold background suppression at the cost of four-fold loss of the signal 
``$\gamma^{\,dir}+jet$'' events.

It is also shown that the imposing of $\Pt^{out}_{CUT}$, $\Pt^{clust}_{CUT}$ together with a usage of jet 
isolation criterion would lead to a substantial improvement of \ptgj balance.

The potentially dangerous role of a new source of background 
to the signal ``$\gamma^{\,dir}+jet$'' events caused by hard
bremsstrahlung photons (``$\gamma-brem$'') is demonstrated. It is shown that at Tevatron energy this new
 irreducible background may be compatible at low $\Ptg$ intervals with the $\pi^0$ contribution
and it may grow faster with $\Ptg$ increasing than the latter one.

\rm\normalsize
\vskip5mm

\noindent
\hspace*{8mm}
To estimate an efficiency of the selection criteria proposed in Section 3.2  we
carried out the simulation
\footnote{ PYTHIA~5.7 version with default CTEQ2L parameterization
of structure functions is used here.}
with a mixture of all QCD and SM subprocesses with large cross sections existing in PYTHIA
\footnote{ISUB=1, 2, 11--20, 28--31, 53, 68 (in notations of PYTHIA)}. 
The events caused by this set of the subprocesses may give a large background 
to the ``$\gamma^{dir}+jet$'' signal events defined by  the subprocesses (1a) and (1b)
\footnote{ISUB=29 and 14 in PYTHIA. A contribution of another possible NLO channel $gg\rrr g\gamma$
(ISUB=115) was found to be still negligible even at Tevatron energies.}
that were also included in this simulation.

Three generations  with the above-mentioned set of subprocesses
were performed. Each of them was done with a different value of
$\pth$ parameter
\footnote{CKIN(3) in PYTHIA} 
that defines a minimal value 
of $\Pt$ appearing in the final state of a hard $2\to 2$ parton level fundamental subprocess
in the case of ISR absence. These values were $\pth=40, 70$ and $100~GeV/c$. 
By 40 million events were generated for each of $\pth$ value. 
The cross sections of the above-mentioned subprocesses define the rates of corresponding physical
events and thus appear in simulation as weight factors.

We selected ``$\gamma^{dir}$-candidate +1 jet'' events  containing 
one $\gamma^{dir}$-candidate (denoted in what follows as ${\tilde{\gamma}}$) and one jet, found by LUCELL,
with $\Pt^{jet}> 30~ GeV/c$.
Here and below, as we work at the PYTHIA particle level of simulation, speaking about the $\gamma^{dir}$-candidate 
we actually mean, apart from  $\gamma^{dir}$,  a set of particles
like electrons, bremsstrahlung photons and also photons from neutral meson decays that may be registered in 
one D0 calorimeter tower of the $\Delta\eta\times\Delta\phi=0.1\times0.1$ size.

Here we consider a set of 17 cuts that are separated into 2 subsets: 6 ``photonic'' cuts and 11
``hadronic'' ones. The first subset consists of the cuts used to select an isolated photon candidate
in some $\Pt^{\tilde{\gamma}}$ interval. The second one includes the cuts 
connected mostly with jets and clusters and are used to select events having
one ``isolated jet'' and limited $\Pt$ activity out of ``${\tilde{\gamma}}+jet$'' system.

The used cuts are listed in Table \ref{tab:sb0}. To give an idea
about their physical meaning and importance we have done an estimation of their possible 
influence on the signal-to-background ratios $S/B$. The letter were calculated after application of each cut.
Their values are presented in Table \ref{tab:sb4} for a case of the most illustrative intermediate interval
of event generation with $\pth=70~GeV/c$. In this table the number in each line corresponds
to the number of the cut in Table \ref{tab:sb0}. 
Three important lines of Table \ref{tab:sb4} are darkened because they
will be often referenced to while discussing the following Tables \ref{tab:sb1}--\ref{tab:sb3}.

The efficiencies $Eff_{S(B)}$ (with their errors) in Table \ref{tab:sb4}  are defined as a ratio
of the number of signal (background) events that passed under a cut
(1--17) to the number of the preselected events (1st cut of this table).
\\[-6mm]
\begin{table}[h]
\caption{List of the applied cuts (will be used also in Tables \ref{tab:sb4}--\ref{tab:sb3}).}
\begin{tabular}{lc} \hline
\label{tab:sb0}
\hspace*{-2.2mm} {\bf 1}. $a)~\Pt^{\tilde{\gamma}}\geq 40 ~GeV/c, 
~~b)~\Pt^{jet}\geq 30 ~GeV/c,$  
\hspace*{9.5mm} {\bf 9}. $\dphi<17^\circ$; \\
\hspace*{4.1mm} $c)~|\eta^{\tilde{\gamma}}|\leq 2.5, ~~~~~~~~~~d)~\Pt^{hadr}\!<7 ~GeV/c^{\;\ast}$;
\hspace*{8.0mm}{\bf 10}. $\Pt^{miss}/\Pt^{\tilde{\gamma}}\!\leq0.10$; \\
{\bf 2}. $\Pt^{isol}\!\leq 5~ GeV/c, ~\epsilon^{\tilde{\gamma}}<15\%$;
\hspace*{3.08cm} {\bf 11}. $\Pt^{clust}<20 ~GeV/c$; \\
{\bf 3}. $\Pt^{\tilde{\gamma}}\geq\pth$;
\hspace*{59mm} {\bf 12}. $\Pt^{clust}<15 ~GeV/c$; \\
{\bf 4}. $\Pt^{isol}_{_ring} \leq 1~ GeV/c^{\;\ast\ast}$;
\hspace*{4.42cm} {\bf 13}. $\Pt^{clust}<10 ~GeV/c$; \\
{\bf 5}. $\Pt^{isol}\!\leq 2~ GeV/c, ~\epsilon^{\tilde{\gamma}}<5\%$;
\hspace*{3.25cm} {\bf 14}. $\Pt^{out}<20 ~GeV/c$; \\
{\bf 6}. $Njet\leq3$;
\hspace*{6.24cm}  {\bf 15}. $\Pt^{out}<15 ~GeV/c$; \\
{\bf 7}. $Njet\leq2$;
\hspace*{6.23cm}  {\bf 16}. $\Pt^{out}<10 ~GeV/c$;\\
{\bf 8}. $Njet=1$;
\hspace*{6.22cm} {\bf 17}. $\epsilon^{jet} \leq 3\%$.\\\hline
\footnotesize{${\;\ast}$ maximal $\Pt$ of a hadron in the ECAL cell containing a 
$\gamma^{dir}$-candidate;}\\
\footnotesize{${\;\ast\ast}$ A scalar sum of $\Pt$ in the ring:
$\Pt^{sum}(R=0.4)-\Pt^{sum}(R=0.2)$.}\\[-3mm]
\end{tabular}
\end{table}
\normalsize

Line number 1 of  Table \ref{tab:sb0} makes primary preselection. It includes and specifies
our first general cut (16) of Section 3.2 as well as the cut connected with ECAL geometry and 
the cut (19) that excludes $\gamma^{dir}$-candidates accompanied by hadrons.

Line number 2 of  Table \ref{tab:sb0} fixes the values of $\Pt^{isol}_{CUT}$ and $\epsilon^{\gamma}_{CUT}$
that, according to (17) and (18), define the isolation parameters of ${\tilde{\gamma}}$.

The third cut selects the events containing $\gamma^{dir}$-candidates with $\Pt$ higher than
$\pth(\equiv CKIN(3))$ threshold
\footnote{see PYTHIA manual \cite{PYT}}. 
We impose the third cut to select the samples of events with
$\Pt^{\tilde{\gamma}}\geq40, 70$ and $100~GeV/c$ as ISR may smear the sharp kinematical cutoff defined by
$CKIN(3)$ \cite{PYT}. This cut reflects an experimental viewpoint when one is interested in
how many events with $\gamma^{dir}$-candidates are contained in some definite interval of $\Pt^{\tilde{\gamma}}$.

The forth cut restricts a value of $\Pt^{isol}_{_ring}=\Pt^{isol}_{R=0.4}-\Pt^{isol}_{R=0.2}$,
where $\Pt^{isol}_{R}$ is a sum of $\Pt$ of all ECAL cells contained in the cone of the radius 
$R$ around the cell fired by $\gamma^{dir}$-candidate \cite{D0_1}, \cite{D0_2}. 

The fifth cut makes tighter the isolation criteria within $R=0.7$ than those imposed onto 
$\gamma^{dir}$-candidate in the second line of Table \ref{tab:sb0}.


%
%

The cuts considered up to now, apart from general preselection cut $\Pt^{jet}\geq30~GeV/c$
used in the first line of Table \ref{tab:sb0},
were connected with photon selection (``photonic'' cuts). 
Before we go further, some words of caution must be said here.
Firstly, we want to emphasize that the starting numbers of the signal ($S$) and background ($B$)
events (first line of Table \ref{tab:sb4}) may be specific only for PYTHIA generator and for the way of
preparing primary samples of the signal and background events described above. So, we want to underline here
that the starting values of $S$ and $B$ in the first columns of Table \ref{tab:sb4} are model dependent.

But nevertheless, for our aim of investigation of efficiency of new cuts 11--17 
(see \cite{9}--\cite{BKS_P5})  the important thing here is that we can use these starting model numbers of
$S$- and $B$-events  for studying a further relative  influence of these cuts on $S/B$ ratio.

The cuts 6--9 are connected with the selection of events having only one jet 
and the definition of jet-photon spatial orientation in $\phi$-plane. The 9-th cut selects 
the events with jet and photon transverse momenta being ``back-to-back'' to each other 
in $\phi-$plane within the angle interval of the $\dphi=17^\circ$ size
\footnote{i.e. within the size of three calorimeter cells}.
%

\begin{figure}
\vskip-12mm
\hspace{-2mm} \includegraphics[height=7cm,width=15cm]{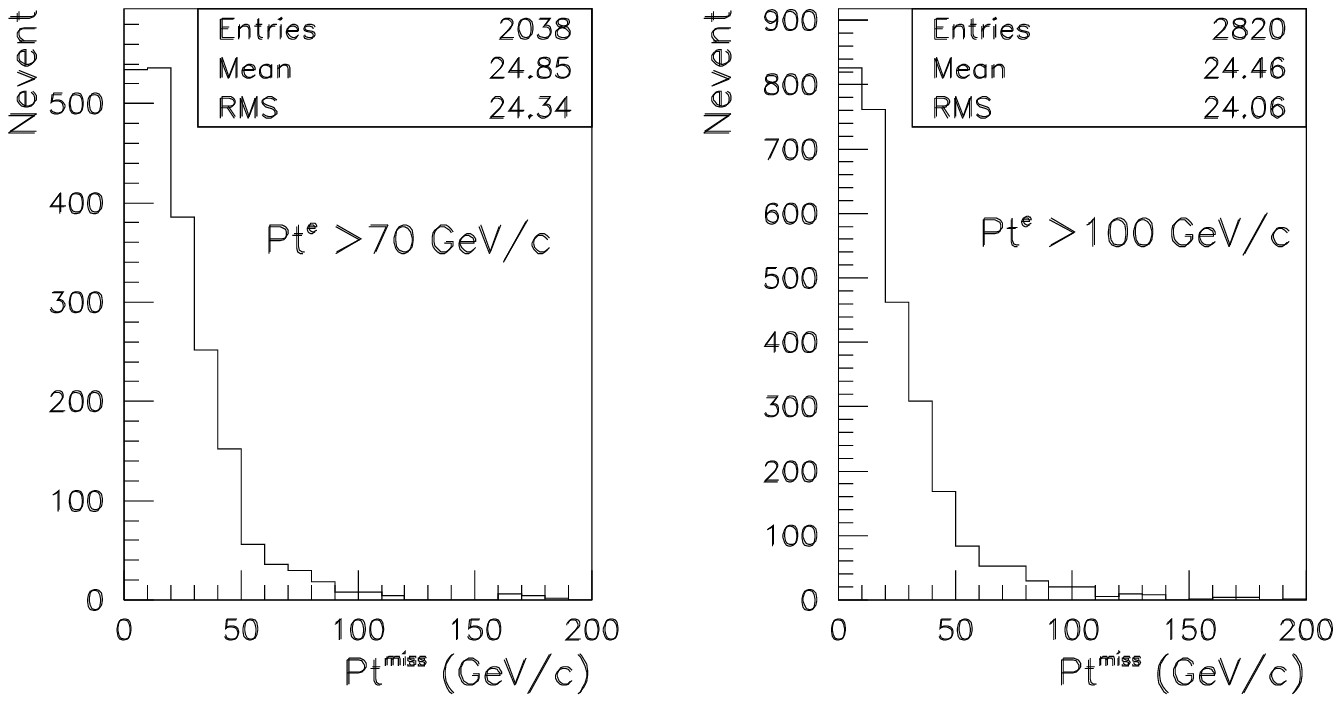} 
\vskip-9mm
\caption{Distribution of events over $\Pt^{miss}$
in  events with energetic $e^\pm$`s appearing as direct photon candidates for 
the cases ~$\Pt^e\geq70~ GeV/c~$ and $~\Pt^e\geq100~ GeV/c$~ 
(here are used events satisfying cuts 1--5 of Table \ref{tab:sb0}).}
\label{fig:ptmiss}
\end{figure}

In line 10 we used the cut on $\Pt^{miss}$ to reduce a background contribution
from the electroweak subprocesses $q\,g \to q' + W^{\pm}$ and $q\bar{~q'} \to g + W^{\pm}$ 
with the subsequent decay $W^{\pm} \to e^{\pm}\nu$ that leads to a substantial $\Pt^{miss}$ value.
It is clear from the distributions over $\Pt^{miss}$ for two
$\Pt^e$ intervals presented in Fig.~\ref{fig:ptmiss} (compare with Fig.~\ref{fig20-22}). 
One can see from the last column of Table \ref{tab:sb4} ``$e^\pm$''
that the cut on $\Pt^{miss}$ reduces strongly (in about 4 times) 
the number of events containing $e^\pm$ as direct photon candidate.

\begin{table}[htbp]
\begin{center}
\vskip-14mm
\small
\vspace*{11mm}
\caption{Values of significance and  efficiencies for $\pth=70~GeV/c$.}
\vskip0.5mm
\begin{tabular}{||c||c|c|c|c|c|c||}                  \hline \hline
\label{tab:sb4}
Cut& $S$ & $B$ & $Eff_S(\%)$ & $Eff_{B}(\%)$  & $S/B$& $e^\pm$ \\\hline \hline
\rowcolor[gray]{\coltab}%
\rowcolor[gray]{\coltab}%
1 & 39340 &   1247005&  100.00$\pm$  0.00& 100.000$\pm$  0.000&  0.03& 17562\\\hline 
2 & 36611 &     51473 &   93.06$\pm$  0.68 &   4.128$\pm$  0.019 &  0.71 & 4402 \\\hline 
3 & 29903 &     18170 &   76.01$\pm$  0.58 &   1.457$\pm$  0.011 &  1.65  &2038 \\\hline 
4 & 26426 &     11458 &   67.17$\pm$  0.53 &   0.919$\pm$  0.009 &  2.31  &1736 \\\hline 
5 & 23830 &      7504 &   60.57$\pm$  0.50 &   0.602$\pm$  0.007 &  3.18  &1568 \\\hline 
6 & 23788 &     7406 &   60.47$\pm$  0.50 &   0.594$\pm$  0.007  & 3.21  &1554 \\\hline 
7 & 23334 &     6780 &   59.31$\pm$  0.49 &   0.544$\pm$  0.007 &  3.44  &1460 \\\hline 
8 & 19386 &     4136 &   49.28$\pm$  0.43 &   0.332$\pm$  0.005 &  4.69  &1142 \\\hline 
9 & 18290 &      3506 &   46.49$\pm$  0.42 &   0.281$\pm$  0.005  & 5.22  &796 \\\hline 
10 &18022 &      3418 &   45.81$\pm$  0.41 &  0.274$\pm$  0.005  & 5.27  &210 \\\hline 
11 & 15812 &     2600 &   40.19$\pm$  0.38 &   0.208$\pm$  0.004  & 6.08  &176 \\\hline 
12 & 13702 &     1998 &   34.83$\pm$  0.35 &   0.160$\pm$  0.004  & 6.86  &130 \\\hline 
13 & 10724 &     1328 &   27.26$\pm$  0.30 &   0.106$\pm$  0.003  & 8.08  &88 \\\hline 
14 & 10636 &      1302 &   27.04$\pm$  0.30 &   0.104$\pm$  0.003  & 8.17  & 86 \\\hline 
15 & 10240 &      1230 &   26.03$\pm$  0.29 &   0.099$\pm$  0.003  & 8.33  & 84 \\\hline 
\rowcolor[gray]{\coltab}%
16 &  8608 &       984 &   21.88$\pm$  0.26 &   0.079$\pm$  0.003  & 8.75  & 64 \\\hline 
\rowcolor[gray]{\coltab}%
17  & 6266 &       622 &   15.93$\pm$  0.22 &   0.050$\pm$  0.002  &10.07  & 52 \\\hline 
\hline 
\end{tabular}
\end{center}
\vskip-4mm
\noindent
\footnotesize{${(\ast)}$ The background ($B$) does not include
the contribution from the ``$e^\pm$ events'' (i.e. in which $e^\pm$ fake
$\gamma$-candidate) that is shown separately in the right-hand column ``$e^\pm$''.}
\vskip0mm
\end{table}
\normalsize

Moving further we see from Table \ref{tab:sb4} that the
cuts 11--16 of Table \ref{tab:sb0} reduce the values of $\Pt^{clust}$ and $\Pt^{out}$ down to the values 
less than $10 ~GeV/c$. The 17-th cut of Table \ref{tab:sb0} imposes the jet isolation requirement. 
It leaves only the events
with jets having the sum of $\Pt$ in a ring surrounding a jet to be less than $3\%$ of $\Pt^{Jet}$.
From comparison of the numbers in 10-th and 17-th lines we make the important conclusion that all these
new cuts (11--17), despite of model dependent nature of starting $S/B$ value in line 10, may, in principle,
lead to the following about two-fold improvement of $S/B$ ratio. 
This improvement is reached by reducing the $\Pt$ activity out of ``$\tilde{\gamma}+1~jet$'' system. 

It is also rather interesting to mention that 
{\it the total effect of ``hadronic cuts'' 6--17 for the case of $\pth=70~GeV/c$
consists of about twelve-fold decrease of background contribution 
at the cost of less than four-fold loss of signal events (what results in about 3.2 times growth of $S/B$
ratio)}. So, in this sense, we may conclude that from the viewpoint of $S/B$ ratio a study of \gpj events 
may be more preferable as compared with a case of inclusive photon production.

Table \ref{tab:sb1} includes the numbers of signal and background events left in three generated event samples
after application of cuts 1--16 and 1--17. They are given for all three intervals of $\Pt^{\tilde{\gamma}}$.
Tables \ref{tab:sb1} and \ref{tab:sb4} are complementary to each other.
The summary of Table \ref{tab:sb4} is presented in the middle section ($\pth=70 ~GeV/c$)
of Table \ref{tab:sb1} where the line ``Preselected'' corresponds to the cut 1 of Table \ref{tab:sb0} 
and, respectively, to the line number 1 of  Table  \ref{tab:sb4} presented above.
The line ``After cuts'' corresponds to the line 16 of  Table \ref{tab:sb4} and 
line ``+jet isolation'' corresponds to the line 17 of  Table \ref{tab:sb4}. 
~\\[-8mm]
\normalsize
\begin{table}[h]
\begin{center}
\small
\caption{Number of signal and background events remained after cuts.}
\vskip.5mm
\begin{tabular}{||c|c||c|c|c|c|c|c|c||}                  \hline \hline
\label{tab:sb1}
\hmm$\pth$\hmm& &$\gamma$ & $\gamma$ &\multicolumn{4}{c|}{  photons from the mesons}  &
\\\cline{5-8}
\Gvc& Cuts&\hmm direct\hmm &\hmm brem\hmm & $\;\;$ $\pi^0$ $\;\;$ &$\quad$ $\eta$ $\quad$ &
$\omega$ &  $K_S^0$ &\hmm $e^{\pm}$\hmm \\\hline \hline
    &Preselected&\hmm18056&\hmm 14466& 152927& 56379& 17292& 14318&\hmm 2890\hmm  \\\cline{2-9}
 40 &After cuts &\hmm 6238&\hmm 686&     824&  396 &   112& 104&\hmm   24\hmm\\\cline{2-9}
    &+ jet isol. &\hmm 3094&\hmm 264&   338&    150&    40& 44&   14\\\hline  \hline
    &Preselected &\hmm39340&\hmm63982&761926&269666&87932& 63499 &\hmm17562  \hmm\\\cline{2-9}
 70 &After cuts&\hmm  8608 &\hmm  424& 320 &146 & 58  &36 &\hmm 64\hmm \\\cline{2-9}   
    &+ jet isol. &\hmm6266 &\hmm 262 & 206 &90 & 40  & 24 &\hmm 52\hmm \\\hline \hline
    &Preselected&\hmm56764 &\hmm111512 &970710 &346349 &117816 &91416 &\hmm38872\hmm\\\cline{2-9}
100 &After cuts&\hmm 11452 &\hmm 280 & 124 &92 & 24  & 24 &\hmm 136\hmm\\\cline{2-9}
    &+ jet isol. &\hmm 9672 &\hmm 204& 92 & 64 & 24  & 20 &\hmm 120\hmm\\\hline \hline
\end{tabular}
\vskip0.2cm
\caption{Efficiency, $S/B$ ratio and significance values in the selected events without jet isolation cut.}
\vskip0.1cm
\begin{tabular}{||c||c|c|c|c|>{\columncolor[gray]{\coltab}}c|c||} \hline \hline
\label{tab:sb2}
$\pth$ \Gvc& $S$ & $B$ & $Eff_S(\%)$  & $Eff_B(\%)$  & $S/B$& $S/\sqrt{B}$
\\\hline \hline
40  & 6238& 2122 & 34.55$\pm$0.51 & 0.831$\pm$0.018 & 2.9 & 135.4 \\\hline
70 & 8608&  984 & 21.88 $\pm$ 0.26 & 0.079 $\pm$ 0.003& 8.8 & 274.4 \\\hline 
100 & 11452& 544 & 20.17 $\pm$ 0.21 & 0.033 $\pm$ 0.001& 21.1 & 491.0 
\\\hline \hline
\end{tabular}
\vskip0.2cm
\caption{Efficiency, $S/B$ ratio and significance values in the selected events with jet isolation cut.}
\vskip0.1cm
\begin{tabular}{||c||c|c|c|c|>{\columncolor[gray]{\coltab}}c|c||}  \hline \hline
\label{tab:sb3}
$\pth$ \Gvc& ~~$S$~~ & ~~$B$~~ & $Eff_S(\%)$ & $Eff_B(\%)$  & $S/B$& $S/\sqrt{B}$
 \\\hline \hline
40  & 3094& 836 &17.14$\pm$0.33 & 0.327$\pm$0.011 &  3.7 & 107.0 \\\hline
70 & 6266& 622 & 15.93 $\pm$ 0.22 & 0.050 $\pm$ 0.002& 10.1 & 251.2 \\\hline
100 & 9672& 404 & 17.04 $\pm$ 0.19 & 0.025 $\pm$ 0.001& 23.9 & 481.2 
\\\hline \hline
\end{tabular}
\end{center}
\vskip-3mm
\end{table}
\normalsize

Table \ref{tab:sb1} is done to show in more detail the origin of $\gamma^{dir}$-candidates.
The numbers in the  ``$\gamma-direct$'' column correspond to the respective  numbers  of 
signal events left in each of $\Pt^{\tilde{\gamma}}$ intervals after application of the cuts defined
in lines 1, 16 and 17 of Table \ref{tab:sb0} (and column ``$S$'' of Table \ref{tab:sb4}). Analogously
the numbers in the ``$\gamma-brem$'' column of Table  \ref{tab:sb1} correspond to the numbers
of events with the photons radiated from quarks
participating in the hard interactions. 
Columns 5--8 of Table \ref{tab:sb1} illustrate the numbers of the ``$\gamma-mes$''  events with photons
originating from $\pi^0,~\eta,~\omega$ and $K^0_S$ meson decays.
In a case of $\Pt^{\tilde{\gamma}}\gt70~GeV/c$ the total numbers of background events,
i.e. a sum over the numbers presented in columns 4--8 of Table \ref{tab:sb1}, 
are shown in the lines 1, 16 and 17 of column ``$B$'' of Table \ref{tab:sb4}.
The other lines of Table \ref{tab:sb1} for $\pth\!=40$ and
$~100 ~GeV/c~$ have the meaning analogous to that described above for $\pth=70 ~GeV/c$.

The last column of Table \ref{tab:sb1} shows the number of preselected events with
$e^\pm$. 

The numbers in Tables \ref{tab:sb2} (without jet isolation cut) and \ref{tab:sb3}
(with jet isolation cut) accumulate in a compact form the final information of 
Tables \ref{tab:sb0} -- \ref{tab:sb1}. 
Thus, for example, the columns $S$ and $B$ of the  line that corresponds to $\pth=70 ~GeV/c$ 
contain the total numbers of the selected signal and background events taken at the level of 16-th  (for Table
\ref{tab:sb2}) and 17-th (for Table \ref{tab:sb3}) cuts from Table \ref{tab:sb4}. 

It is seen from Table \ref{tab:sb2}  that in the case of Selection 1 the ratio $S/B$ grows   
from 2.9 to 21.1 while $\Pt^{\tilde{\gamma}}$ increases from
$\Pt^{\tilde{\gamma}}\geq 40 ~GeV/c$ to $\Pt^{\tilde{\gamma}}\geq 100 ~GeV/c$ interval.

The jet isolation requirement (cut 17 from Table \ref{tab:sb0})
noticeably improves the situation at low $\Pt^{\tilde{\gamma}}$ (see Table \ref{tab:sb3}).
After application of this criterion the value of $S/B$ increases from 2.9 
to 3.7 at $\Pt^{\tilde{\gamma}}\geq 40 ~GeV/c$ 
and only from 21.1 to 23.9 at $\Pt^{\tilde{\gamma}}\geq 100 ~GeV/c$.
Remember on this occasion the conclusion  that the sample of events
selected with our criteria has a tendency to contain more events with an isolated jet
as $\Pt^{\tilde{\gamma}}$ increases 
\footnote{see Sections 5--7 and Appendix 2}.
%
%
%

Let us underline here that, in contrast to other types of background, ``$\gamma-brem$'' background
has an irreducible nature. Thus, the number of ``$\gamma-brem$'' events
should be carefully estimated for each $\Pt^{\tilde{\gamma}}$ interval using the particle level
of simulation in the framework of event generator like PYTHIA.
They are also have to be taken into account in experimental analysis
of the prompt photon production data at high energies.

Tables \ref{tab:bg_or_gr} and \ref{tab:bg_or_ms} 
shows  the relative contributions of fundamental QCD subprocesses  (having the largest cross sections)
$qg\to qg$, $qq\to qq$, $gg\to q\bar{q}$ and $gg\to gg$ 
\footnote{ISUB=11, 12, 28, 53 and 68 (see \cite{PYT})}
that define the main production of ``$\gamma\!-\!brem$'' and `$\gamma\!-\!mes$''
background in event samples selected with  criteria 1--13 of Table \ref{tab:sb0} 
in three $\Pt^{\tilde{\gamma}}$ intervals.

Accepting the results of simulation with PYTHIA, we found from the event listing analysis that
in the main part of selected ``$\gamma\!-\!brem$'' events
these photons are produced in the final state of the fundamental $2\to2$ subprocess
\footnote{i.e. from lines 7, 8 in Fig.~3}.
Namely, they are mostly radiated from the outgoing quarks 
in the case of the first three sets of subprocesses ($qg\to qg$, $qq\to qq$ and $gg\to q\bar{q}$).
They may also appear as a result of string breaking in a final state of 
$gg\to gg$ scattering. But this subprocess,
naturally, gives a small contribution into ``$\tilde{\gamma}+jet$'' events production.

~\\[-12mm]
\begin{table}[h]
\begin{center}
\vskip-3mm
\caption{Relative contribution (in per cents) of different QCD subprocesses into
the ``$\gamma\!-\!brem$'' events production.}
\normalsize
\vskip.1cm
\begin{tabular}{|c||c|c|c|c|}                  \hline \hline
\label{tab:bg_or_gr}
$\Ptg$& \multicolumn{4}{c|}{fundamental QCD subprocess} \\\cline{2-5}
 \Gvc & $qg\to qg$ & $qq\to qq$ & $gg\to q\bar{q}$& $gg\to gg$  
\\\hline \hline
 40--70   & 62.1$\pm$6.6 & 31.8$\pm$4.0 &  3.3$\pm$1.0 &  2.8$\pm$0.9  \\\hline 
 70--100  & 52.3$\pm$7.7 & 42.4$\pm$6.4 &  3.8$\pm$1.4 &  1.5$\pm$0.9 \\\hline 
 $>100$   & 41.8$\pm$6.0 & 56.9$\pm$7.2 &  1.3$\pm$0.7 &  ---  \\\hline\hline  
\end{tabular}
\end{center}
\vskip-2mm
\end{table}

\begin{table}[h]
\begin{center}
\vskip-7mm
\caption{Relative contribution (in per cents) of different QCD subprocesses into
the ``$\gamma\!-\!mes$'' events production.}
\normalsize
\vskip.1cm
\begin{tabular}{|c||c|c|c|c|}                  \hline \hline
\label{tab:bg_or_ms}
$\Ptg$& \multicolumn{4}{c|}{fundamental QCD subprocess} \\\cline{2-5}
 \Gvc & $qg\to qg$ & $qq\to qq$ & $gg\to q\bar{q}$& $gg\to gg$ 
\\\hline \hline
 40--70   & 59.3$\pm$5.2 & 34.8$\pm$3.5 &  2.9$\pm$0.7 &  2.4$\pm$0.7 \\\hline 
 70--100  & 48.6$\pm$8.0 & 47.3$\pm$7.8 &  0.5$\pm$0.5 &  0.5$\pm$0.5 \\\hline 
 $>100$   & 41.8$\pm$6.4 & 53.9$\pm$7.6 &  1.8$\pm$0.9 &  0.7$\pm$0.5  \\\hline\hline 
\end{tabular}
\end{center}
\vskip-3mm
\end{table}

It may be noted also from the first two columns of Tables \ref{tab:bg_or_gr} and \ref{tab:bg_or_ms}
that the most of ``$\gamma\!-\!brem$'' and ``$\gamma\!-\!mes$'' background events ($93\%$ at least) 
originate from $qg\to qg$ and $q_i q_j\to q_i q_j$, 
$q_i\bar{q_i}\to q_j\bar{q_j}$ subprocesses. Tables  \ref{tab:bg_or_gr} and \ref{tab:bg_or_ms}
show also a tendency of increasing the contribution from the subprocess $q_i q_j\to q_i q_j$ and
$q_i\bar{q_i}\to q_j\bar{q_j}$ (given in the second columns of tables) with growing $\Pt^{\tilde{\gamma}}$.

\begin{figure}[htbp]
\vspace{-0.8cm}
\hspace{3.7cm} \includegraphics[width=10cm,height=8cm]{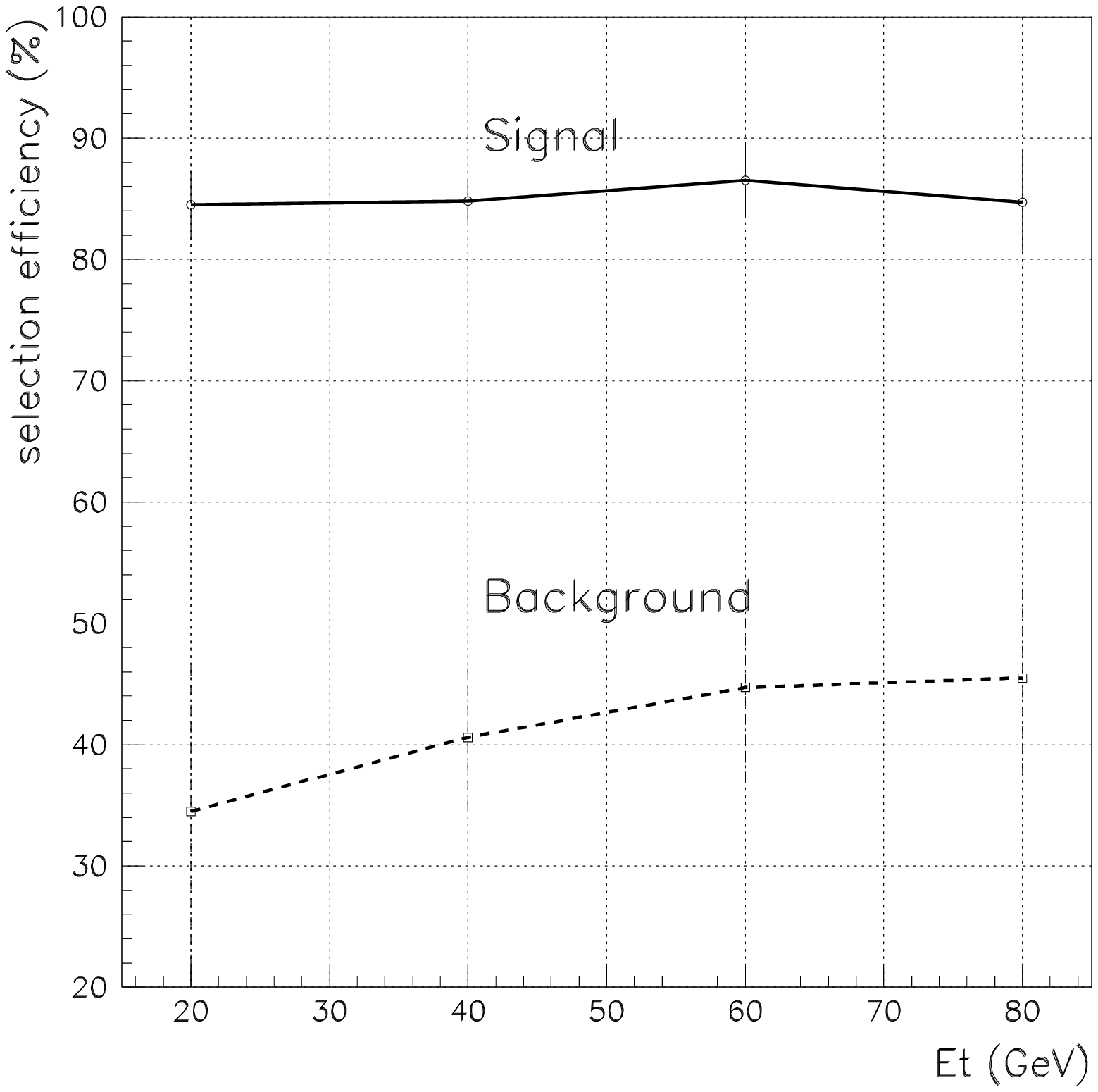}
\vspace{-0.7cm}
\caption{\hspace*{0.0cm} Selection efficiencies for 
photons from \gpj process and photon candidates from QCD background obtained after cut
on the number of 2-D clusters in the central preshower: $N_{clust}^{PS}\leq 1$.}
\label{fig:PSncl}
\vspace{-0.0cm}
\hspace{2.7cm} \includegraphics[width=13.0cm,height=12.0cm]{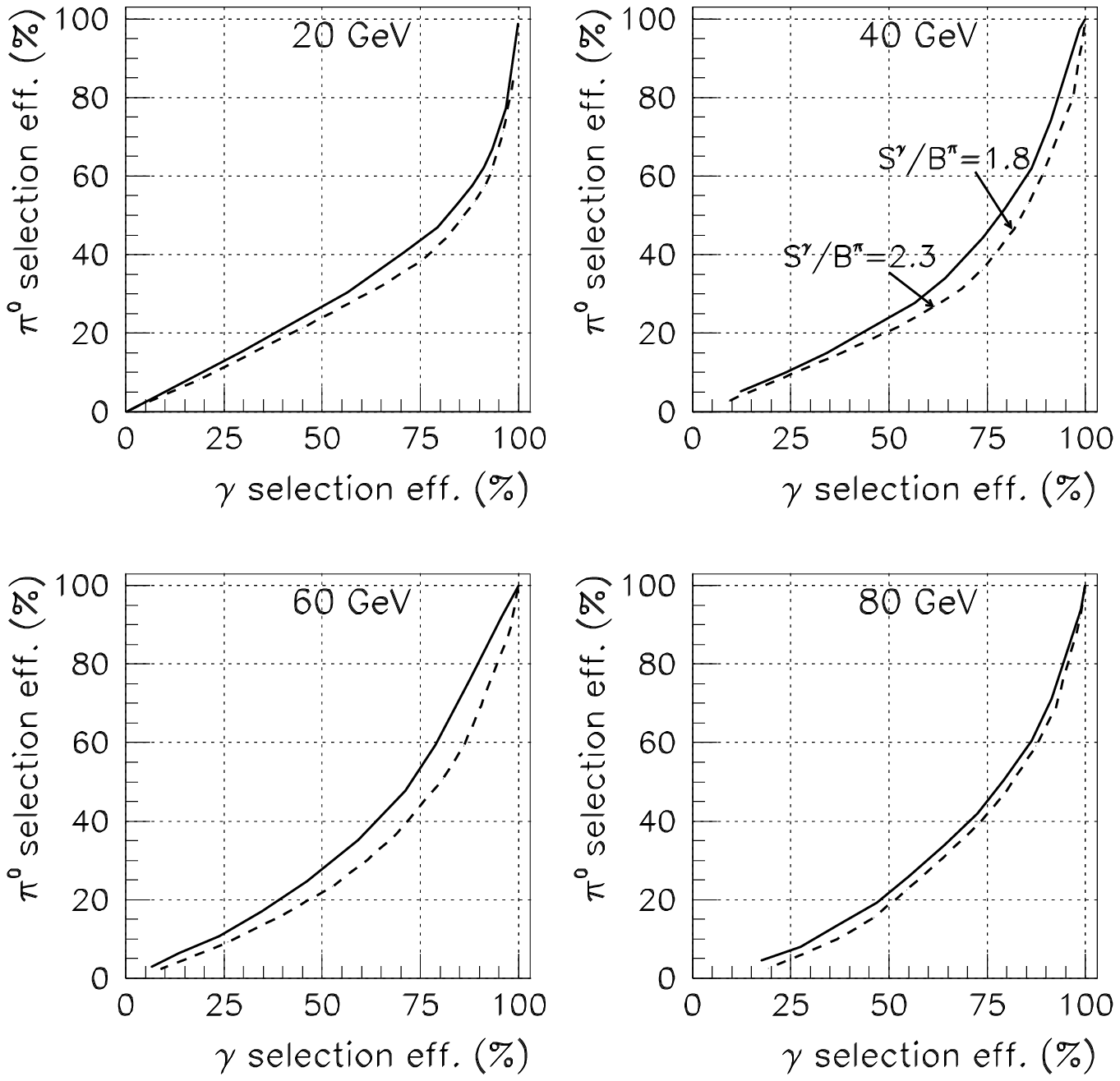}
\vspace{-0.7cm}
\caption{\hspace*{0.0cm} 
Selection efficiency of single photon via selection efficiency of $\pi^0$ obtained
by using two set of quantities, measured in the preshower detector:
three energy weighted widths of the single layer clusters (full line) and
the same plus three ratios of energy deposited in the hottest strip to the total energy 
of SLC clusters $E^{max}_{strp}/E_{SLC}$ (dashed line). 
Four $\Pt$ values, 20, 40, 60, 80, are considered on the plots above.}
\label{fig:PSeff}
\end{figure}

The values of signal-to-background ratios in Tables \ref{tab:sb2}, \ref{tab:sb3} 
are obtained without any detector effects. But these numbers can be noticeably increased
if we take into account information from the preshower detector
\footnote{Central (CPS) and forward (FPS) preshower detectors are placed at
$|\eta|<1.1$ and $1.2<|\eta|<2.5$, respectively, and have a similar 3-layered architecture
with set of triangular scintillator strips in every layer.}.
First of all, photons in the signal \gpj events have the distribution over number
of preshower 3-dimensional
\footnote{because they are built from 3 layers rotated in the space by some angles with respect 
to each other} 
clusters $N_{clust}^{PS}$ different from one
for photon candidates in the QCD background events. Selection efficiencies for
the signal and background events after application of the cut $N_{clust}^{PS}\leq 1$ is shown
in Fig.~\ref{fig:PSncl}  for $|\eta^{\tg}|<0.9$. Relatively big
numbers of $N_{clust}^{PS}$
in the QCD background may be explained by the facts that besides $\pi^0$'s we have
a contribution from events with multiphoton decays of $\eta, K^0_s$ and $\omega$ mesons
and that despite the strong isolation criteria photon candidates from the background events
still have a hadron accompaniment.

Additional rejection can be obtained after analysis of energy distributions among the strips
of each of three single layer clusters (SLC). They are again different for the signal and background 
events. As parameters for the discrimination one can take the energy weighted widths of three SLC's
and ratios of energy deposited in the hottest strip to the total energy of SLC cluster 
$E^{max}_{strp}/E_{SLC}$. 
The selection efficiencies for single $\gamma$'s and $\pi^0$'s (as a most difficult case
from the point of view of discrimination) are presented in Fig.~\ref{fig:PSeff}
\footnote{A consideration of the full QCD background left after our selection cuts (see cuts 
$1-16$ of Table \ref{tab:sb0} plus requirement $N_{clust}^{PS}\leq 1$ above) is very difficult because
of a pure statistics. Obviously should decrease ``$\gamma-mes$'' background selection due to 
the ``$\eta, K^0_s, \omega$ events'' contribution and probably due to an admixture of 
hadron accompaniment around $\gamma$-candidates in those events.}.

Thus, the total effect of data analysis in the preshower detector can lead to additional
increase in the $S/B$ of order of $3-4$
\footnote{these factors are caused mainly by the single photon selection efficiency and
$\Pt^{\tg}$ interval.}.

From  Tables \ref{tab:sb1} -- \ref{tab:sb3} we have seen that the cuts
listed in Table \ref{tab:sb0} (having rather moderate values of
$\Pt^{clust}_{CUT}$ and  $\Pt^{out}_{CUT}$) allow to suppress
the major part of the background events.
The influence of these two cuts  on: \\[1pt]
\hspace*{10mm} (a) the number of selected events (for $L_{int}=300\,pb^{-1}$);\\
\hspace*{10mm} (b) the signal-to-background ratio $S/B$;\\
\hspace*{10mm} (c) the mean value of 
$F\equiv (\Pt^{\tilde{\gamma}}\!-\!\Pt^{Jet})/\Pt^{\tilde{\gamma}}$ and
its  standard deviation value $\sigma (F)$\\[1pt]
is presented in Tables 1--12 of Appendix 3 for their variation in a wide range.

Let us emphasize that the tables of Appendix 3 include, in contrast to Appendix 2, the results
obtained after analyzing three generated samples (described in the beginning of this section)
of {\it signal and background} events. 
These events were selected with the cuts of Table \ref{tab:sb0}.

Namely, the cuts 1--10 of Table \ref{tab:sb0} were applied for preselection of 
``$\tilde{\gamma}+1~jet$'' events. 
The jets in these events as well as clusters were found by use of only one jetfinder LUCELL
(for the whole $\eta$ region $|\eta^{jet}|<4.2$).

Tables 1--4 of Appendix 3 correspond to the simulation with
$\pth=40 ~GeV/c$. Analogously, the values of $\pth=70 ~GeV/c$ and $\pth=100 ~GeV/c$ were used for
Tables 5--8  and Tables  9--12 respectively.
The  rows and  columns of Tables 1--12 illustrate, respectively, the influence of
$\Pt^{clust}_{CUT}$ and $\Pt^{out}_{CUT}$ on the quantities
mentioned above in the points (a), (b), (c).

First of all, we see from Tables 2, 6 and 10 of Appendix 3 that
a noticeable reduction
of the background take place while moving along the table diagonal from the right-hand bottom corner to the
left-hand upper one, i.e. with reinforcing $\Pt^{clust}_{CUT}$ and $\Pt^{out}_{CUT}$. 
So, we see that for $\pth=40 ~GeV/c$  the value of 
$S/B$ ratio changes in the table cells along the diagonal
from $S/B=2.2$ (in the case of no limits on these two variables), to $S/B=2.9$ for the
cell with $\Pt^{clust}_{CUT}=10~ GeV/c$ and $\Pt^{out}_{CUT}=10 ~GeV/c$.
Analogously, for $\pth=100 ~GeV/c\,$ the value of $S/B$ changes in the same table cells
from 10.0 to 29.5 (see Table 10 of Appendix 3)
\footnote{even better results produces a combined application of stronger cuts
$\Pt^{clust}_{CUT}=5~ GeV/c$ and $\Pt^{out}_{CUT}=5 ~GeV/c$ (see Appendix 3)}.

The second observation from Appendix 3. The restriction of $\Pt^{clust}_{CUT}$ and
$\Pt^{out}_{CUT}$ improves the calibration accuracy. Table 3 shows that in the interval 
$\Pt^{\tilde{\gamma}}\gt40~GeV/c$
the mean value of the fraction $F(\equiv (\Pt^{\tilde{\gamma}}\!-\!\Pt^{Jet})/\Pt^{\tilde{\gamma}})$
decreases from 0.049 (the bottom right-hand corner) to 0.024
for the table cell with $\Pt^{clust}_{CUT}=10~ GeV/c$ and $\Pt^{out}_{CUT}=10 ~GeV/c$.
At the same time, the both cuts lead to a noticeable decrease of
the Gaussian width $\sigma (F)$ (see Table 4 and also Tables 8, 12).  
For instance, for $\pth=40 ~GeV/c$ ~$\sigma (F)$ drops by about a factor of two: from 0.159 to 0.080.
It should be also noted that Tables 4, 8 and 12 demonstrate that for any fixed
value of $\Pt^{clust}_{CUT}$  
further improvement in $\sigma (F)$ can be achieved  by limiting $\Pt^{out}$ 
(e.g. in line with $\Pt^{clust}_{CUT}=10~GeV/c$
$\sigma (F)$ drops by a factor of 2 with variation of $\Pt^{out}$ from $1000$ to $5~GeV/c$).

The explanation is simple. The balance~ equation (\ref{eq:sc_bal}) contains 2 terms on the right-hand
side ($1-cos\dphi$) and $\Db/\Pt^{\tilde{\gamma}}$.
The first one is negligibly small in a case of Selection 1 and tends to decrease with growing 
$\Pt^{\tilde{\gamma}}$ (see tables in Appendix 2). So, we see that in this case
the main source of the disbalance in  equation (\ref{eq:sc_bal}) is the term $\Db/\Pt^{\tilde{\gamma}}$.
This term can be diminished by decreasing $\Pt$ activity beyond the jet,
i.e. by decreasing $\Pt^{out}$.

\begin{figure}[h]
\vspace{-1.6cm}
\hspace{.7cm} \includegraphics[width=14cm,height=15.5cm]{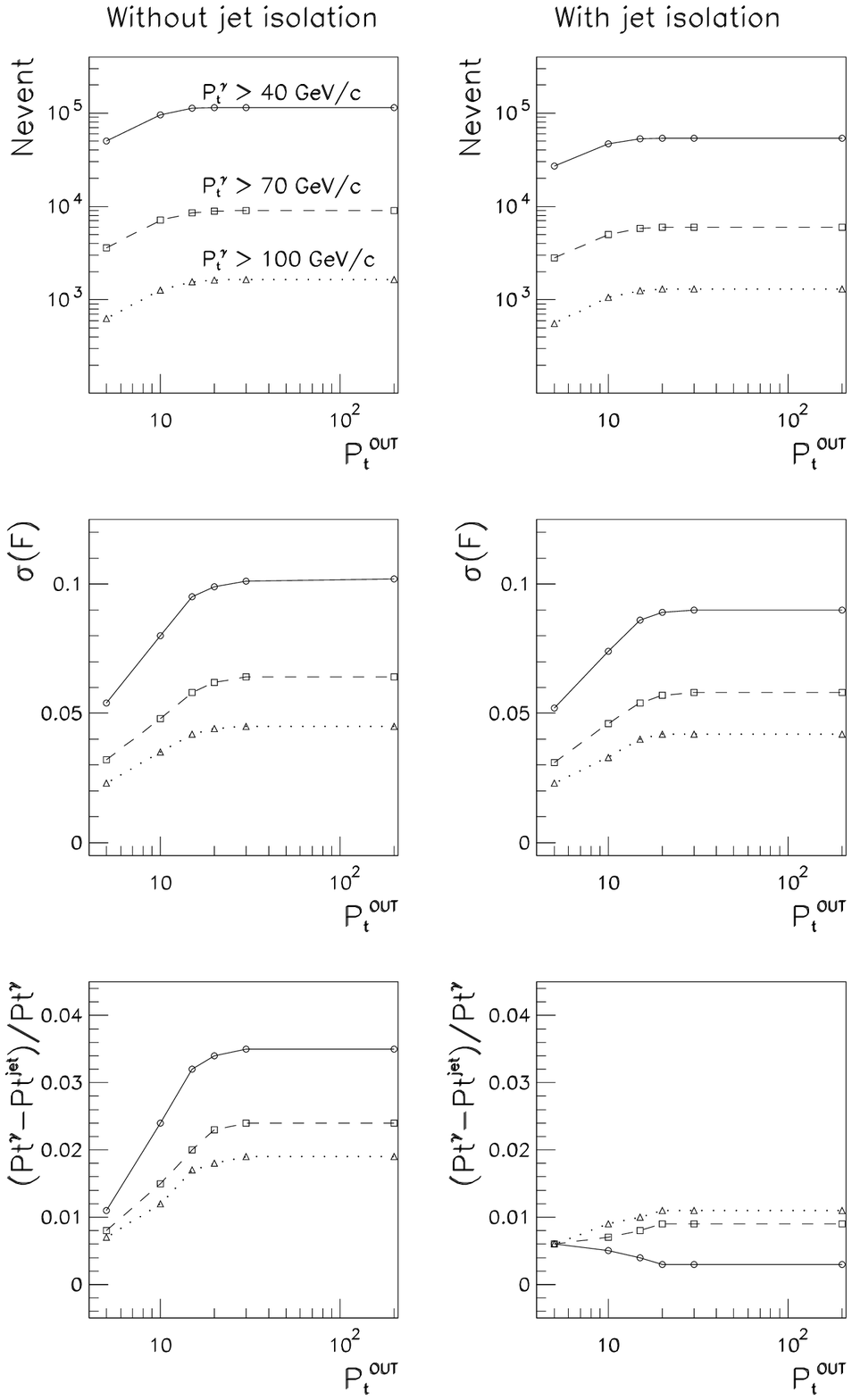}
\vspace{-0.7cm}
\caption{\hspace*{0.0cm} Number of events (for $L_{int}=300\,pb^{-1}$),
 mean value of $(\Pt^{\tilde{\gamma}}-\Pt^{Jet})/\Pt^{\tilde{\gamma}}$
($\equiv F$) and its standard deviation $\sigma(F)$ distributions over $\Pt^{out}$
for the cases of non-isolated (left-hand column) and isolated (right-hand column) jet and for
three  intervals: $\Pt^{\tilde{\gamma}} > 40, 70$ and $100 ~GeV/c$. $\Pt^{clust}_{CUT}=10 ~GeV/c.$
}
\label{fig:mu-sig}
\end{figure}

The behavior of the number of selected events (for $L_{int}=300\,pb^{-1}$),
 the mean values of $F=(\Pt^{\tilde{\gamma}}\!-\!\Pt^{Jet})/\Pt^{\tilde{\gamma}}$ and
its standard deviation $\sigma (F)$ as a function of $\Pt^{out}_{CUT}$ 
(with fixed $\Pt^{clust}_{CUT}=10~GeV/c$)
are also displayed in Fig.~\ref{fig:mu-sig} for events with non-isolated 
(left-hand column) and isolated jets (right-hand column, see also Tables 13--24 of Appendix 3).

{\it Thus, we can conclude that application of two criteria introduced
in Section 3.2, i.e. $\Pt^{clust}_{CUT}$ and $\Pt^{out}_{CUT}$,
results in two important consequences: significant background reduction
and essential improvement of the calibration accuracy.
}

The numbers of events (for $L_{int}= 300 ~pb^{-1}$)
for different $\Pt^{clust}_{CUT}$ and $\Pt^{out}_{CUT}$
are given in the cells of Tables 1, 5 and 9 of Appendix 3. One can see that even with such strict
$\Pt^{clust}_{CUT}$ and $\Pt^{out}_{CUT}$ values as, for example, $10 ~GeV/c$ for both
we would have a sufficient number of events
(about 100 000, 7 000 and 1 300 for $\Pt^{\tilde{\gamma}}\geq40 ~GeV/c$,
$\Pt^{\tilde{\gamma}}\geq70 ~GeV/c$ and  $\Pt^{\tilde{\gamma}}\geq100 ~GeV/c$, respectively)
with low background contamination ($S/B=2.9,~8.8$ and $21.1$)
and a good accuracy of the $\Pt^{\tilde{\gamma}}-\Pt^{Jet}$ balance:
$F=2.4\%, 1.5\%$ and $1.2\%$, respectively, for the case of Selection 1.

In addition, we also present Tables 13--24 of Appendix 3.
They contain the information analogous to that in Tables 1 -- 12
but for the case of isolated jets with $\epsilon^{jet}<3\%$.
From these tables we see that with the same cuts
 $\Pt^{clust}_{CUT}=\Pt^{out}_{CUT}=10 ~GeV/c$ one can expect about\\
47 000, 5 000 and 1000 events for $\Pt^{\tilde{\gamma}}\geq40 ~GeV/c$,
$\Pt^{\tilde{\gamma}}\geq70 ~GeV/c$ and  $\Pt^{\tilde{\gamma}}\geq100 ~GeV/c$, 
respectively, with a much more better fractional $\Pt^{\tilde{\gamma}}-\Pt^{Jet}$ balance:
$F=0.5\%, 0.7\%$ and $0.1\%$.

Let us mention that all these PYTHIA results give us an indication of a tendency
and may serve as a guideline for further full GEANT simulation that would allow to come to 
a final conclusion. 

To conclude this section we would like to stress, firstly, that, as is seen from
Tables~\ref{tab:sb1}, the  ``$\gamma-brem$'' background
defines a dominant part of the total background. 
One can see from Table ~\ref{tab:sb1}
that $\pi^0$ contribution being about the same as
``$\gamma-brem$'' at $\pth>40~GeV/c$ becomes three times less than
``$\gamma-brem$'' contribution at $\pth>100~GeV/c$. We would like to emphasize here
that this is a strong prediction of the PYTHIA generator that has to be compared with
predictions of another generator like HERWIG, for example.
%
%

Secondly, we would like to mention also that, as it is seen from Tables \ref{tab:sb4} and
\ref{tab:sb1},  the photon isolation and selection cuts 1--6, usually used in 
the study of inclusive  photon production (see, for instance, \cite{CDF1}, \cite{D0_1}, 
\cite{D0_2}), increase the $S/B$ ratio up to 3.20 only (for $\Pt^{\tilde{\gamma}}\geq70 ~GeV/c$).
The other cuts 6--17, that select events with a clear \gpj topology and
limited $\Pt$ activity beyond \gpj system, lead to quite a significant improvement of
$S/B$ ratio by a factor of three (to $S/B=10.07$).

The numbers in the tables of Appendix 3 were obtained with inclusion of the contribution
from the background events. The tables show that their account does not spoil the \ptgj
balance in the event samples preselected with the cuts 1--10 of Table \ref{tab:sb0}. 
The estimation of the number of these background events would be important
for the gluon distribution determination (see Section 9).

\normalsize
\def\baselinestretch{1.0}

%
\section{\gpj EVENT RATE ESTIMATION FOR GLUON DISTRIBUTION DETERMINATION AT THE TEVATRON RUN~II.}

\it\small
\hspace*{9mm}
The number of \gpj events suitable for measurement of gluon distribution in different
$x$ and $Q^{\;2}$ intervals at Run~II is estimated. It is shown that with $L_{int}=3~fb^{-1}$
it would be possible to collect about one million of these events. This number
would allow to cover a new kinematical area not studied in any previous experiment
($10^{-3}\lt x\lt 1.0$ with $1.6\cdot 10^{3}\leq Q^2\leq2\cdot10^{4} ~(GeV/c)^2$).
This area in the region of small $x\geq10^{-3}$ has $Q^2$ by about one order of magnitude higher than 
reached at HERA now.
\rm\normalsize
\vskip4mm

As many of theoretical predictions for production of new particles
(Higgs, SUSY) at the Tevatron are based on model estimations of the gluon density behavior at
low $x$ and high $Q^2$, the measurement of the proton gluon density
for this kinematic region directly in Tevatron experiments
would be obviously useful. One of the promising channels for this measurement, as was shown in ~\cite{Au1},
is a high $\Pt$ direct photon production $p\bar{p}(p)\rightarrow \gamma^{dir} + X$.
The region of high $\Pt$, reached by UA1 \cite{UA1}, UA2 \cite{UA2}, CDF \cite{CDF1} and
D0 \cite{D0_1} extends up to $\Pt \approx 80~ GeV/c$ and recently up to $\Pt=105~GeV/c$ \cite{D0_2}. 
These data together with the later ones (see references in \cite{Fer}--\cite{Fr1} and recent
E706 \cite{E706} and UA6 \cite{UA6} results) give an opportunity for tuning the form of gluon 
distribution (see \cite{Au2}, \cite{Vo1}, \cite{Mar}). The rates and estimated cross sections 
of inclusive direct photon production at the LHC were given in \cite{Au1} (see also \cite{AFF}).

Here for the same aim we shall consider 
the process $p\bar{p}\rightarrow \gamma^{dir}\, +\, 1\,Jet\, + \,X$
defined in the leading order by two QCD subprocesses (1a) and (1b)
(for experimental results see \cite{ISR}, \cite{CDF2}).

Apart from the advantages, discussed in Section 8  in connection with the background suppression 
(see also \cite{Ber}--\cite{Hu2}),
the ``$\gamma^{dir}+1~Jet$'' final state may be easier for physical analysis
than inclusive photon production process ``$\gamma^{dir}+X$''  if we shall look at this problem from
the viewpoint of extraction of information on the gluon distribution in a proton.
Indeed, in the case of inclusive direct photon production the
cross section is given as an integral over the products of a fundamental $2\to2$ parton subprocess
cross sections and the corresponding parton distribution functions $f_a(x_a,Q^2)$ (a = quark or gluon), 
while in the case of $p\bar{p}\rightarrow \gamma^{dir}+1~Jet+X$ for $\Pt^{Jet}\,
\geq \,30\, GeV/c$ (i.e. in the region where ``$k_t$ smearing effects''
\footnote{This terminology is different from ours, used in Sections 2 and 9, as we denote by ``$k_t$''
only the value of parton intrinsic transverse momentum.}
are not important, see \cite{Hu3}) the cross section is
expressed directly in terms of these distributions (see, for example,
\cite{Owe}): \\[-16pt]
\begin{eqnarray}
\frac{d\sigma}{d\eta_1d\eta_2d\Pt^2} = \sum\limits_{a,b}\,x_a\,f_a(x_a,Q^2)\,
x_b\,f_b(x_b,Q^2)\frac{d\sigma}{d\hat{t}}(a\,b\rightarrow c\,d),
\label{gl:4}
\end{eqnarray}
\vskip-3mm
\noindent
where \\[-9mm]
\begin{eqnarray}
x_{a,b} \,=\,\Pt/\sqrt{s}\cdot \,(exp(\pm \eta_{1})\,+\,exp(\pm \eta_{2})).
\label{eq:x_def}
\end{eqnarray}
\vskip-1mm

The designation used above are as the following:
$\eta_1=\eta^\gamma$, $\eta_2=\eta^{Jet}$; ~$\Pt=\Pt^\gamma$;~ $a,b=q,\bar{q},g$; 
$c,d=q,\bar{q},g,\gamma$. Formula (\ref{gl:4}) and the knowledge of 
$q, \,\bar{q}$ distributions 
allow the gluon  distribution $f_g(x,Q^2)$
to be determined after account of selection efficiencies for jets and  $\gamma^{dir}-$candidates 
as well as after subtraction of the background contribution, 
left after the used selection cuts 1--13 of Table \ref{tab:sb0}
(as it was discussed in Section 8 keeping in hand this physical application).

In the previous sections a lot of details connected with the
structure and topology of these events and the features of objects appearing
in them were discussed. Now with this information in mind we are
in position to discuss an application of the \gpj
event samples, selected with the previously proposed cuts, for estimating
the rates of the gluon-based  subprocess (1a) in different $x$ and $Q^2$ intervals.

Table~\ref{tab:q/g-1} shows percentage of ``Compton-like'' subprocess
(1a) (amounting to $100\%$ together with (1b)) in the samples of events selected with cuts
(16)--(22) of Section 3.2 for $\Pt^{clust}_{CUT}=10~GeV/c$ for different $\Ptg$
and $\eta^{jet}$ intervals: Central (CC) ($|\eta^{jet}|\lt0.7$)
\footnote{see also tables of Appendix 1 containing lines ``$29sub/all$''},
Intercryostat (IC) $0.7\lt|\eta^{jet}|\lt1.8$ and End (EC) $1.8\lt|\eta^{jet}|\lt2.5$ parts of calorimeter.
We see that the contribution of Compton-like subprocess grows by about $5-6\%$ with $|\eta^{jet}|$
enlarging and drops with growing $\Pt^{jet}(\approx\Ptg$ in the sample of the events collected with the cuts 
$1-13$  of Table \ref{tab:sb0}).
\\[-10mm]
\begin{table}[h]
\begin{center}
\caption{The percentage of Compton-like process  $q~ g\rrr \gamma +q$.}
\normalsize
\vskip.1cm
\begin{tabular}{||c||c|c|c|c|}                  \hline \hline
\label{tab:q/g-1}
Calorimeter& \multicolumn{4}{c|}{$\Pt^{Jet}$ interval ($GeV/c$)} \\\cline{2-5}
    part   & 40--50 & 50--70 & 70--90 & 90--140   \\\hline \hline
CC         & 84     &  80   &  74&  68  \\\hline
IC         & 85     &  82   &  76&  70  \\\hline
EC         & 89     &  85   &  82&  73  \\\hline
\end{tabular}
\end{center}
\end{table}
\normalsize
~\\[-10mm]


In Table~\ref{tab:q/g-2} we present 
distribution of the number of events that are caused by the $q~ g\rrr \gamma +q$ subprocess,
in various intervals of the $Q^2 (\equiv(\Ptg)^2)$
\footnote{see \cite{PYT}}
and $x$ (defined according to (\ref{eq:x_def})). These events have 
passed the following cuts ($\Pt^{out}$ was not limited):\\[-8mm]
\begin{eqnarray}
\Ptg>40~ GeV/c,\quad |\eta^{\gamma}|<2.5,\quad \Pt^{Jet}>30~ GeV/c,\quad |\eta^{Jet}|<4.2,\quad \Pt^{hadr}>7~ GeV/c,
\label{l1}
\nonumber
\end{eqnarray}
~\\[-15mm]
\begin{eqnarray}
\Pt^{isol}_{CUT}=4\;GeV/c, \;
{\epsilon}^{\gamma}_{CUT}=7\%, \;
\dphi<17^{\circ}, \; 
\Pt^{clust}_{CUT}=10\;GeV/c. \;
\label{l2}
\end{eqnarray}

\begin{table}[h]
\begin{center}
\vskip0.2cm
\caption{Number of~ $g\,q\to \gamma^{dir} \,+\,q$~
events at different $Q^2$ and $x$ intervals for $L_{int}=3~fb^{-1}$.}
\label{tab:q/g-2}
\vskip-0.2cm
\begin{tabular}{|lc|r|r|r|r|r|r|r|}                  \hline
 & $Q^2$ &\multicolumn{6}{c|}{ \hspace{-0.9cm} $x$ values of a parton} &All $x$  \\\cline{3-9}
& $(GeV/c)^2$  &$.001-.005$ & $.005-.01$ & $.01-.05$ &$.05-.1$ & $.1-.5$ &$.5-1.$
&$.001-1.$     \\\hline
&\hmm\hmm 1600-2500\hmm  & 8582  & 56288  &245157  &115870  &203018  &  3647  &632563  \\\hline
&\hmm\hmm 2500-4900\hmm  &  371  & 13514  &119305  & 64412  &119889  &  3196  &320688 \\\hline
&\hmm\hmm 4900-8100\hmm  &    0  &   204  & 17865  & 13514  & 26364  &  1059  & 59007\\\hline
&\hmm\hmm 8100-19600\hmm &    0  &     0  &  3838  &  5623  & 11539  &   548  & 21549\\\hline
\multicolumn{8}{c|}{}&{\bf 1 033 807}\\\cline{9-9}
\end{tabular}
\end{center}
\end{table}

The analogous information for events with the charmed quarks in the initial state
$g\,c\to \gamma^{dir} \,+\,c$ is presented in Table~\ref{tab:q/g-3}. 
The simulation of the process $g\,b\to \gamma^{dir}\,+\,b$ $\;$ has shown that the rates
for the $b$-quark are 8 -- 10 times smaller than for
the $c$-quark. These event rates are also given in Appendix 1 for different $\Ptg$ intervals
in the lines denoted by ``$Nevents(c/b)$''
\footnote{Analogous estimation for LHC energy was done in \cite{BKS_GLU} and \cite{MD1}.}. 

\begin{table}[h]
\begin{center}
\vskip-.2cm
\caption{Number of~ $g\,c\to \gamma^{dir} \,+\,c$~
events at different $Q^2$ and $x$ intervals for $L_{int}=3~fb^{-1}$.}
\label{tab:q/g-3}
\vskip0.2cm
\begin{tabular}{|lc|r|r|r|r|r|r|r|}                  \hline
 & $Q^2$ &\multicolumn{6}{c|}{ \hspace{-0.9cm} $x$ values of a parton} &All $x$\\\cline{3-9}
& $(GeV/c)^2$  &$.001-.005$ & $.005-.01$ & $.01-.05$ &$.05-.1$ & $.1-.5$ &$.5-1.$
&$.001-1.$     \\\hline
&\hmm\hmm 1600-2500\hmm  &  264  &  2318  & 21236  & 11758  & 14172  &    58  & 49805 \\\hline
&\hmm\hmm 2500-4900\hmm  &   13  &   332  &  9522  &  6193  &  7785  &    40  & 23885  \\\hline
&\hmm\hmm 4900-8100\hmm  &    0  &     4  &   914  &  1055  &  1648  &    16  &  3637\\\hline
&\hmm\hmm 8100-19600\hmm &    0  &     0  &   142  &   329  &   612  &     8  &  1092 \\\hline
\multicolumn{8}{c|}{}&{\bf 78 419}\\\cline{9-9}
\end{tabular}
\end{center}
\end{table}

~\\[10mm]
\begin{flushleft}
\begin{figure}[h]
   \vskip-30mm
   \hspace{-1mm} \includegraphics[width=.55\linewidth,height=9.0cm,angle=0]{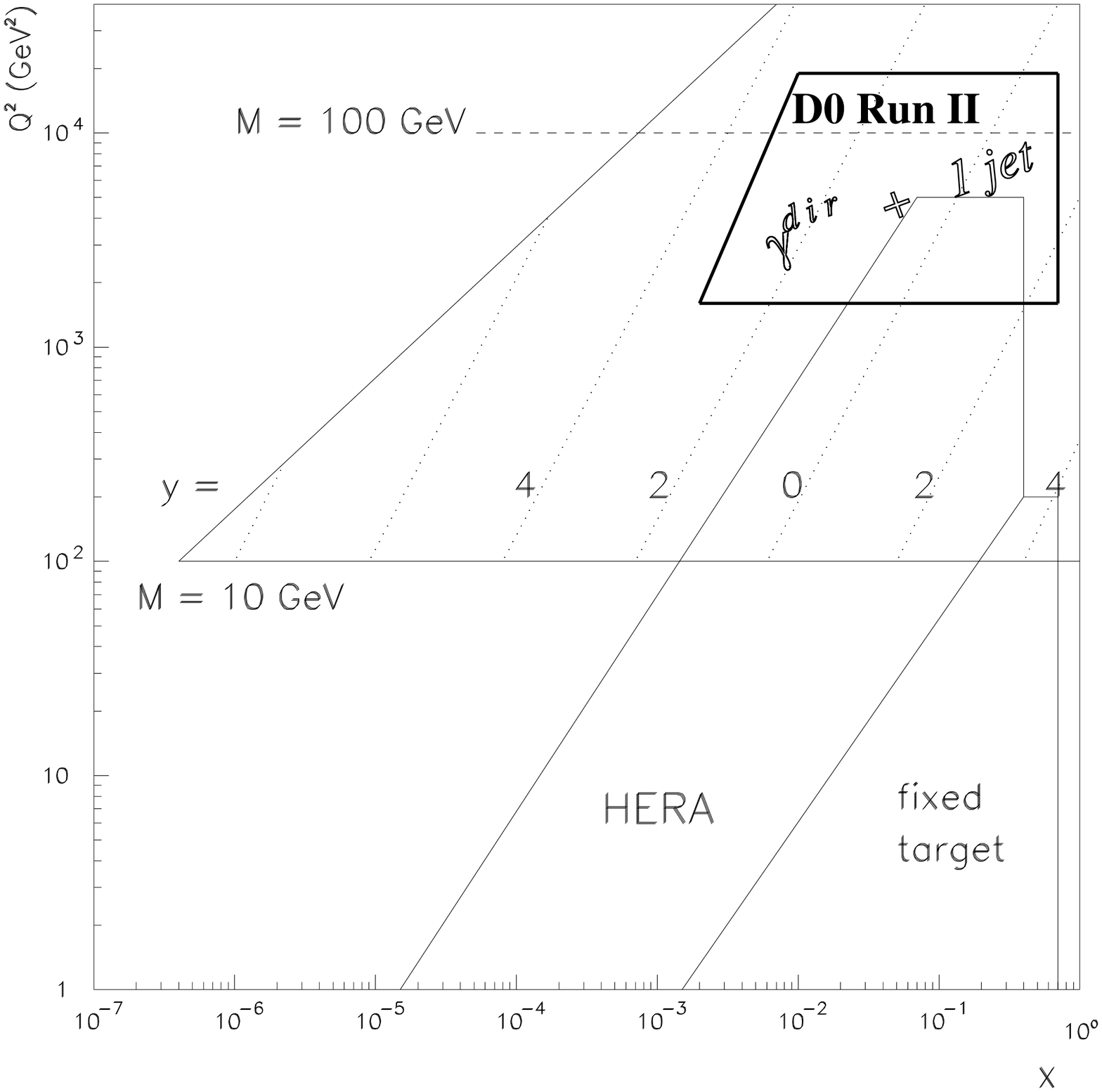}
\label{fig:q/g}
\end{figure}
\end{flushleft}
\begin{flushright}
\vskip-108mm
\parbox[r]{.42\linewidth}
{ \vspace*{0.1cm}
 \hspace{8mm} Fig.~17 shows in the widely used $(x,Q^2)$
kinematic plot (see \cite{Sti} and also in  \cite{Hu3})
what area can be covered by studying the process $q~ g\rrr \gamma +q$ at Tevatron.
The distribution of number of events in this area is given by Table~\ref{tab:q/g-2}.
From this figure and Table~\ref{tab:q/g-2} it becomes clear that with integrated luminosity
$L_{int}=3~fb^{-1}$ it would be possible to study the gluon distribution with a good statistics 
of \gpj events in the region of  $10^{-3}\lt x\lt 1.0$ with $Q^2$ by about 
one order of magnitude higher than reached at HERA now.
It is worth emphasizing that extension of the experimentally reachable
region at the Tevatron to the region of lower $Q^2$ overlapping with the area
covered by HERA would also be of great interest.}
\end{flushright}

{\vskip0.1cm
\hspace*{-.5cm} {Fig.~17: \footnotesize {The  $(x,Q^2)$ kinematic region for
studying~ $p\bar{p}\to \gamma+Jet$ process at Tevatron Run~II.}}
}

~\\[1pt]

~\\[-2mm]

%
\section{SUMMARY.}

We have done an attempt here to consider, following 
\cite{9}--\cite{BKS_GLU}, the physics of high $\Pt$ direct photon and jet 
associative production in proton-antiproton collisions
basing on the predictions of PYTHIA generator and the models implemented there.
This work may be useful for two practical goals: for absolute jet energy scale determination
and for gluon distribution measurement at Tevatron energy.

The detailed information provided in the PYTHIA event listings allows to track the origin of
different particles (like photons) and of objects (like clusters and jets) 
that appear in the final state. 
So, the aims of this work was to explore at the particle level as much as
possible this information for finding out what effect may be produced
by new variables, proposed in \cite{9}--\cite{BKS_P5} for selection of \gpj events,  
and the cuts on them for solution of the mentioned above practical tasks.

For the first problem of the jet energy determination an important task is to select the events
that may be caused (with a high probability) by the $q\bar{q}\to g+\gamma$ and 
$qg\to q+\gamma$~ fundamental parton subprocesses of direct photon production. 
To take into account a possible effect
of initial state radiation (its spectra are presented in different \ptg ~intervals 
in Section 5) we used here the $\Pt$-balance equation (see (\ref{eq:vc_bal})) written for
an event as a whole. It allows to express \ptgj fractional disbalance (see (\ref{eq:sc_bal})) 
through new variables
\cite{9}--\cite{BKS_P5} that describe the $\Pt$ activity {\it out} of \gpj system. They are $\Pt^{out}$ and
$\Pt^{clust}$, i.e. $\Pt$  of mini-jets or clusters that are additional to the main jet in event.
The latter is the most ``visible'' part of $\Pt^{out}$.

The sources of the \ptgj disbalance are investigated. It is shown that the limitation of $\Pt$
of clusters, i.e. $\Pt^{clust}$,  can help to decrease this disbalance.
Analogously, the limitation of $\Pt$ activity of all detectable particles ($|\eta_i|\lt4.2$)
 beyond the \gpj system, 
i.e. $\Pt^{out}$, also leads to a noticeable \ptgj disbalance reduction (see Sections 7,8).

It is demonstrated that in the events selected by means of simultaneous restriction  from above
of the $\Pt^{clust}$ and $\Pt^{out}$ activity the values of \Ptgj are well balanced with each other.
 The samples of these \gpj events gained
in this way are of a large enough volume for the jet energy scale determination in the interval
$40\lt\Ptg\lt140~GeV/c$ (see Tables 1--12 of Appendix 3). 

It is worth mentioning that the most effect for improvement of \Ptgj balance can be reached by applying 
additionally the jet isolation criterion defined in \cite{9}--\cite{BKS_P5}. 
As it can be seen from
Tables 13--18 of Appendix 2 and Tables 13--24 of Appendix 3, the application of this criterion
allows to select the events having the \ptgj disbalance at the particle level less than $1\%$.
%
%
Definitely, the detector effects may worsen the balance determination
due to a limited accuracy of the experimental measurement
\footnote{We are planning to present the results of full GEANT simulation with 
the following digitization and reconstruction of signals by using the corresponding
D0 packages in the forthcoming papers.}.

We present also PYTHIA predictions for the dependence of the distributions of the number of 
selected \gpj events on \ptg~ and $\eta^{jet}$
(see Section 5 and also tables of Appendix 2 with account of $\Pt^{clust}$ 
variation). 

The corrections to a jet $\Pt$
the measurable values of $\Pt^{jet}$ that have take into account the contribution 
from neutrinos belonging to a jet are presented for different $\Pt^{Jet}(\approx\Ptg$ for 
the selected events$)$ intervals
in the tables of Appendix 1. It is shown in Section 4 that a cut on $\Pt^{miss}\lt10~GeV/c$ allows 
to reduce the neutrino contribution to the value of
$\la\Pt^{Jet}_{(\nu)}\ra_{all\; events}=0.1~GeV/c$. 

At the same time, as it is shown in  
\cite{QCD_talk}, and discussed in Section 8 (see also \cite{BKS_P5}), 
this cut noticeably decreases the number of the background $e^\pm$-events 
in which $e^\pm$ (produced in the $W^\pm\to e^\pm\nu$ weak decay) may be registered as direct photon.


The possibility of the background events (caused by QCD subprocesses of $qg, gg, qq$ scattering) 
suppression  was studied in Section 8. Basing on the introduced selection
criteria that include 17 cuts (see Table \ref{tab:sb0} of Section 8),
the background suppression relative factors and the values of signal event selection
efficiencies  are estimated (see Tables \ref{tab:sb4}-\ref{tab:sb3}). 

It is shown that after applying the first 6 ``photonic'' cuts 
(that may be used, for example, for selecting events with inclusive photon production and lead to
$S/B$ ratio equal to $3.2$ for $\Ptg>70 ~GeV/c$, see Table \ref{tab:sb4}) 
the use of the next 11 ``hadronic'' cuts of Table \ref{tab:sb0} may lead to further essential improvement 
of $S/B$ ratio (by factor of 3.2 for the same $\Ptg>70 ~GeV/c$ where $S/B$ becomes $10.1$.

It is important to underline that this improvement is achieved by applying ``hadronic'' cuts
that select the events having clear \gpj topology at the particle level and also having rather ``clean'' area
(in a sense of limited $\Pt$ activity) beyond a \gpj system.
In this sense and taking into account the fact that these ``hadronic'' cuts lead 
to an essential improvement of \ptgj balance,
one may say that the cuts on $\Pt^{clust}$ and $\Pt^{out}$, considered here,
do act quite effectively to select the events caused by leading order diagrams (see Fig.~1) and do suppress the
contribution of NLO diagrams, presented in Figs.~2, 4.

The consideration of the cuts, connected with detector effects 
(e.g., based the preshower usage) may lead to  further noticeable improvement of $S/B$ ratio.

Another interesting predictions of PYTHIA is about the dominant contribution of
``$\gamma$-brem'' events into the total background at Tevatron energy, as in was already mentioned in Section 8
(see also \cite{BKS_P5} and \cite{QCD_talk}).  As the ``$\gamma$-brem'' background has irreducible nature
its careful estimation is an important task and requires the analogous estimation with another generator. 

To finish the discussion of the jet calibration study let us mention that the main results on this subject 
are summed up in Tables 1--12 (Selection 1) and 13--24 (Selection 2 with jet isolation criterion)
of Appendix 3 and Fig.~\ref{fig:mu-sig}.

It should be emphasized that numbers presented in all mentioned tables and figures were found
within the PYTHIA particle level of simulation. They may depend on the used generator and on the particular
choice of a long set of its parameters
\footnote{We have already mentioned that we are planning
to perform analogous analysis by help of another generator like HERWIG, for example.
The comparison of predictions of different generators (PYTHIA, HERWIG, etc.) with the experimental
results is a part of a work in any experiment.}
as well as they may change after account of the results of the full GEANT-based simulation.


It is also shown that the samples of the \gpj events, gained with the cuts used for the jet energy calibration,
can provide an information suitable also for determining the
gluon distribution inside a proton in the kinematic region (see Fig.~17) that includes
$x$ values as small as accessible at HERA  \cite{H1}, \cite{ZEUS}, but
at much higher $Q^2$ values (by about one order of magnitude):
$10^{-3}\leq x \leq 1.0$ with $1.6\cdot10^3\leq Q^2\leq2\cdot10^4 ~(GeV/c)^2$.
The number of  events, based on the gluonic process (1a), that may be collected 
with $L_{int}=3~fb^{-1}$ in different $x$- and $Q^2$- intervals of this new kinematic region
for this goal are presented in Table \ref{tab:q/g-2}
(all quarks included) and in Table \ref{tab:q/g-3} (only for charm quarks).
%
%

~\\[-1mm]
\section{ACKNOWLEDGMENTS.}                                         
\normalsize
\rm
We are greatly thankful to D.~Denegri who initiated our interest to study the physics of
\gpj processes, for his permanent support and fruitful suggestions.
It is a pleasure for us to express our deep recognition for helpful discussions to P.~Aurenche,
M.~Dittmar, M.~Fontannaz, J.Ph.~Guillet, M.L.~Mangano, E.~Pilon,
H.~Rohringer, S.~Tapprogge, H.~Weerts and J.~Womersley. Our special gratitude is
to J.~Womersley also for supplying us with the preliminary version of paper [1],
interest in the work and encouragements.

~\\[1mm]


\def\baselinestretch{.95}


\end{document}